\RequirePackage{fix-cm}
\documentclass[smallextended]{svjour3}       
\smartqed  
\usepackage{graphicx}

\usepackage{amsmath}
\usepackage{amsfonts}
\usepackage{amssymb}
\usepackage{mathrsfs}
\usepackage{bm}
\usepackage{graphicx}
\usepackage{xcolor}
\usepackage{epsfig}
\usepackage{setspace}
\usepackage{cite}
\usepackage{pdfpages}
\usepackage{hyperref}
\usepackage{afterpage}
\usepackage[toc,page]{appendix}
\usepackage{cancel}
\usepackage{array,multirow,makecell}

 \newcommand{\be}{\begin{equation}}
 \newcommand{\ee}{\end{equation}}

 \newcommand{\lambdabar}{{\mkern0.75mu\mathchar '26\mkern -9.75mu\lambda}}

\begin{document}

\title{Phase-space modelling of solid-state plasmas}
\subtitle{A journey from classical to quantum}


\author{Giovanni Manfredi \and Paul-Antoine Hervieux \and J\'er\^ome Hurst
}


\institute{G. Manfredi \at
              Universit\'e de Strasbourg, CNRS, Institut de Physique et Chimie des Mat\'eriaux de Strasbourg, UMR 7504, F-67000 Strasbourg, France \\
              \email{giovanni.manfredi@ipcms.unistra.fr}           
           \and
           P.-A. Hervieux \at
              Universit\'e de Strasbourg, CNRS, Institut de Physique et Chimie des Mat\'eriaux de Strasbourg, UMR 7504, F-67000 Strasbourg, France\\
              \email{paul-antoine.hervieux@ipcms.unistra.fr}
                         \and
           J. Hurst \at
              Department of Physics and Astronomy, Uppsala University, Box 516, SE-75120 Uppsala, Sweden\\
              \email{jerome.hurst@physics.uu.se}
}

\date{Received: date / Accepted: date}

\maketitle

\begin{abstract}
Conduction electrons in metallic nano-objects ($\rm 1\,nm = 10^{-9}\, m$) behave as mobile negative charges confined by a fixed positively-charged background, the atomic ions. In many respects, this electron gas displays typical plasma properties such as screening and Langmuir waves, with more or less pronounced quantum features depending on the size of the object. To study these dynamical effects, the mathematical artillery of condensed-matter theorists mainly relies on wave function $\psi(\bm{r},t)$ based methods, such as the celebrated Hartree-Fock equations.
The theoretical plasma physicist, in contrast, lives and breaths in the six-dimensional phase space, where the electron gas is fully described by a probability distribution function $f(\bm{r},\bm{p},t)$ that evolves according to an appropriate kinetic equation.
Here, we illustrate the power and flexibility of the phase-space approach to describe the electron dynamics in small nano-objects. Starting from classical and semiclassical scenarios, we progressively add further features that are relevant to solid-state plasmas: quantum, spin, and relativistic effects, as well as collisions and dissipation.
As examples of  applications, we study the spin-induced modifications to the linear response of a homogeneous electron gas and the nonlinear  dynamics of the electrons confined in a thin metal films of nanometric dimensions.

\end{abstract}


\section{Introduction -- Plasmas and nanophysics}
\label{sec:intro}

\begin{figure*}
	\centering \includegraphics[scale=0.35]{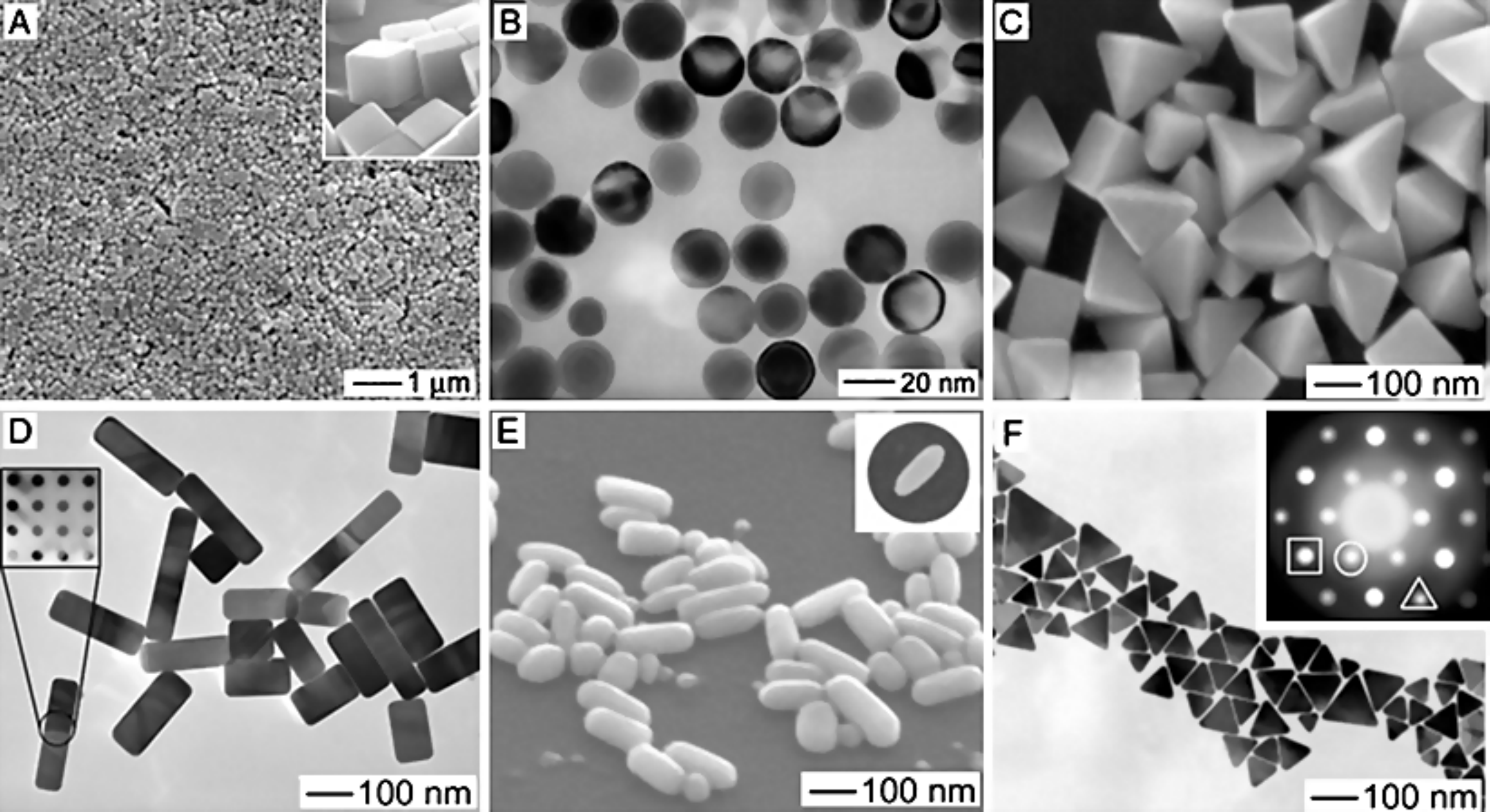}
    \caption{Some examples of silver nano-objects with different shapes and sizes. From \cite{Cobley2009}. }\label{nano_fig}
\end{figure*}

The physics of metallic nano-objects has stimulated a vast amount of scientific interest in the last two decades, both for fundamental research and for potential technological applications to nanophotonics \cite{Stockman2011,Moreau2012}, physical chemistry \cite{Daniel2004}, and even biology and medicine \cite{Hainfeld2004,TatsuroEndo2006}.
Metallic nano-objects are mesoscopic systems composed of a relatively small number of atoms, typically between a few tens and several millions. The typical size of these systems ranges from a few to several hundred nanometres ($\rm 1\,nm = 10^{-9}\, m$), with properties that  are intermediate  between  those  of  molecules  and  bulk solids. They can be synthesized in different geometries, ranging from spherical nanoparticles, to thin films, nanorods, cubes or pyramids (see Fig. \ref{nano_fig}).

Metallic nano-objects also present a fundamental interest as large objects that still display some quantum features \cite{Scholl2012,Luo2013,Tame2013,Raza2013}. Quantum effects arise because at metallic densities ($n \approx 10^{28}\, \rm m^{-3}$) the electrons are so closely packed together that their wave functions overlap even at room temperature, but also because of finite-size effects and the related presence of surfaces where the density varies significantly across very short distances (less than 1~nm).

Many of the optical properties of metallic nano-objects are mediated by their conduction electrons. The conduction electrons in a metal are delocalized over comparatively large distances, so that they can be treated, at least to first approximation, as an electron plasma confined by a fixed positively-charged background, the atomic ions. Such plasma is known to display collective effects, giving rise to typical plasma properties such as pronounced resonances near the plasma frequency $\omega_{p} = \sqrt{n e^{2}/m \epsilon_{0}}$. Due to the very large electron density, the plasma period $\tau_p=2\pi/\omega_{p}$ lies in the femtosecond range ($1\rm\, fs = 10^{-15}\,\rm s$). Thus, it is not surprising that the study of collective electron resonances in nano-objects became possible following the development of ultrafast spectroscopy techniques (``pump-probe experiments") in the femtosecond regime \cite{Voisin2000,Bigot2000}. Today, plasmonic resonances are actively investigated for applications to such diverse areas  as biomedicine \cite{Salata2004,Loomba2013} and  high-harmonic generation \cite{Butet2010,Singhal2010}.

\begin{figure}
	\centering \includegraphics[scale=0.3]{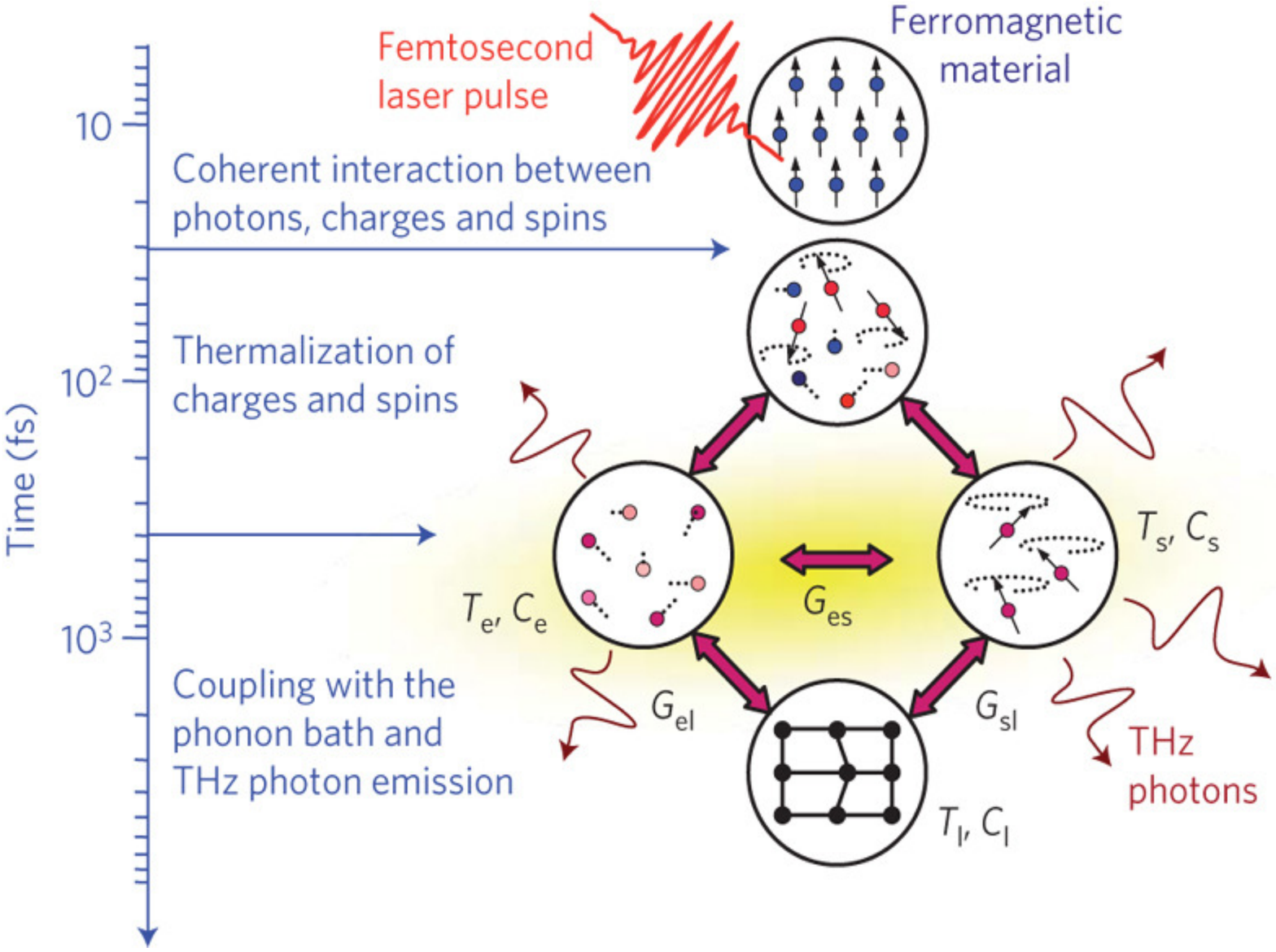}
    \caption{Schematic diagram of the different physical processes and the associated timescales following the excitation of a ferromagnetic nano-object by a femtosecond laser pulse. From \cite{Bigot2009}.}\label{time_scale_fig}
\end{figure}

\paragraph{Spectroscopy experiments.}
Pump-probe experiments involve a sequence of two ultrafast laser pulses. The first pulse (the pump) is more intense and excites the electron dynamics, while the second (the probe) is weaker and serves as a measure the time-dependent response.
A typical experimental scenario is depicted schematically in Fig. \ref{time_scale_fig} for the case of a ferromagnetic nano-object, where the  charge and the spin dynamics are closely intertwined.
In the first few tens of femtoseconds, the electric field of the laser couples coherently to the electron charges (direct coupling between the laser magnetic field and the electron spins occurs only at very high intensities \cite{Bigot2009}). Some of the laser energy is absorbed by the electrons, which are then driven out of equilibrium.
After the laser pulse is switched off,  the electrons evolve coherently under the action the self-consistent mean field.
A typical response mode is the so-called  {\textit surface plasmon}, which is a rigid-body oscillation of the entire electron cloud around the fixed ion lattice. For an ideal spherical nanoparticle, the oscillation frequency of the surface plasmon mode is equal too $\omega_{p} / \sqrt{3}$.
The surface plasmon eventually damps away by coupling to the single-particle modes, an effect that is akin to the standard Landau damping of ordinary plasmas.
Such Landau damping, which occurs on a very short time scale ($10-50~\rm fs$), was observed experimentally in gold nanoparticles \cite{Lamprecht1999} and was studied theoretically in several works \cite{Kreibig1995,Molina2002,Shahbazyan2016}.

However, electrons possess not only a charge, but also a spin -- i.e., a magnetic moment -- which is a crucial property for ferromagnetic nano-objects. A remarkable result, first observed by Beaurepaire and Bigot  in nickel films \cite{Beaurepaire1996,Bigot2009}, is the ultrafast loss of magnetization occurring during the first $100\, \rm fs$ that follows the laser pulse. Over twenty years after its discovery, there is no general agreement about the underlying causes of this ultrafast demagnetization, which was attributed to various mechanisms such as the spin-orbit interaction \cite{Krieger2015,Stamenova2016} (a semi-relativistic effect, as we shall see) or the  superdiffusive electron transport induced by the laser field \cite{Battiato2010}.

During these initial ultrafast processes, the ionic background remains frozen and the electron energy distribution is nonthermal due to the laser excitation.
For longer times ($t > 50~\rm fs$, see Fig. \ref{time_scale_fig}), the laser energy  is redistributed amongst the electrons through electron-electron collisions and spin-flip processes, leading to the internal thermalization of the electron gas. During this stage the electron population can still be much hotter  (up to several thousand degrees) than the ion lattice, which remains close to room temperature.
Finally, on the picosecond timescale, electron-phonon scattering leads to the exchange of energy between the electrons and the ion lattice
and the subsequent thermalization of the full system.

\paragraph{Theory and modelling.}
From a theoretical point of view, the description of the electronic dynamics in metallic nano-structures is a very complex challenge. Exact approaches based on the $N$-body Schr\"odinger equation are necessarily limited to a very small number of particles. Although such few- or even single-electron systems can nowadays be realized in the laboratory, in most practical situations a great many electrons are involved \cite{Pereira2004,Muller2004}. In that case, self-consistent effects arising from the Coulomb interactions (between all the electrons) play a crucial role on the dynamics. Several theoretical and computational studies, which treat the many-body dynamics in an approximate way, focused on the linear and nonlinear electron response. Earlier works were based on macroscopic phenomenological models \cite{Guillon2003,Aeschlimann2000,Rethfeld2002} that employed Boltzmann-type equations within the framework of the Fermi-liquid theory \cite{Pines1995}. Studies based on microscopic models are more recent and limited to relatively small systems, due to their considerable computational complexity. In the quantum regime, the ultrafast electron dynamics in metallic clusters was studied by Calvayrac et al. \cite{Calvayrac2000} and more recently by Teperik et al. \cite{Teperik2013} using  time-dependent density functional theory.
The many-particle quantum dynamics of the electron gas in a thin metal film was studied
by Schwengelbeck et al. \cite{Schwengelbeck2000} within the framework of the time-dependent Hartree-Fock  approximation.
All the above-mentioned methods, being essentially quantum, are based on the evolution of a set of wave functions $\psi_j(\bm{r},t)$, each obeying a Schr\"odinger-like equation.

In this work, we will review a possible alternative that relies on the use of phase-space models inspired from classical plasma physics, for which the system is governed by a probability distribution function $f(\bm{r},\bm{p},t)$ that evolves according to a kinetic equation. Indeed, the semiclassical limit of the above-mentioned quantum models is the self-consistent Vlasov-Poisson system, largely employed in plasma physics.

The Vlasov-Poisson model itself was used by several authors to model semi-classically the electron dynamics in metal clusters \cite{Calvayrac2000,Daligault2003,Fomichev1999} and in thin metal films \cite{Zaretsky2004,Manfredi2005film}.
These works were later extended to the quantum regime using the Wigner phase-space description \cite{Jasiak2009}. The Wigner representation is a way to express standard quantum mechanics in a classical phase-space language and is suitable to treat both single-particle and many-particle systems. It is often more intuitive than the standard Schr\"odinger approach, especially for problems where semiclassical considerations are important. For these reasons, it
is used in many areas of quantum physics, including quantum optics \cite{Smithey1993}, semiclassical analysis \cite{Heller1976,Dittrich2010}, electronic transport \cite{Bertoni1999}, nonlinear electron dynamics \cite{Jasiak2009}, and quantum plasma theory \cite{Haas2011}.
It is also the starting point for the construction of quantum hydrodynamic equations, which are approximate models obtained by taking velocity moments of the Wigner function. Such models were used in the past to study the electron dynamics in molecular systems \cite{Brewczyk1997}, metal clusters and
nanoparticles \cite{Domps1998,Banerjee2000,Manfredi2012}, thin metal films \cite{Crouseilles2008}, quantum plasmas \cite{Shukla2006,Shukla2010}, and semiconductors \cite{Haas2009}.

The above studies included the electron charge, but not its spin. However, it is well known that spin effects (particularly the Zeeman interaction and the spin-orbit coupling) can play a decisive role in nanometric systems such as semiconductor quantum dots \cite{Puente2000,Serra2001} or diluted magnetic semiconductors \cite{Morandi2009,Morandi2010}.
The coupling between the spin degrees of freedom and the electron's orbital motion is of the utmost importance in many experimental studies involving magnetized nano-objects, such as the above-mentioned laser-induced ultrafast demagnetization \cite{Bigot2009}.

Phase-space models based on the Boltzmann equation \cite{Thomas1970}, and the corresponding fluid models \cite{Snider1967}, were derived in the past to describe the dynamics of a gas where the constituents  possess internal degrees of freedom (internal angular momentum). However, in these models the spin is not treated \textit{ab-initio} as a fundamental quantity, but is rather incorporated into the transport equations to ensure the correct conservation properties. More recently, a few theoretical models that include the spin in the Wigner formalism were also developed. One approach \cite{Zamanian2010nj} consists in defining a scalar probability distribution that evolves in an extended phase space, where the spin is treated as a classical two-component variable (related to the two angles on a unit-radius sphere) on the same footing as the position or the momentum. This approach was used to derive a Wigner equation that incorporates spin effects through the Zeeman interaction \cite{Zamanian2010nj}. Semiclassical \cite{Zamanian2010} and hydrodynamic \cite{Asenjo2012} spin equations were also derived from those models, as well as other relativistic effects.

An alternative approach is to use a matrix form for the phase-space  distribution function \cite{Arnold1989}, which originates from the $2 \times 2$ density matrix for spin-1/2 particles. Using this approach, the corresponding Wigner equations were derived from the full Dirac theory \cite{Bialynicki-Birula2014}.
In their semiclassical limit, these equations give rise to a matrix spin-Vlasov equation, which treats the electron motion in a classical fashion while preserving the intrinsically quantum character of the spin degrees of freedoms \cite{Hurst2014,Hurst2017}. This approach was recently used to study the generation of spin currents in ferromagnetic thin films \cite{Hurst2018}.

Both approaches (extended phase space and matrix Wigner function) are equivalent from the mathematical point of view. However, the extended phase-space approach leads to cumbersome hydrodynamic equations that are in practice very hard to solve, either analytically  or numerically, even in the non-relativistic limit. The matrix technique separates clearly the orbital motion from the spin dynamics and leads to  simpler and more transparent hydrodynamic models.
From a computational point of view, the extended phase space method is more apt to be simulated using particle-in-cell codes, because the corresponding distribution function is transported along classical trajectories in the extended phase space  (which is eight-dimensional: three positions, three velocities, and two spin angular coordinates).
In contrast, the matrix Wigner function methods is more naturally amenable to grid-based Vlasov codes, because the corresponding distribution function only depends  on the six variables of the ordinary phase space.

\paragraph{Summary.}
In this review, we will first recall the basic concepts, parameters and physical mechanisms characterizing the electron plasma in a solid nano-object (Sec. \ref{sec:basic}). This will be followed by a very short review of wave-function-based methods (Hartree-Fock and density functional theory), which are the golden standard of computer simulations in condensed matter physics (Sec. \ref{sec:wavefunction}). Then, we will introduce the Wigner phase-space representation  of quantum mechanics (Sec. \ref{sec:wigner}) and its extension to spin-1/2 fermions (Sec. \ref{sec:wignermagnetic}). One of the attractive features of phase-space methods is that they can incorporate dissipative effects more naturally than wave-function-based methods -- this is described in Sec. \ref{sec:collisions}. All the above concepts are illustrated by two examples. First (Sec. \ref{sec:linear}), we study the linear response of a homogeneous electron gas including spin effects and derive the corresponding spin-dependent dispersion relation, for both Maxwell-Boltzmann and Fermi-Dirac equilibria. Second (Sec. \ref{sec:films}), we summarize a series of computational studies on the nonlinear electron dynamics in thin metal films carried out in our research group over the last 15 years, culminating in the recent observation of spin currents in a ferromagnetic nickel film \cite{Hurst2018}.

\section{Basic concepts, parameters, mechanisms}
\label{sec:basic}

In this paper, we focus our attention on the theoretical description of the electron dynamics in metallic nano-objects. Metals are condensed-matter systems with the specificity of having a half-filled conduction band. The electrons that belong to the conduction band are not attached to a particle nucleus, but are rather delocalized in the material and behave, to first approximation, as a non-interacting electron gas.
This property was exploited by Drude \cite{Drude1900} at the beginning of the twentieth century to derive approximate estimations of the electric and thermal conductivities of a metal. A more accurate understanding may be achieved by treating electrons as a one component plasma, i.e., by considering the interactions between the electrons.
Because of the large density of solid-state objects, such system display prominent quantum properties, thus deserving the name of ``quantum plasmas".

In a fermion gas, quantum effects become important when the temperature of the gas is comparable to, or smaller than, the Fermi temperature, defined as:
\be
T_{{F}} = \frac{\hbar}{2m k_{{B}}} \left( 3 \pi^{2} n\right)^{2/3}.
\label{fermi_temperature}
\ee
For metals, $T_F \approx 10^4 \, \rm K$, and therefore conduction electrons are in the quantum regime even at room temperature. Indeed the Fermi-Dirac distribution deviates drastically from the classical Maxwell-Boltzmann distribution for temperatures much lower than the Fermi temperature.
Quite often it is a sufficiently good approximation to assume that the electron temperature is equal to zero. In that case, all energy levels up to the Fermi energy $E_F=k_B T_F$ are occupied, whereas all levels with $E> E_F$ are empty, and the electron gas is said to be fully degenerate.

One also defines the Fermi velocity:
\be
v_{F} = \sqrt{\frac{2E_F}{m}}  = \frac{\hbar}{m} \left( 3 \pi^{2} n\right)^{1/3},
\label{fermi velocity}
\ee
and with it, the Thomas-Fermi screening length:
$\lambda_{TF} = v_F/\omega_p$.
This is the quantum analog of the classical Debye length $\lambda_D = \sqrt{\epsilon k_B T_e/(ne^2)}$, and represents the typical distance over which the Coulomb force is screened.

\begin{figure}
	\centering \includegraphics[scale=0.5]{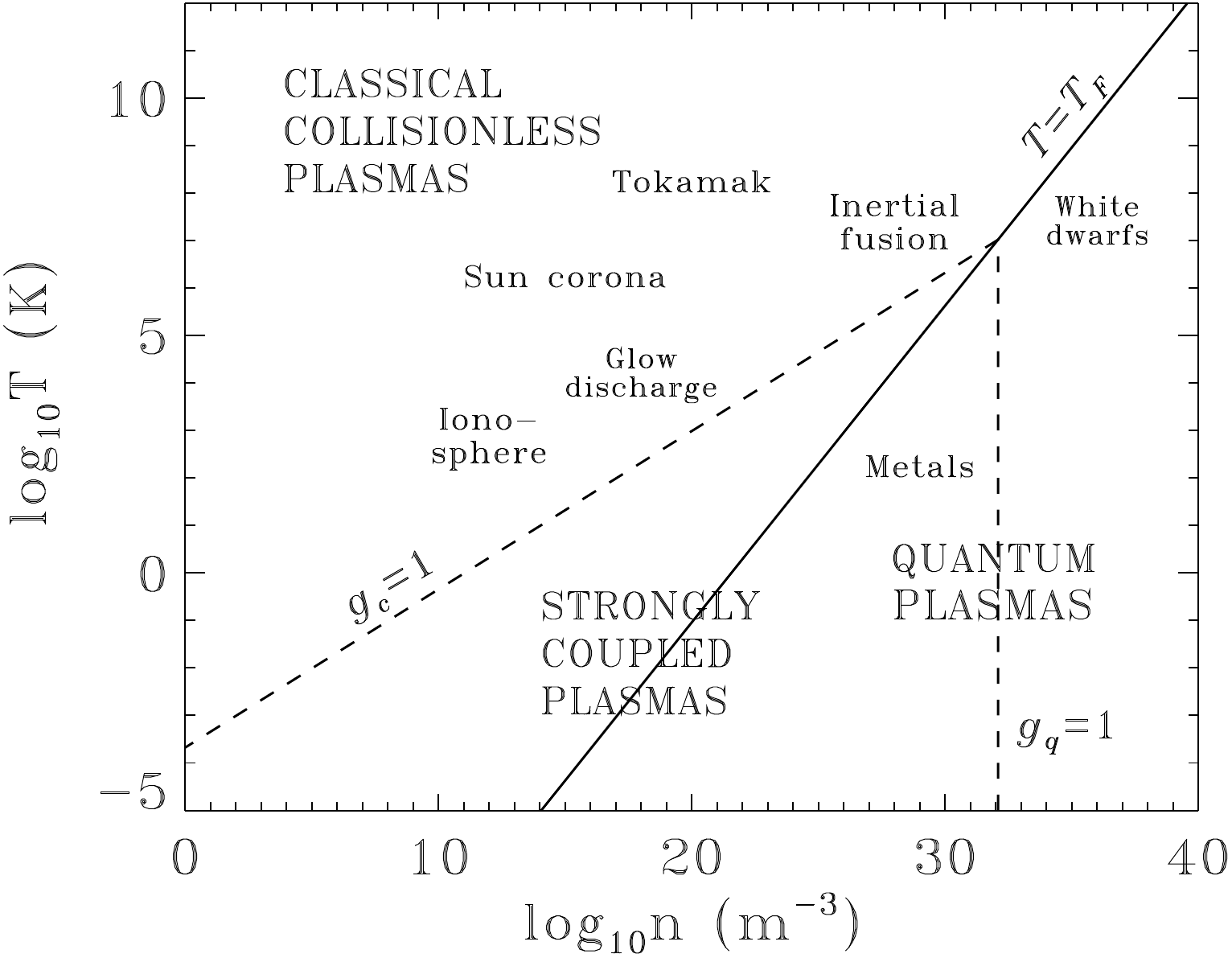}
    \caption{Phase diagram of the various physical regimes as a function of the density and temperature of the electron plasma. } \label{fig:logn-logT}
\end{figure}

In the fully degenerate regime, the coupling parameter can be written as:
\be g_q  = \frac{E_{int}}{E_F} \sim \frac{e^2 m }{\hbar^2 \varepsilon ~n^{1/3}} \sim \left(\frac{1}{n \lambda_{TF}^3}\right)^{2/3} \sim \left(\frac{\hbar
\omega_p}{E_F}\right)^2 \sim \frac{r_s}{a_0},
\label{gq}
\ee
where $E_{int}$ is the typical Coulomb interaction energy between two electrons situated at a distance $d=n^{-1/3}$, $r_s$ is the Wigner-Seitz radius $r_{s}=(3/4\pi n)^{1/3}$ (i.e., the typical volume occupied on average by one electron), and $a_0=\hbar^2\varepsilon_0/(\pi m e^2)$ is the Bohr radius.
The coupling parameter discriminates between the collisionless or weakly correlated regime ($g_q \ll 1$), where two-body collisions are unimportant and the system can be described in the mean-field approximation, and the collisional or strongly correlated regime ($g_q \approx 1$), where two-body correlations cannot be neglected.
From Eq. (\ref{gq}), the collisionless regime can be interpreted in various ways as the regime where: (i) the number of electrons in a Thomas-Fermi volume is large, (ii) the plasmon energy $\hbar\omega_p$ is small compared to the Fermi energy, or (iii) the Wigner-Seitz radius is small compared to the Bohr radius. The latter ratio is the one that is usually employed in solid-state and nano-physics and is often tabulated in standard textbooks.
The quantum coupling parameter $g_q$ should be compared to its classical counterpart:
\be
g_c = \frac{e^2 n^{1/3}}{\varepsilon_0 k_B T_e}. \label{gc}
\ee
Note that a quantum plasma becomes \textit{less} strongly correlated at high densities, whereas the opposite is true for a classical plasma \cite{Manfredi2005} (this is because of the density-dependence of the Fermi energy).

Finally, relativistic effects, in the quantum regime, may be quantified by the ratio $v_F/c$. This parameter is small ($\approx 10^{-3}$) for electrons in metals, but becomes of order unity for $n \approx 10^{35}\,\rm m^{-3}$, which is compatible with the density of compact astrophysical objects like white dwarfs and neutron stars.
Nevertheless, some low-order relativistic effects -- in particular, the spin-orbit coupling --  may still be important for magnetic nano-objects, where the spin dynamics plays a crucial role. This will be discussed in Sec. \ref{sec:wignermagnetic}.

An overview of the typical parameters for gold nano-objects is provided in Table \ref{tab:goldparam}.

The three dimensionless parameters $g_c$, $g_q$ and $T_e/T_F$ depend on the density and the temperature of the electron gas, and determine four different regions in the $(n,T)$ plane, represented in Fig. \ref{fig:logn-logT}. We notice that electrons in metals are situated in the quantum and strongly coupled regime and indeed a quick estimation shows that for solid-state densities the coupling parameter is of order unity (see Table \ref{tab:goldparam}). Nevertheless, the mean-field approximation is still used in condensed-matter physics, in the form of the Hartree and Hartree-Fock equations, which will be reviewed briefly in Sec. \ref{sec:wavefunction}.

This is in part due to the fact that  electron-electron (e-e) collisions are mitigated in degenerate (quantum) plasmas by the so-called ``Pauli blocking" effect, which is a consequence of the exclusion principle \cite{Manfredi2005}.
Without going into the details, this effect stipulates that, in strongly degenerate plasmas, the electrons below the Fermi energy cannot undergo any collisions, since most quantum states are already occupied. The collision rate can be heuristically estimated as the inverse of the electron lifetime at finite temperature $k_B T_e/\hbar$, multiplied by the number of electrons above the Fermi energy $\sim T_e/T_F$, which yields:
$\nu_{ee} \sim k_B T_e^2/(\hbar T_F)$, or in dimensionless units:
\be
\frac{\nu_{ee}}{\omega_p} \sim
g_q^{-1/2}\, \left(\frac{T_e}{T_F}\right)^2 \label{nuee} .
\ee
Hence, the collision rate can be small even when the coupling parameter is of order unity, provided the temperature is low enough. This contrasts with the classical case, where the collision rate scales as $\nu_{ee}/\omega_p \sim g_c^{3/2}$.
This rough estimation yields an e-e collision time $\tau_{ee}=\nu_{ee}^{-1}$ of the order of the picosecond, thus much larger than the typical period of plasma oscillations.
However, in current pump-probe experiments, the electrons absorb the laser energy very quickly, so that the ``temperature" $T_e$ appearing in Eq. \eqref{nuee} should actually refer to the average kinetic energy of the resulting out-of-equilibrium electron gas, which can easily attain a few thousand degrees. This brings the theoretical e-e collision time down to 100\,fs or less, depending on the intensity of the laser, which is consistent with the experiments \cite{Fourment2014}.
At the end of this phase, the electrons have thermalized to a temperature much higher than the lattice, which has not yet had the time to interact with the electron gas.

\begin{table}
\caption{Typical parameters for electrons in gold at room temperature. The values show that the relevant time, space, and energy scales are respectively, the femtosecond, the nanometer, and the electron-volt.}
\label{tab:goldparam}
\begin{tabular}{lll}
\hline\noalign{\smallskip}
Parameter & Value & Units  \\
\noalign{\smallskip}\hline\noalign{\smallskip}
$r_{s}$ & 0.16 & nm \\
$n$ & $59$ & $\rm nm^{-3}$ \\
$T$ & 300 & K \\
$2\pi/\omega_{p}$ & 0.46 & fs \\
$T_F$ & 64\,000 & K \\
$E_F$ & 5.6 & eV \\
$\hbar \omega_{p}$ & 9.1& eV \\
$v_F$ & $1.4$ & $\rm fm/ns$ \\
$\lambda_{TF}$ & 0.1 & nm \\
$\tau_{ee}$ & $\approx 50$ & fs \\
$\tau_{e-ph}$ & $\approx 5$ & ps \\
$g_{q}$ & 5.5 & -- \\
$r_s/a_0$ & 3 & -- \\
$v_F/c$ & $0.0047$ & -- \\
\noalign{\smallskip}\hline
\end{tabular}
\end{table}

On even longer timescales, the electrons couple to the phonons, i.e., vibrations of the ion lattice. A detailed microscopic description of this phenomenon is very complex, so it it useful to resort to empirical macroscopic models. The two-temperature model (TTM) is a simple description of electron-phonon (e-ph) interactions that considers two thermal baths for the electrons and the lattice, with temperatures $T_e$ and $T_{l}$, which interact through a coupling constant $G$, and obey the evolution equations:
\begin{eqnarray}
C_e(T_e)\,\frac{\partial T_e}{\partial t} &=& \kappa \nabla^2 T_e - G(T_e -T_{l}) + P(t), \label{eq:TTM1}\\
C_l\,\frac{\partial T_l}{\partial t} &=& G(T_e -T_{l}) . \label{eq:TTM2}
\end{eqnarray}
where $\kappa$ is the heat conductivity, $C_e$ and $C_l$ are respectively the electron and lattice heat capacities, $G$ is the e-ph coupling constant, and $P(t)$ is the power absorbed by the electron gas following the laser excitation.
For gold, typical measured values of these parameters are \cite{Ekici2008}: $G=2 \times 10^{16} \,\rm W m^{-3} K^{-1}$, $C_l = 2.1 \times 10^6 \,\rm J m^{-3} K^{-1}$, and $C_e = 70\, T_e \,\rm J m^{-3} K^{-1}$. We note that $C_e$ is rather close to the heat capacity of an ideal Fermi gas: $C_{\rm Fermi}={\pi^2 \over 2} n k_B \frac{T_e}{T_F}$ ($=62.8\, T_e \,\rm J m^{-3} K^{-1}$ for the density of gold).
Since $C_l \gg C_e$, the electrons respond much more quickly than the ions to the e-ph coupling. Neglecting the heat conductivity, the typical timescale is given by $\tau_{e-ph} \sim C_e/G \approx 1\,\rm ps$.
At the end of this phase, the electron and lattice temperature are equilibrated. Finally, on even longer timescales ($\approx 1\,\rm ns$) any excess temperature is evacuated into the external environment.

\begin{figure}
	\centering \includegraphics[scale=0.55]{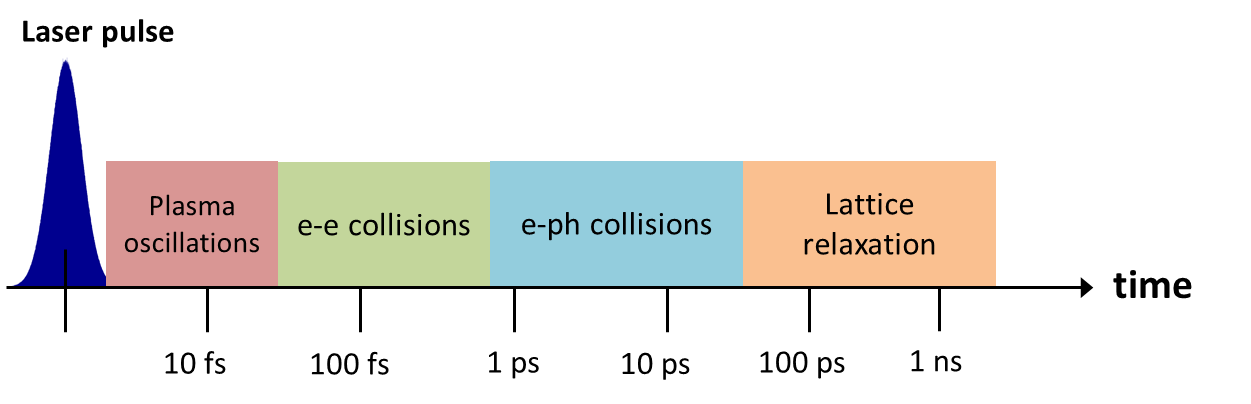}
    \caption{Pictorial view of the different timescales involved in the electron dynamics following the initial laser excitation.} \label{fig plasma condition n and T}
    \label{fig:timescales}
\end{figure}

The various timescales, from the ultrashort laser pulse to the lattice relaxation are represented pictorially in Fig. \ref{fig:timescales}.

In summary, there exists an early stage of the laser-induced electron dynamics, lasting around 50\,fs, which is essentially collisionless. This early stage can be described in the phase space by a Vlasov equation, or its quantum counterpart, the Wigner equation, as we will show in the remainder of this article.
For longer times, collisions (either e-e or e-ph) should be taken into account. This is an issue where phase-space-based models have an advantage with respect to wave-function-based ones. Indeed, there is a large literature on dissipative kinetic equations for classical plasmas (Boltzmann, Fokker-Planck, Lennard-Balescu,\dots), which can inspire useful extensions to the quantum regime.
In contrast, wave-function-based methods, being essentially Hamiltonian, face more difficulty in incorporating dissipative effects.


\section{Wave-function-based methods: Hartree, Hartree-Fock, DFT}
\label{sec:wavefunction}

The dynamics of $N$ interacting bodies is a fascinating and challenging problem in physics. Classically, solving the exact $N$-body problem involves integrating the equations of motion of the $N$ particles using some time-stepping technique. The computational cost of such an operation grows like $N^2$. Sometimes, using special tricks as in hierarchical tree codes, this can be brought down to $N \log N$. For systems where the number of particles is very large, like plasmas, this is still a formidable problem, but it can be attacked using modern molecular dynamics techniques.

The quantum $N$-body problem is even worse, because the complexity of the Hilbert space of a system made of $N$ particles grows exponentially with $N$. Indeed, a quantum system is described by the $N$ body wave function
\be
\Psi ^{N} = \Psi \left(\bm{r}_{1},\bm{r}_{2},\cdots,\bm{r}_{n},t \right),
\label{N body wave function}
\ee
where $\bm{r}_{i}$ is the position of the $i$-th particle. This evolves according to the Schr\"odinger equation
\be
i \hbar \frac{\partial \Psi ^{N}}{\partial t}
=
- \frac{\hbar^{2}}{2m} \sum _{i=1}^{N}\left[ \bm{\nabla}^{2}_{i} \Psi^{N}  +  \sum _{k=1,k\neq i}^{N} \frac{1}{2} \mathcal{V}\left(  |\bm{r}_{i}-\bm{r}_{k}| \right) \Psi  ^{N}\right],
\label{N body schrodinger equation}
\ee
where $ \mathcal{V}(\bm{r}) = \frac{e^{2}}{4 \pi \epsilon_{0}} \frac{1}{|\bm{r}|} $ is the Coulomb potential.
The wave function lives in the $3N$-dimensional configuration space. If we use 10 points for each direction, then we need $10^{3N}$ points to code the whole wave function, which is a huge number even for small systems made of a few dozen particles.

It is clear, therefore, that some kind of approximation is mandatory if we want to model the dynamics of nano-objects that contain hundreds or thousands of electrons. In this section, we will briefly review the standard methods that are used in condensed-matter theory to describe the electron dynamics. These fall in two broad categories: (i) Hartree and Hartree-Fock methods and (ii) time-dependent density functional theory (TD-DFT). Both are based on the propagation of some reduced one-body wave functions.

\subsection{The mean-field approach: Hartree equations}\label{subsec:wave_hartree}
The approximation leading to the Hartree equations is the same that is used classically to obtain the Vlasov equation from the $N$-body Liouville equation, namely, neglecting two-body (and higher-order) correlations. Mathematically, this consists in factoring the $N$ body wave function into $N$ single-particle wave functions
\be
\Psi^{N} = \Psi_{1} \left(\bm{r}_{1},t\right)\Psi_{2} \left(\bm{r}_{2},t\right) \cdots \Psi_{N} \left(\bm{r}_{N},t\right).
\label{factorization wave function}
\ee
This approximation was first considered by Hartree in 1927 \cite{Hartree1928}, in the context of atomic physics, to describe the self-consistent effect of the atomic electrons on the Coulomb potential of the nucleus.
As we saw in the preceding section, this mean-field approximation is valid only when the corresponding coupling parameter $g_q$ is small.

Using Eqs. \eqref{N body schrodinger equation} and \eqref{factorization wave function}, one obtains that the single-particle wave functions obey the following Hartree equations:
\begin{align}
&i \hbar \frac{\partial \Psi_{\alpha}(\bm{r},t)}{\partial t}
=
- \frac{\hbar ^{2}}{2m} \bm{\nabla}^{2} \Psi_{\alpha}(\bm{r},t)  + \underbrace{\frac{e^{2}}{4 \pi \epsilon_{0}} \left[ \sum_{\alpha '=1 }^{N} \int  \frac{  | \Psi_{\alpha '}(\bm{r'},t) |^{2}}{|\bm{r}- \bm{r'}|} d\bm{r'} \right]}_{V_{H}} \Psi_{\alpha}(\bm{r},t) \nonumber \\
&~- \underbrace{\frac{e^{2}}{4 \pi \epsilon_{0}}  \int  \frac{  | \Psi_{\alpha }(\bm{r'},t) |^{2}}{|\bm{r}- \bm{r'}|} d\bm{r'}}_{V_{sic}} \Psi_{\alpha}(\bm{r},t).
\label{mean field Schrodinger equation}
\end{align}
where the quantum states of each particle are labelled by the wave functions \\ $\left\{ \Psi _{\alpha} \left(\bm{r}, t \right), ~\alpha = 1, \cdots , N \right\}$.
The Hartree potential $V_{H}$ is the self-consistent potential created by the ensemble of all electrons.
The term $V_{sic}$ is known as the self-interaction correction (SIC) and takes into account the fact that an electron should not interact with itself. The SIC is often neglected for classical plasmas, because its contribution goes like $1/N$, but in small nano-objects it has sometimes to be taken into account. For instance, the SIC correction may be important to enforce the correct asymptotic behavior of the Coulomb potential ($\sim 1/r$) at long distances \cite{Ullrich2000}.
Here, we shall neglect the SIC correction, so that Eq. \eqref{mean field Schrodinger equation} can be rewritten as a set of Schr\"odinger-Poisson equations:
\begin{align}
\left\{
\begin{array}{lcl}
\displaystyle
i\hbar \frac{\partial  \Psi_{\alpha}(\bm{r},t) }{\partial t}
 &=& \displaystyle - \frac{\hbar^{2}}{2m} \bm{\nabla}^{2}  \Psi_{\alpha}(\bm{r},t) - e V_{H} (\bm{r},t)\Psi_{\alpha}(\bm{r},t),\\ \\
 \displaystyle
\bm{\nabla}^{2}V_{H}(\bm{r},t)  &=& \displaystyle \frac{e}{\epsilon _{0}}  \sum_{\alpha=1}^{N} \left| \Psi_{\alpha}(\bm{r},t)  \right|^{2}.
\end{array}
\right.
\label{schrodinger poisson equ}
\end{align}

The simplification achieved through the Hartree equations is enormous, because now we only need to solve $N$ equations for a wave function evolving in three-dimensional (3D) space. Using 10 points per dimension, we only need $10^3 N$ points to encode the wave functions, instead of $10^{3N}$ for the full $N$-body problem.

\subsection{The exchange interaction: Hartree-Fock equations}\label{subsec:wave_hf}
However, the Hartree equations suffer from two problems. Firstly, they violate the Pauli exclusion principle, which stipulates that two fermions cannot be in the same quantum state. Consequently, the total wave function should be antisymmetric with respect to the exchange of two particles, for instance exchanging particles 1 and 2:
\be
\Psi \left(\bm{r}_{2},\bm{r}_{1},\cdots,\bm{r}_{N},t \right) = -\Psi \left(\bm{r}_{1},\bm{r}_{2},\cdots,\bm{r}_{N},t \right).
\label{antisymmetry}
\ee
This property creates some special (purely quantum) correlation  between the fermions, which is known as the \emph{exchange interaction}. In 1930, Slater and Fock \cite{Fock1930} proposed an exact method to describe the exchange interactions. They introduced the Slater determinant \cite{Slater1929} to write the $N$-body wave function as follows:
\begin{align}
\Psi^{N} &= \frac{1}{\sqrt{N!}}
\begin{vmatrix}
\Psi_{1} \left(\bm{r}_{1},t\right) &
\Psi_{2} \left(\bm{r}_{1},t\right) &
\cdots &
\Psi_{N} \left(\bm{r}_{1},t\right) \\
\Psi_{1} \left(\bm{r}_{2},t\right) &
\Psi_{2} \left(\bm{r}_{2},t\right) &
\cdots &
\Psi_{N} \left(\bm{r}_{2},t\right) \\
\vdots &
\vdots &
 & \\
\Psi_{1} \left(\bm{r}_{N},t\right) &
\Psi_{2} \left(\bm{r}_{N},t\right) &
\cdots &
\Psi_{N} \left(\bm{r}_{N},t\right)
\end{vmatrix}.
\label{slater determinant}
\end{align}

In this case, the total wave function is no longer a product of one-body wave functions, but it satisfies the antisymmetry property required by the Pauli principle. Using Eq. \eqref{slater determinant} into Eq. \eqref{N body wave function}, one obtains the Hartree-Fock (HF) equations:
\begin{align}
i \hbar \frac{\partial \Psi_{\alpha}(\bm{r},t)}{\partial t}
&=
- \frac{\hbar ^{2}}{2m} \bm{\nabla}^{2} \Psi_{\alpha}(\bm{r},t)  + \frac{e^{2}}{4 \pi \epsilon_{0}} \left[ \sum_{\alpha '=1}^{N} \int  \frac{  | \Psi_{\alpha '}(\bm{r'},t) |^{2}}{|\bm{r}- \bm{r'}|}d\bm{r'} \right] \Psi_{\alpha}(\bm{r},t) \nonumber \\
 &~ - \frac{e^{2}}{4 \pi \epsilon_{0}}  \sum_{\alpha '=1}^{N}  \Psi_{\alpha'}(\bm{r},t)\int  \frac{   \Psi_{\alpha '}^{\ast}(\bm{r'},t) \Psi_{\alpha }(\bm{r'},t)  }{|\bm{r}- \bm{r'}|}d\bm{r'}.
\label{hartree fock equation}
\end{align}
The above HF equations differ from the Hartree equations by the last term in Eq. \eqref{hartree fock equation}, which encodes the effect of the exchange interactions.
Note that the exchange term cannot be represented as a Poisson equation like the Hartree term. It is inherently nonlocal, in the sense that it couples all wave functions together, whereas in the Hartree case the wave functions are only coupled through the partial densities  $|\Psi_{\alpha}|^2$ appearing in Poisson's equation \eqref{schrodinger poisson equ}. This fact renders the HF equations considerably more difficult to solve numerically.

The second limitation, inherent to both the Hartree and the HF methods, is that two-body correlations are not taken into account. This drawback has been overcome since the advent of density-functional theory (DFT), first for the ground state in the 1960s, and then for time-dependent problems in the 1980s.

\subsection{Time-dependent density-functional theory (TD-DFT)}\label{subsec:wave_dft}
A big step forward in the treatment of many-body problems was achieved thanks to the results of Hohenberg and Kohn \cite{Hohenberg1964} and Kohn and Sham \cite{Kohn1965}, which can be summarize in two theorems. The first Hohenberg-Kohn theorem states that the ground-state properties of a many-electron system are exactly and uniquely determined by the electron density. This is a huge progress, as the extremely
complex $N$-body problem (wave function depending on $3N$ spatial coordinates) has been reduced to finding the correct electron ground-state density (3 spatial coordinates).
The second Hohenberg-Kohn theorem states that the ground-state density can be found by minimizing a certain energy functional:
\be
E[n] = T[n] + E_{ext}[n] + E_{int}[n] = T[n]+ E_{ext}[n] + E_H[n]+ E_X[n] + E_C[n] ,
\label{enfunctional}
\ee
where $T[n]$ is the kinetic energy functional, $E_{ext}[n]$ is the external energy, and $E_{int}[n]$ is the Coulomb interaction energy. The interaction energy can be decomposed into the Hartree energy $E_H[n]$, the exchange energy $E_X[n]$, and the correlation energy $E_C[n]$ -- the latter containing all higher-order correlations beyond the mean field (Hartree) and the Pauli principle (exchange). The above expression \eqref{enfunctional} is in principle \emph{exact}, in the sense that we know that such functionals exist. However, some of the terms are not known and thus need to be approximated.

As an example, the old Thomas-Fermi theory \cite{Thomas1926} of the atomic electron gas can be viewed as an early precursor of modern DFT. In the Thomas-Fermi theory, exchange and correlations are neglected and the external potential is that of the atomic nucleus, $V_{ext}(\bm{r})=-Ze^2/4\pi\epsilon_0 |\bm{r}|$. Then, we have for the external energy functional: $E_{ext}[n]=\int n(\bm{r})V_{ext}(\bm{r}) d{r}$, and for the Hartree energy:
\be
E_H[n] = \int V_H[n]\, n \, d\bm{r}=
\frac{e^2}{4 \pi \epsilon_0} \int\int  \frac{n(\bm{r}) n(\bm{r'})}
{|\bm{r}- \bm{r'}|}\, d\bm{r}\, d\bm{r'}.
\label{EH}
\ee
Finally, Thomas and Fermi chose a semiclassical functional for the kinetic energy of an ideal fermion gas at zero temperature:
\be
T[n] = {3 \over 10}\,(3\pi^2)^{2/3}\,{\hbar^2 \over {m}}\int n^{5/3} d\bm{r}.
\label{KinEnergy}
\ee
Minimizing the total energy functional $E[n]$ with the constraint $\int n\,d\bm{r}=N$, yields the usual Thomas-Fermi equation for the ground-state density.

Apart from neglecting exchange and correlation effects, the main drawback of the Thomas-Fermi approach is its poor approximation of the kinetic energy functional.
To improve on this approximation, Kohn and Sham \cite{Kohn1965} suggested that one uses the kinetic energy $T_s[n]$ of a fictitious non-interacting electron gas, with the same density $n$ as the original interacting one, which evolves in an effective potential $V_{ eff}[n]= -e V_H[n]+ V_{ext}+V_{X}[n]+V_{C}[n]$.  Each term of the effective potential is obtained as a functional derivative of the corresponding energy functional: $V_{ k}[n]=\delta E_k[n]/\delta n$. The Kohn-Sham (KS) equations are then:
\be
- \frac{\hbar^{2}}{2m} \bm{\nabla}^{2}  \phi_{j}(\bm{r})+ V_{ eff}[n(\bm{r})] \phi_{j}(\bm{r}) = \epsilon_j \phi_{j}(\bm{r}),
\label{Kohn-Sham}
\ee
where the $\phi_j$ are the single-particle wave functions of the fictitious non-interacting system.
The electron density is then obtained as: $n(\bm{r}) = \sum_j p_j |\phi_{j}(\bm{r})|^2$, where the $p_j$ are occupation probabilities.
The KS equations are potentially exact, provided one knows all the terms of the effective potential  $V_{eff}[n]$. Unfortunately, this is not true for the exchange and correlation potentials, for which we need to find approximate expressions.

This caveat notwithstanding, the KS accomplish the truly impressive task of representing \emph{in an exact way} the full $N$-body problem through a set of one-body wave functions. In addition, unlike the HF equations, the KS equations are local and only coupled to each other via the electron density $n$. Their computational complexity is thus similar to that of the Hartree equations. Thus, the KS equations can be seen as a way to make the Hartree approach exact by including the appropriate exchange and correlation functionals.
As the latter functionals are not known exactly, all the art of DFT is to find the best way to approximate them. The simplest choice is the so-called local density approximation (LDA) \cite{Kohn1965}, whereby the exchange and the correlation functionals depend locally on the electron density.
For instance, the LDA approximation for the exchange potential is:
\be
V_{X} = -\frac{e^{2}}{4 \pi\epsilon_{0}}\left( \frac{3}{\pi} \right)^{1/3} n^{1/3} .
\label{eq:xlda}
\ee
Many other sophisticated approximations have been developed over the years (such as the generalized gradient approximation, GGA), making DFT methods a cornerstone of computational materials science and theoretical chemistry \cite{Jones2015}.

A time-dependent version of DFT (TD-DFT) was developed by Runge and Gross \cite{Runge1984} in the 1980s. The Runge-Gross theorem stipulates that, for the same initial $N$-particle state, two external potentials differing only by a time-dependent function $c(t)$ cannot give rise to the same density $n(\bm{r},t)$.
Using this theorem, one can construct the time-dependent KS equations:
\be
- \frac{\hbar^{2}}{2m} \bm{\nabla}^{2}  \phi_{j}(\bm{r},t)+ V_{eff}[n(\bm{r})] \phi_{j}(\bm{r},t) = i \hbar \frac{\partial \phi_{j}(\bm{r},t)}{\partial t} .
\label{Kohn-Sham-td}
\ee
Compared to their static counterpart, the time-dependent KS have the additional difficulty of requiring a time-dependent approximation for the functionals. The simplest choice is to use the same functional as in the ground state, but allowing a time dependence in the density. This is known as the adiabatic local-density approximation (ALDA). Note also that, by setting $V_X=V_C=0$ in the KS equations, one recovers exactly the time-dependent Hartree equations \eqref{mean field Schrodinger equation}.

Finally, we stress that, since the time-dependent KS equations \eqref{Kohn-Sham-td} have the same mathematical form as a set of nonlinear Schr\"odinger equations, they can be used to construct a phase-space formalism by taking the Wigner transform of the KS wave function.

\section{Quantum mechanics in the phase space}
\label{sec:wigner}

In this section, we summarize the main properties of Wigner's phase-space formulation of quantum mechanics, which was first introduced by Eugene Wigner in 1932 to study quantum corrections to classical statistical mechanics \cite{Wigner1932}. The goal was to link the wave function that appears in the Schr\"{o}dinger equation to a pseudo-probability distribution defined in the classical phase space. This pseudo-probability distribution changes in time according to an evolution equation (Wigner equation) that is somewhat similar to the classical Liouville equation.

A mathematically rigorous treatment of the Wigner formulation is based on the Weyl transformation \cite{WeylHermann1928,Cohen2013}, which is a general method to transform operators defined in the Hilbert space into phase-space functions.
We have developed these arguments at some length in a previous publication \cite{Hurst2017}, to which we address the reader for further details. Here, we will summarize the main properties of the Wigner approach for particles in electric and magnetic fields, with and without spin.

\subsection{The Wigner equation for a scalar electric potential}\label{subsec:wignerscalar}

The density matrix of a quantum mixture of $N$ particles is given by:
\begin{align}
\rho = \sum_{\alpha =1}^{N} p_{\alpha} \left| \Psi_{\alpha} \rangle \langle \Psi_{\alpha} \right|,
\label{density matrix quantum mixture of state}
\end{align}
where $p_{\alpha}$ is the probability for one particle to be in the state $\Psi_{\alpha}$. The wave functions are supposed to obey a set of Schr\"odinger-like equations such as the time-dependent Hartree equations \eqref{schrodinger poisson equ} or  Kohn-Sham equations \eqref{Kohn-Sham-td} discussed in Sec. \ref{sec:wavefunction}.

The Wigner function of the system is defined as
\begin{eqnarray}
f\left(\bm{r},\bm{p}, t \right) &=& \frac{1}{\left( 2 \pi \hbar \right)^{3}} \int d\bm{\lambda} \exp \left( \frac{i}{\hbar} \bm{\lambda} \cdot \bm{p} \right) \left\langle \bm{r} - \frac{1}{2}\bm{\lambda} \, \vline\, \rho \, \vline \, \bm{r} + \frac{1}{2}\bm{\lambda} \right\rangle \nonumber\\
&=& \frac{1 }{\left(2 \pi \hbar\right)^{3}}  \sum _{\alpha = 1}^{N} p_{\alpha} \int d\lambda \, e ^{\frac{i \bm{p} \cdot \bm{\lambda} }{\hbar}}\, \Psi _{\alpha}^{*}\left(\bm{r} + \frac{\bm{\lambda}}{2}, t \right)\Psi _{\alpha}\left(\bm{r} - \frac{\bm{\lambda}}{2}, t \right).
\label{wignerfunction}
\end{eqnarray}
The  Wigner function evolves in time according to the following Wigner equation
\be
\frac{\partial f}{\partial t} + \frac{1}{m}\bm{p}\cdot \bm{\nabla}f =
\frac{ie}{\hbar}\frac{1}{\left( 2 \pi \hbar \right)^{3}} \int d\bm{\lambda}\, d\bm{p'} e^{\frac{ i \left(\bm{p}- \bm{p'} \right) \cdot \bm{\lambda}}{\hbar}} \left[ V\left( \bm{r}_{+} \right) - V\left( \bm{r}_{-}  \right) \right] f(\bm{r},\bm{p'},t),
\label{wignerequation}
\ee
where the subscripts $\bm{\pm}$ denote the shifted positions $\bm{r}_{+} = \bm{r} \pm \bm{\lambda} /2$ and $V(\bm{r},t)$ indicates a generic scalar potential, which can be the Hartree potential or the effective potential of TD-DFT.

The Wigner equation \eqref{wignerequation} is completely equivalent to the Schr\"odinger-like equations from which it is derived. Therefore, it provides a useful way to cast TD-DFT in a phase-space formalism without any loss of generality.
In addition, the Wigner function can be used to compute all macroscopic quantities in the same way as a classical probability distribution. For instance, the particle and current densities:
\begin{eqnarray}
n(\bm{r},t) &=& \int f(\bm{r},\bm{p},t) \,d\bm{p},\\
\bm{J}(\bm{r},t) &=& \int f(\bm{r},\bm{p},t)\, {\bm{p} \over m} \,d\bm{p}.
\end{eqnarray}

Finally, the Wigner approach is particularly useful to obtain the semiclassical limit.
By developing the integral term in (\ref{wignerequation}) up to order
$O(\hbar^2)$ we obtain
\be
\frac{\partial f}{\partial t} + \frac{\bm{p}}{m} \cdot \bm{\nabla} f +e \bm{\nabla} V \cdot \bm{\nabla_{p}} f = O(\hbar^2).
\label{eq:vlasov}
\ee
The Vlasov equation is thus recovered in the formal semiclassical
limit $\hbar \to 0$.

\subsection{The Wigner equation in a magnetic field}\label{subsec:wignermagnetic}
So far, we did not include any magnetic effects. This approximation may be justified in some cases, for instance if we are only interested in plasmonic excitations. However, magnetic interactions cannot be escaped if we want to include the spin degrees of freedom in our treatment.
The introduction of a magnetic field in the Wigner formalism is not trivial. In the presence of magnetic fields, one should use the kinetic momentum operator $\widehat{\bm{\pi}} = \widehat{\bm{p}} -q \widehat{\bm{A}}$, instead of $\widehat{\bm{p}}$ (with $q=-e$ for an electron), where $\widehat{\bm{A}}$ is the vector potential operator. However, it can be proven that simply substituting $\bm{p}$ with $\bm{\pi} \equiv m\bm{v}$ in the definition \eqref{wignerfunction} will not work, as the resulting Wigner function is not gauge invariant.

A gauge-independent definition of the Wigner function was first introduced by Stratonovich \cite{Stratonovich1957}:
\begin{align}
f\left(\bm{r},\bm{v}, t \right) &=
\left(\frac{ m}{2 \pi \hbar}\right)^{3} \int d\bm{\lambda} \exp \left[ \frac{i \bm{\lambda}}{\hbar} \cdot \left(m \bm{v} - e \int_{-1/2}^{1/2} d\tau \bm{A}\left(\bm{r} + \tau \bm{\lambda} \right) \right) \right] \left\langle \bm{r} - \frac{\bm{\lambda}}{2} \vline\, \rho \,\vline\, \bm{r} + \frac{\bm{\lambda}}{2} \right\rangle,
\label{one body dis function exp with magnetic field}
\end{align}
where the momentum $\bm{p}$ was replaced by $m \bm{v} - e \int_{-1/2}^{1/2} d\tau \bm{A}\left(\bm{r} + \tau \bm{\lambda}\right)$.

After some rather convoluted algebra, one finally obtains the gauge-invariant Wigner equation for a spinless particle interacting with an electromagnetic field:
\be
\frac{\partial f}{\partial t}
+
\frac{1}{m}\left(\bm{\pi} + \bm{\Delta \widetilde{\pi}}  \right)\cdot \bm{\nabla}f
-
\frac{e}{m} \left[ m\widetilde{\bm{E}} + \left(\bm{\pi} + \bm{\Delta \widetilde{\pi}}  \right) \times \widetilde{\bm{B}} \right]_{i} \partial_{\pi_{i}} f = 0,
\label{wigner equation with B}
\ee
where $\bm{\Delta \widetilde{\pi}} $ depends on the magnetic field and corresponds to a quantum shift of the velocity
\begin{align}
\bm{\Delta \widetilde{\pi}} = -i\hbar e \partial_{\bm{\pi}} \times \left[ \int^{1/2}_{-1/2} d\tau \,\tau \bm{B} \left( \bm{r} + i\hbar \tau \partial_{\bm{\pi}} \right) \right]
\label{def shift vitesse quantique}
\end{align}
and  $\widetilde{\bm{E}}$, $\widetilde{\bm{B}}$  are written in terms of the electric and magnetic fields
\begin{align}
\widetilde{\bm{E}} &= \int^{1/2}_{-1/2} d\tau \bm{E} \left( \bm{r} + i\hbar \tau \partial_{\bm{\pi}} \right),~~~~
\widetilde{\bm{B}} =  \int^{1/2}_{-1/2} d\tau  \bm{B} \left( \bm{r} + i\hbar \tau \partial_{\bm{\pi}} \right). \label{def E B quantique}
\end{align}
This form of the Wigner equation was first proposed by Serimaa et al. \cite{Serimaa1986}, where the authors also discuss the case where the electromagnetic fields are quantized. In the classical limit, it is straightforward to see that: $\widetilde{\bm{E}} = \bm{E}$, $\widetilde{\bm{B}} = \bm{B}$, and $\bm{\Delta \widetilde{\pi}}=0$, so that the Wigner equations becomes
\be
\frac{\partial f}{\partial t} + \bm{v} \cdot \bm{\nabla} f -\frac{e}{m} \left(\bm{E} + \bm{v} \times \bm{B} \right) \cdot \bm{\nabla_{v}} f =0 \, ,
\label{vlasov with B no spin}
\ee
thus recovering the classical Vlasov equation with the Lorentz force.

\section{Spin and relativistic effects}\label{sec:wignermagnetic}

\subsection{Semi-relativistic Pauli equation}\label{subsec:pauli}

In the previous chapter, we omitted all mentions to the spin of the electrons. The spin is an intrinsic property of any elementary particle as much as the mass or the charge. It was first discovered in 1922 thanks to the experiments of Stern and Gerlach \cite{Gerlach1922} and was interpreted as an internal angular momentum of the electron.
From a theoretical point of view, the spin appears naturally in the Dirac equation, which is the relativistic extension of the Schr\"odinger equation for spin $1/2$ particles. For an electron interacting with an external electromagnetic field, the Dirac equation reads as:
\begin{align}
i \hbar \frac{\partial \Psi^{\textrm{D}} \left( \bm{r},t\right) }{\partial t} &=
\left[ c \bm{\alpha} \cdot \left( \bm{p} + e \bm{A}\left( \bm{r},t\right) \right) + \beta m c^{2} - e V \left( \bm{r},t\right)  \right] \Psi^{\textrm{D}}  \left( \bm{r},t\right),
\label{dirac equation}
\end{align}
where $V \left( \bm{r},t\right) $ and $\bm{A}\left( \bm{r},t\right)$ are, respectively, the scalar and vector potentials. The operators $\bm{\alpha}$ and $\beta$ are $4 \times 4$ matrices
\begin{align}
\bm{\alpha} =
\begin{pmatrix}
0 & \bm{\sigma} \\
\bm{\sigma} &0
\end{pmatrix}, ~~~~
\beta =
\begin{pmatrix}
\sigma_{0} & 0 \\
0 & -\sigma_{0}
\end{pmatrix},
\end{align}
where $ \bm{ \sigma} = (\sigma_x, \sigma_y, \sigma_z)$ is the vector of the $2 \times 2$ Pauli matrices
\be
\sigma_x = \begin{pmatrix}
0 & 1 \\ 1&0 \end{pmatrix}, \,\,\,
\sigma_y = \begin{pmatrix}
0 & -i \\ i&0 \end{pmatrix}, \,\,\, \label{eq:paulimatrices}
\sigma_z = \begin{pmatrix}
1 & 0 \\ 0&-1\end{pmatrix},
\ee
and $\sigma_{0}$ is the $2 \times 2$ identity matrix. Therefore the wave functions $ \Psi^{\textrm{D}}\left( \bm{r},t\right)$ that obey to the Dirac equation \eqref{dirac equation} are four-component objects called bispinors.

The Dirac equation contains much more information than just the spin, as it deals at the same time with the dynamics of particles (electrons) and antiparticles (positrons),
described respectively by the two upper (lower) components of the Dirac wave function.
In our case, since we are only interested in the low-energy phenomena occurring in condensed-matter physics, we would like to discard all effects related to electron-positron pair formation.

For this purpose, it is helpful to use the unitary Foldy-Wouthuysen transformation \cite{Foldy1950}, which enables one to separate the electron and the positron dynamics in the Dirac equation. This transformation is exact in the field-free case, and leads to a
semi-relativistic expansion in $1/c$ (where $c$ is the speed of light) for a particle interacting with an electromagnetic field \cite{Strange1998}.

At second order in $1/c$, the Dirac Hamiltonian transforms into the following semi-relativistic Pauli Hamiltonian \cite{Dixit2013}:
\begin{eqnarray}
\hat{H} &=& mc^2-eV+\frac{(\hat{\bm{p}}+e\bm{A})^2}{2m}+
\frac{e\hbar}{2m}\boldsymbol{\sigma}\cdot\bm{B} +\frac{(\hat{\bm{p}}+e\bm{A})^4}{8m^3c^2}  \label{Eq:hamiltonian} \\
&+& \frac{e\hbar^2}{8m^2c^2} \nabla\cdot\bm{E}+\frac{e\hbar}{8m^2c^2}\boldsymbol{\sigma} \cdot[\bm{E}\times(\hat{\bm{p}}+e\bm{A})-(\hat{\bm{p}} +e\bm{A})\times\bm{E}]
 \nonumber
\end{eqnarray}
where the electromagnetic fields are defined as usual as: $\bm{E}=-\nabla V-\partial_t\bm{A}$ and $\bm{B}=\nabla\times\bm{A}$, and $\hat{\bm{p}}=-i\hbar\nabla$.
The first term on the right-hand side is the rest-mass energy of the electron; the next two terms are the standard nonrelativistic Hamiltonian in the presence of an electromagnetic field; the fourth term is the Pauli spin term (Zeeman effect); the $(\hat{\bm{p}}+e\bm{A})^4$ term is the first relativistic correction to the electron mass (expansion of the Lorentz factor $\gamma$ to second order); the $\nabla \cdot \bm{E}$ term is the Darwin term; and the last two terms represent the spin-orbit coupling (SOC).
The wave function $\Psi = {^{t}(}\Psi^{\uparrow} , \Psi^{\downarrow})$ is a spinor, the upper and lower components describing respectively the spin-up and the spin-down electrons, and it obeys the spinorial Schr\"odinger equation:
\(
i\hbar\partial_t \Psi = \hat{H}\, \Psi
\).
Higher order extensions of the Foldy-Wouthuysen expansion can be found in \cite{Hinschberger2012}.

The Zeeman and spin-orbit effects are of paramount importance to describe the magnetic properties of ferromagnetic nano-objects. For instance, there are strong indications that the SOC plays a crucial role in many experiments where the magnetisation is excited with optical pulses \cite{Krieger2015}.
Thus, in the following, we will retain the Zeeman and SOC effects, but neglect the relativistic correction to the electron mass (which would generate awkward fourth-order gradients in the Schr\"odinger equation) and the Darwin term. The latter is a manifestation of the so-called Zitterbewegung, i.e., a quivering motion of the electron around its mean path \cite{Dixit2013}, which is due to the interference between the positive and negative energy states in the Dirac equation. This term could be reintroduced in our treatment without much difficulty.

\subsection{The Wigner equation with spin}\label{subsec:wigspin}

For spinless particles, the Wigner function is a scalar function related to the density matrix by Eq. \eqref{one body dis function exp with magnetic field}. In the case of spin-${1/2}$ particles, both the Wigner function $\mathcal{F}$ and the density matrix $\rho $ are $2\times 2$ matrices:
\begin{equation}
\mathcal{F} =
\begin{pmatrix}
f ^{\uparrow \uparrow} & f ^{\uparrow \downarrow}\\
f ^{\downarrow \uparrow } & f ^{\downarrow \downarrow}
\end{pmatrix}
~~~~\textrm{and}~~~~
\rho =
\begin{pmatrix}
\rho^{\uparrow \uparrow} & \rho^{\uparrow \downarrow}\\
\rho^{\downarrow \uparrow } & \rho^{\downarrow \downarrow}
\end{pmatrix},
\label{matrice densitee}
\end{equation} where $\uparrow , \downarrow $ denote respectively the spin-up and spin-down components.

It is convenient to project the matrix $\mathcal{F}$ onto the Pauli basis set \cite{Barletti2003,Morandi2011}
\begin{equation}
\mathcal{F}= \frac{1}{2}\sigma _0 f_0 + \frac{1}{2}\bm{f} \cdot \bm{\sigma},
\label{change basis wigner function}
\end{equation}
where
\begin{equation}
f _{0} = {\rm Tr} \left\{ \mathcal{F} \right\}  = f ^{\uparrow \uparrow} + f ^{\downarrow \downarrow}, ~~~~
\bm{f}   =  {\rm Tr} \left( \mathcal{F} \bm \sigma \right).
\label{def f0 f_vec}
\end{equation}
Here  ${\bm \sigma}=(\sigma_x, \sigma_y, \sigma_z)$ are the Pauli matrices and $\bm{f} =(f_x, f_y, f_z)$.

With this definition, the particle density $n$ and the spin polarization $\bm{S} $ of the electron gas are easily expressed as moments of the pseudo-distribution functions  $f_0$ and $\bm f$:
\begin{eqnarray}
n(\bm{r},t)
&=&\sum _{\mu}  \left|  \Psi_{\mu}^\dagger (\bm{r},t) \right| ^2
=
\int f_{0} (\bm{r},\bm{v}, t) d\bm{v}, \label{def n} \\
\bm S (\bm{r},t)
&=&
\frac{\hbar}{2} \sum_{\alpha}  \Psi_{\alpha}^\dagger(\bm{r},t)\bm \,{\bm\sigma}\,  \Psi_{\alpha}(\bm{r},t)
= \frac{\hbar}{2}
\int \bm f (\bm{r},\bm{v}, t)  d\bm{v}.\label{def S}
\end{eqnarray}
In this representation, the Wigner functions have a clear physical interpretation: $f_{0}$ is related to the  electron density (in the phase space), whereas $f_{i}$ ($i=x,y,z$) is related to the spin polarization density in the direction $i$.
In other words, $f_{0}$ represents the probability to find an electron at one point of the phase space at a given time, while $f_{i}$ represents the probability that the spin polarization of such electron is directed along in the $i$-th direction.
One can prove the following interesting bound:
\begin{equation}
\left| \bm S (\bm r,t) \right| \leq n(\bm r,t)\frac{\hbar}{2}.  \label{S leq n}
\end{equation}
Equation \eqref{S leq n} is a direct consequence of the property of the density matrix: $Tr \left( \rho^{2} \right) \leq 1$. The equality sign holds true for a pure state or for a system where all the spins are aligned along the same direction (fully spin-polarized electron gas).

The full (quantum) evolution equation for the matrix Wigner function $\mathcal{F}$ was derived in \cite{Hurst2014} and is mathematically very complicated. A better physical insight can be gained from its semiclassical limit, i.e., keeping only terms of order $O(\hbar)$. One obtains:
\begin{align}
&\frac{\partial f_{0}}{\partial t}
+
\bm{v} \cdot \bm{\nabla} f_{0}  - \frac{e}{m} \left( \bm{E} + \bm{v} \times \bm{B} \right) \cdot \bm{\nabla_{v}} f_{0} + \frac{\mu_{B} }{2mc^{2}} \left( \bm{E} \times \bm{\nabla} \right)_{i} f_{i} \nonumber \\
&~
- \frac{\mu_{B}}{m} \bm{\nabla} \left[ B_{i} - \frac{1}{2c^{2}}\left( \bm{v} \times \bm{E} \right)_{i} \right] \cdot \bm{\nabla_{v}} f_{i}  - \frac{\mu_{B} e}{2m^{2}c^{2}} \left[ \bm{E} \times \left( \bm{B} \times \bm{\nabla_{v}} \right) \right]_{i} f_{i} = 0.
\label{vlasov equation f0 avec spin orbit} \\ \nonumber \\
&
\frac{\partial f_{i}}{\partial t}
+
\bm{v} \cdot \bm{\nabla} f_{i}  - \frac{e}{m} \left( \bm{E} + \bm{v} \times \bm{B} \right) \cdot \bm{\nabla_{v}} f_{i} + \frac{\mu_{B} }{2mc^{2}} \left( \bm{E} \times \bm{\nabla} \right)_{i} f_{0} \nonumber \\
&~
- \frac{\mu_{B}}{m} \bm{\nabla} \left[ B_{i} - \frac{1}{2c^{2}}\left( \bm{v} \times \bm{E} \right)_{i} \right] \cdot \bm{\nabla_{v}} f_{0}  - \frac{\mu_{B} e}{2m^{2}c^{2}} \left[ \bm{E} \times \left( \bm{B} \times \bm{\nabla_{v}} \right) \right]_{i} f_{0}  \nonumber \\
&~
- \frac{2\mu_{B}}{\hbar} \left\{ \left[ \bm{B} - \frac{1}{2 c^{2}} \left(\bm{v} \times \bm{E} \right) \right] \times \bm{f} \right\}_{i} = 0.
\label{vlasov equation f avec spin orbit}
\end{align}
where the factor $\hbar$ is hidden in the definition of the Bohr magneton  $\mu_{B} = e \hbar /(2m)$, which precedes all quantum corrections in Eqs. \eqref{vlasov equation f0 avec spin orbit}-\eqref{vlasov equation f avec spin orbit}. Such quantum corrections  couple  the orbital motion with the spin terms, through the Zeeman effect or the spin-orbit interactions. There are no quantum corrections to the orbital electron dynamics, because they would only appear at second order in $\hbar$. Therefore, the orbital motion of the electrons is purely classical and determined by the Lorentz force. In contrast, the spin degree of freedom is treated as a quantum variable.

In summary, the Eqs. \eqref{vlasov equation f0 avec spin orbit} and \eqref{vlasov equation f avec spin orbit} represent the semi-relativistic (order $c^{-2}$) and semi-classical (order $\hbar$) form of the kinetic equations for particles with spin $1/2$, which we shall term the ``spin-Vlasov equations" in the following.

\subsection{Self-consistency}\label{subsec:self}

Equations \eqref{vlasov equation f0 avec spin orbit}-\eqref{vlasov equation f avec spin orbit} can be used, in a mean-field approach, to described the self-consistent spin dynamics of an ensemble of interacting charged particles. In this case, the electric and the magnetic field are solutions of the Maxwell equations:
\begin{align}
\begin{array}{lcl}
\displaystyle
\bm{\nabla} \cdot \bm{E} &=& \displaystyle \frac{\rho}{\epsilon_{0}} - \frac{\bm{\nabla} \cdot \bm{P} }{\epsilon_{0}},  \\
\bm{\nabla} \cdot \bm{B} &=& \displaystyle 0, \\
\bm{\nabla} \times \bm{E} &= &\displaystyle-\frac{\partial \bm{B}}{\partial t},  \\
\bm{\nabla} \times \bm{B} &=&\displaystyle \mu_{0} \bm{j} + \mu_{0} \epsilon_{0}\frac{\partial \bm{E}}{\partial t}  + \mu_{0} \frac{\partial \bm{P}}{\partial t} + \mu_{0}\bm{\nabla} \times \bm{M}.
\end{array}
\label{maxwell equations}
\end{align}
where $\bm{P}$ and $\bm{M}$ are, respectively, a polarization and a magnetization vector. These terms arise because, in the Foldy-Wouthuysen procedure, both the Dirac Hamiltonian and the wave functions are expanded in powers of $1/c$. The expansion of the wave function to order $c^{-2}$, entails  that the charge and current densities are modified, as was shown using a variational method in \cite{Dixit2013,Manfredi2013}. The resulting source terms are the following:
\begin{align}
\rho   &= -e \int f_{0} d\bm{v},\label{def density}  \\
\bm{j} &= -e \left[ \int  \bm{v} f_{0} d\bm{v} + \frac{\bm{E} \times \bm{M} }{2mc^{2}}   \right], \label{def current generalized} \\
\bm{M} &= - \mu_{B}\int \bm{f} d\bm{v}, \label{def magnetization}  \\
\bm{P} &=  -\frac{\mu_{B}}{2c^{2}} \int \bm{v} \times \bm{f} d\bm{v} .
\label{def spin polarization}
\end{align}

The spin-Vlasov equations \eqref{vlasov equation f0 avec spin orbit}-\eqref{vlasov equation f avec spin orbit},  coupled to the Maxwell equations \eqref{maxwell equations}, constitute a self-consistent set of equations to describe the spin dynamics of an interacting electron gas in the mean-field approximation.
Such mean-field approach can also be extended, in the spirit of TD-DFT (see Sec. \ref{subsec:wave_dft}), to include exchange and correlation effects by adding suitable potentials and fields that are functionals of the electron density \cite{Manfredi2010}.

\subsection{Scalar representation in an extended phase space}\label{subsec:extphasespace}

There exists another way to include the spin in a phase-space Wigner formalism, which is mathematically different but physically identical to the approach described in Sec. \ref{subsec:wigspin}. The basic idea \cite{Zamanian2010nj,Marklund2010} is to define an extended phase-space distribution $g$ that depends not only on the position $\bm{r}$ and the velocity $\bm{v}$, but also on a unitary vector $\bm{s}$ (defined by two angles on the unit-radius sphere) which represents the spin direction:
\be
g(\bm{r}, \bm{v}, \bm{s}, t) = \frac{1}{4\pi}\sum_{\alpha,\beta=1}^{2}
(\delta_{\alpha, \beta} + \bm{s} \cdot \bm{\sigma}_{\alpha,\beta})\,
\mathcal{F}_{\beta\alpha}(\bm{r}, \bm{v},t),
\label{wigfun_extended}
\ee
where  $\mathcal{F}_{\beta\alpha}$ is the matrix Wigner function as in Eq. \eqref{matrice densitee}, $\delta_{\alpha, \beta}$ is the Kronecker delta function, and $\bm{\sigma}_{\alpha,\beta}$ represents the $(\alpha,\beta)$ element of one of the Pauli matrices $\bm{\sigma}=(\sigma_x, \sigma_y, \sigma_z)$ defined in Eq. \eqref{eq:paulimatrices}.

The distribution $g(\bm{r}, \bm{v}, \bm{s}, t)$ is a scalar function that evolves in the extended phase space, which is therefore eight-dimensional (8D): three positions, three velocities, and two angles for the spin.
This is in contrast with the approach of Sec. \ref{subsec:wigspin}, where the phase space is the usual 6D one ($\bm{r}, \bm{v}$), but the distribution function is a $2\times 2$ matrix.
The correspondence relations between our distribution functions $f_{0}(\bm{r}, \bm{v}, t)$ and $f_{i}(\bm{r}, \bm{v}, t)$ and the scalar distribution $g(\bm{r}, \bm{v}, \bm{s}, t)$ used by Zamanian et al. \cite{Zamanian2010nj} can be written as:
\begin{align}
f_{0} &= \int g\, d\bm{s},~~~~~~ \bm{f} = 3\int \bm{s}\, g \,d\bm{s}.
\end{align}

The semi-relativistic theory in the extended phase space, equivalent to the matrix approach of Eqs. \eqref{vlasov equation f0 avec spin orbit} and \eqref{vlasov equation f avec spin orbit}, was derived in \cite{Asenjo2012} and a fully relativistic extension of this scalar theory was established later in \cite{Ekman2017}.

The semi-relativistic scalar kinetic equations reads as follows:
\begin{eqnarray}
\frac{\partial g}{\partial t} &+& \bm{w} \cdot \nabla g -{e\over m} \left(\bm{E} + \bm{w} \times\bm{B}\right) \cdot \nabla_{\bm{v}} g + {{2\mu_B} \over \hbar}\, \bm{s} \times \overline{\bm{B}}  \cdot \nabla_{\bm{s}}\, g \\
&+& {\mu_B\over m}\,(\bm{s}+\nabla_{\bm{s}}) \cdot \left( \partial^i \overline{\bm{B}}  \, \partial_v^i g \right)
+ \frac{e\hbar^2 }{8m^3 c^2}\, \partial^i(\nabla\cdot \bm{E})\, \partial_v^i g ,
\end{eqnarray}
where the summation over repeated superscripts $i=x,y,z$ is understood and we have defined the quantities
\[
\bm{w}= \bm{v} - {\mu_B \over {2mc^2}}\,\bm{E}\times(\bm{s}+ \nabla_{\bm{s}}), \,\,\,\,\,\,\,\,\,\,
\overline{\bm{B}} = \bm{B}-\frac{\bm{v} \times \bm{E}}{2c^2}.
\]
The deformed magnetic field $\overline{\bm{B}}$  is the same expression that appears in the Eqs. \eqref{vlasov equation f0 avec spin orbit}-\eqref{vlasov equation f avec spin orbit} and corresponds to the first relativistic correction to the Thomas precession \cite{Thomas1926,Dragan2013}.

The deformed velocity $\bm{w}$ is related to the spin-orbit  correction of the velocity operator. Indeed, in the Heisenberg picture, the velocity operator $\bm{\hat{V}}$ is determined by the evolution equation of the position operator $\bm{\hat{r}}$
\be
\bm{\hat{V}} = \frac{1}{i\hbar} \left[ \bm{\hat{r}} , \hat{H} \right] = \frac{\bm{\hat{\pi}}} {m} - \frac{\mu_B}{ 2mc^{2}} \bm{E} \times \bm{\sigma},
\label{velocity operators}
\ee
where we used the Pauli Hamiltonian \eqref{Eq:hamiltonian}. The related phase-space function is
\begin{align}
\bm{V} = \bm{v} - \frac{\mu_B}{2mc^{2}} \bm{E} \times \bm{\sigma},
\label{velocity phase space function}
\end{align}
which is also the velocity appearing in the Eqs. \eqref{vlasov equation f0 avec spin orbit}-\eqref{vlasov equation f avec spin orbit}.

From a computational point of view, it is interesting to note that the matrix and the scalar formalism of the spin-Vlasov equations lend themselves better to two different families of numerical methods. The scalar formalism is better adapted to particle-in-cell (PIC) methods, because the two  degrees of freedom related to the spin only act as two additional labels attached to each particles. The computational cost is therefore only marginally greater than that required for spinless particle.
In contrast, the scalar approach is hardly applicable to grid-based Vlasov codes, as it would require to mesh the extended 8D phase space. Grid-based codes are better suited for the matrix formalism, because the phase space is at most 6D and the only drawback compared to spinless particles is that one has to advance in time four, instead of one, distribution functions.

\section{Beyond the mean field: collisions and relaxation to equilibrium}
\label{sec:collisions}

The mean-field approach described so far is accurate to treat the electron dynamics
on very short timescales ($<100\rm\, fs$), see Fig. \ref{fig:timescales}. On longer timescales
(0.1--1ps), the laser energy is redistributed among the electrons via electron-electron (e-e) collisions, and finally delivered to the ion lattice via electron-phonon (e-ph) collisions.

One of the advantages of phase-space-based methods is that effects going beyond the mean-field approximation (e.g., two-body collisions) can be incorporated with relative ease in the governing equations. This is not the case for wave-function-based methods (Hartree-Fock, DFT), which have an essentially Hamiltonian nature reflected in the unitary propagation of the wave function.

Collisions are dynamical correlations that cannot be included in adiabatic correlation functionals in TD-DFT. There have been some attempts to include potentially dissipative effects in TD-DFT. The most accomplished of such attempts is the so-called time-dependent current-density-functional theory (TD-CDFT) \cite{Vignale97}, which uses the electron current density $\bm{j}(\bm{r},t)$ as the basic building block, instead of the density $n(\bm{r},t)$. However, the equations of TD-CDFT are mathematically very complicated and not of easy implementation in practical situations.

For phase-space methods, the construction of dissipative terms can rely on the experience acquired in plasma physics. Generally speaking, the relevant kinetic equation (Vlasov, Wigner) may be augmented by a collision term of the form:
\[
\left(\frac{\partial f}{\partial t}\right)_{coll} .
\]

We also note that it is conceptually harder to include collisions in fully quantum
models. A significant constraint is that non-unitary corrections to
the Wigner equation should be written in Lindblad form
\cite{Lindblad1976}, which guarantees that the evolved Wigner function always
corresponds to a positive-definite density matrix. This constraint is not always satisfied for the models described below.

In the following, we will discuss three different collision integrals for e-e and e-ph collisions, based on a relaxation term, a Boltzmann-like approach, and a Fokker-Planck approach.

\paragraph{Relaxation methods.} To model e-e
collisions, a relaxation term can be added to the right-hand side of
the Vlasov or Wigner equation \cite{Manfredi2005film}:
\be \left(\frac{\partial f}{\partial t} \right)_{\rm rel} \equiv
-\nu_{ee}(T_e)(f-f_{eq}), \ee
where $\nu_{ee}$ is the average e-e collision rate and
$f_{eq}(\bm{r}, \bm{v})$ is a Fermi-Dirac distribution. The idea behind
this model is that the electron distribution will eventually
relax, on a time scale of the order $\nu_{ee}^{-1}$, towards a
Fermi-Dirac equilibrium $f_{eq}$ with total energy equal to
that of the  electron distribution $f(\bm{r}, \bm{v},t=0^+)$ {\em after} the initial excitation.
For electrons
near the Fermi surface, the e-e collision rate can be written as
\cite{Pines1995}:
\be \nu_{ee}(T_e) = {\rm const.} \times (k_B T_e)^2, \label{nuee1}
\ee
where $T_e$ is the instantaneous electron temperature. The proportionality constant can be estimated  from experimental considerations or extracted from first-principles simulations \cite{Domps1998}.

Equation \eqref{nuee1} conserves the total number of particles, but not the total momentum or energy. It can only ensure that the distribution function relaxes to the correct equilibrium distribution, but its relevance for the nonequilibrium transient is questionable. In order to conserve energy and momentum, the function $f_{eq}[n,\bm{u},T]$ should be a local equilibrium with density $n(\bm{r},t)$, mean velocity $\bm{u}(\bm{r},t)$, and temperature $T(\bm{r},t)$ computed instantaneously from the electron distribution $f(\bm{r}, \bm{v},t)$. Note that the task of computing the local temperature is not trivial for a Fermi-Dirac distribution.

\paragraph{Boltzmann-like methods.}
A Boltzmann-like electron-electron collision integral that respects
Pauli's exclusion principle was devised long ago by \"Uhling and Uhlenbeck \cite{UU}:
\be\left(\frac{\partial f}{\partial t} \right)_{\rm UU}=\int
\frac{d^3 {\bf p_2} d\Omega}{(2\pi\hbar)^3}
~\sigma(\Omega)|v_{12}| (f_1 f_2 \overline{f}_3 \overline{f}_4 -
f_3 f_4 \overline{f}_1 \overline{f}_2)~, \label{vuu} \ee
where $v_{12}$ is the relative velocity of the colliding particles
1 and 2, $\sigma(\Omega)$ is the differential cross section
depending on the scattering angle $\Omega$, and indices 3 and 4
label the outgoing momenta, $f_i = f(\vec{r}, \vec{p}_i , t)$ and
$\overline{f}_i = 1-f_i/2$. This collision term is similar to the
well-known classical Boltzmann collision term but for Pauli
blocking factors $\overline{f}_i \overline{f}_j$. As seen in Sec. \ref{sec:basic}, Pauli blocking plays a major role for electronic systems. At zero electron temperature, all
collisions are Pauli-blocked and the collisional mean free path of
the electrons becomes infinite. However, after a strong excitation, the
distribution function is no longer a Fermi-Dirac one and collisions become possible. The effect
of the above e-e collision term on the semiclassical Vlasov
dynamics in metal clusters was investigated numerically in
\cite{Domps1998}.

\paragraph{Fokker-Planck methods.}
By coupling to the ionic lattice vibrations (phonons), the electrons progressively relax to a thermal
distribution with a temperature equal to that of the lattice
$T_l$. This behavior can be described phenomenologically using the two-temperature model illustrated in Sec. \ref{sec:basic}. The typical relaxation time for metal nano-objects is $\tau_{e-ph} \approx 1\,\rm ps$.
In addition, the lattice acts as an external environment for the electrons, leading to the loss of
quantum coherence over a timescale $\tau_{dec}$ (decoherence time).
The relaxation and decoherence times correspond, respectively, to
the decay of diagonal and nondiagonal terms in the density matrix
describing the electron population.

Such effects can be modeled, in the Wigner representation, by a classical Fokker-Planck (FP) term \cite{Jasiak2010}:
\be
\left(\frac{\partial f}{\partial t}\right)_{\rm
e-ph} = D_r \,\nabla^2_r f+D \,\nabla^2_v f + \gamma \nabla_v \cdot\left({\bf v}\, G[f]\right),
\label{eq:fokkerplanck}
\ee
where $D_{r,v}$ are diffusion coefficients in real space and
velocity space respectively, $\gamma$ is the e-ph relaxation rate, and $G[f]$ is a functional that depends on the statistics
and on the dimensionality of the system.

For classical particles obeying an exclusion principle, the corresponding FP equation was derived by
Kaniadakis et al. \cite{Kania}. For instance, $G[f]=f$ for particles obeying Maxwell-Boltzmann statistics and $G[f]=f(1-f)$ for fermions in 3D. For fermions in 1D, which is the case of the electrons in thin films that we shall consider later, we assume that the electron distribution always remains a Fermi-Dirac one in the transverse directions $(v_y, v_z)$. Then, integrating the collision term \eqref{eq:fokkerplanck}
along these two directions, one obtains the expression $G[f]=f_0[1-\exp(-f/f_0)]$, where $f_0=\frac{3}{4}\frac{n_0}{ v_F}\frac{T_l}{T_F}$, $T_l$ is the lattice temperature, and $f(x,v_x,t)$ is now the 1D distribution function.
The preceding expression must be changed slightly to account for the fact that the Wigner
distribution can be negative: $G[f]=f_0[1-\exp(-|f|/f_0)]{\rm sgn}(f)$.

It can be proven that
$(\partial{f}/\partial{t})_{\rm e-ph} =0$ when the electron distribution is given by a spatially homogeneous 1D Fermi-Dirac distribution ($\mu$ is the chemical potential):
\begin{equation}
f_{eq}(v_x) = \frac{3}{4}\frac{n_0}{ v_F}\frac{T_l}{T_F}\ln
\left[1+\exp\left(-\frac{m v_x^2/2 -\mu}{k_B \,T_l}\right) \right], \label{fd}
\end{equation}
provided $D_v$ and $\gamma$ satisfy the relation: $D_v=\gamma k_B T_l/m$. Thus, the FP term ensures that the electron distribution relaxes to a classical Fermi-Dirac distribution with a temperature equal
to that of the lattice. The latter constitutes a perfect reservoir with
infinite heat capacity, so that $T_l$ remains constant.

As seen in Sec. \ref{sec:basic}, the electron relaxation rate can be written as $\gamma \equiv \tau_{e-ph}^{-1} \propto G/C_e$, where $G$ is the electron-phonon coupling constant appearing in the two-temperature equations \eqref{eq:TTM1}-\eqref{eq:TTM2} and $C_e$ is electron heat capacity.
For an ideal Fermi gas, the heat capacity depends on the temperature as $C_e(T_e)=
\pi^2n_0 k_B(T_e/2T_F)$. In a dynamical simulation, the electron temperature is
a time-dependent quantity that can be computed self-consistently from the Wigner distribution function $f$.

We note that a Vlasov or Wigner equation endowed with the e-ph collision term \eqref{eq:fokkerplanck} can be viewed as the microscopic counterpart of the electron equation \eqref{eq:TTM1} in the phenomenological two-temperature model.
This can be proven easily in the case of a 1D Maxwell-Boltzmann distribution, using the  diffusion operator in the phase space:
\be
\left(\frac{\partial f}{\partial t}\right)_{coll} = D_x \,\frac{\partial^2 f}{\partial x^2} + D_v \,\frac{\partial^2 f}{\partial v^2} + \gamma\, \frac{\partial(vf)}{\partial v}.
\label{eq:fooker-lindblad}
\ee
The above collision term is in the Lindblad form whenever $D_x\,D_v\ge \gamma^2\hbar^2/16m^2$ \cite{Lindblad1976,Zurek2003}.
Taking the first and second velocity moments of the Vlasov or Wigner equation (there are no quantum corrections to these orders), neglecting all electromagnetic fields, and defining the kinetic electron temperature in the usual way: $k_B T_e= m\langle(v-\langle v \rangle)^2\rangle$, where $\langle A \rangle=\int\int f A dx dv$ denotes the phase-space average, one obtains:
\be
\frac{\partial T_e}{\partial t}= \frac{2D_v m}{k_B} -2\gamma T_e \, .
\ee
Then, taking $D_v=\gamma k_B T_l/m$ and $\gamma=\tau_{e-ph}^{-1}=G/(2C_e)$, we get
\be
C_e\,\frac{\partial T_e}{\partial t}= G  (T_l - T_e) \, ,
\ee
which is the electron temperature equation \eqref{eq:TTM1} in the TTM with $\kappa=P = 0$.

Finally, we would like to mention the recent work of Daligault \cite{Daligault2016}, who derived a quantum form of the Landau/Fokker-Planck collision operator.

\section{Application I: Spin corrections to longitudinal plasma waves}
\label{sec:linear}

The linear response of a homogeneous spinless electron gas has been studied theoretically for a long time, starting with the pioneering works of Vlasov \cite{vlasov1938} and Landau \cite{Landau1946}.
The main linear mode is an oscillation at the plasma frequency $\omega_{p} = \sqrt{e^{2} n_{0} / m \epsilon_0}$ (``plasmon") ,  corresponding to the collective motion of the electrons immersed in a neutralizing background of positive ions, which are assumed to be fixed because of the large mass difference between the two species.

\begin{figure}
\centering
 \includegraphics[scale=0.65]{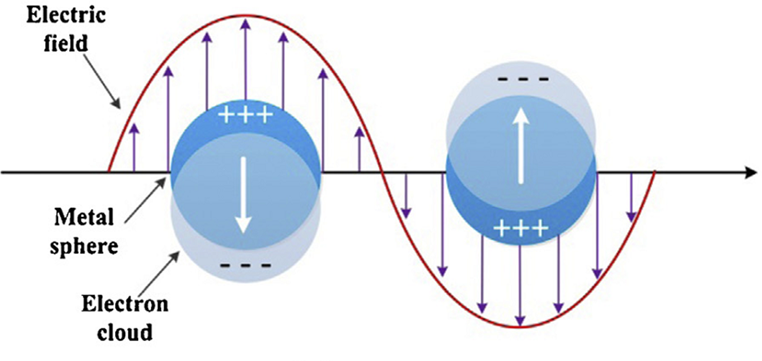}
 \caption{Schematic illustration of a surface plasmon resonance resulting from the collective oscillations of delocalized electrons in response to an external electric field. From \cite{Sun16}. } \label{plasmon_fig}
\end{figure}

In spherical nano-objects, a typical example of plasmonic oscillation is the localized surface plasmon (LSP), which originates from a displacement of the electron gas that creates a net charge imbalance (see Fig. \ref{plasmon_fig}). The resulting Coulomb force pulls the electrons back inside the system but, due to their inertia, they will travel further away, thus recreating a new Coulomb force in the opposite direction. After a few cycles, the plasma oscillations are usually damped away through Landau damping, which results from the mixing of single-electron oscillations at slightly different frequencies.
LSPs have been observed in all sorts of nano-objects \cite{Scholl2012,Luo2013,Tame2013,Raza2013,Tanjia2018} and are one of the basic features of the fast-developing field of plasmonics \cite{manfredi2018}.

Here, our aim is to show a possible signature of the electron spin polarization on the plasmonic oscillations.

\subsection{General linear response theory}

In this section, we will derive the dielectric function of the spin-Vlasov equations coupled to the relevant Maxwell equations and then find the associated dispersion relation.
In general, this is a very complicated task. Here, we simplify the problem by assuming that the equilibrium ground-state distribution is homogeneous and isotropic and that the electron spins are polarized along the $z$ direction only (collinear approximation). Therefore, the diagonal components of the Wigner matrix for the spin-up and spin-down equilibria can be written as (only one arrow is used, to simplify the notation):
\be
f^{\uparrow  (0)}(|\bm{v}|) = \mathcal{G} \left(m |\bm{v}|^{2}/2 + \mu_{B} B \right),
\,\,\,\,\,\,\,\,\,
f^{\downarrow  (0)}(|\bm{v}|) = \mathcal{G} \left(m |\bm{v}|^{2}/2 - \mu_{B} B \right),
\label{eq:equilibrium}
\ee
where $\mathcal{G}$ can be a Maxwell-Boltzmann or a Fermi-Dirac distribution and $\bm{B}= B \bm{e}_z$ is a uniform magnetic field parallel to the $z$ axis.
From Eq. \eqref{def f0 f_vec}, one has: $f_0^{(0)}=f^{\uparrow(0)}+f^{\downarrow(0)}$ and $f_z^{(0)}=f^{\uparrow(0)}-f^{\downarrow(0)}$.

The uniform magnetic field in Eqs. \eqref{eq:equilibrium} creates a difference between the spin-up and spin-down  distributions; hence, it creates a net spin polarization at equilibrium. However, as it is homogeneous in space, it has no influence on the ensuing dynamics.
The equilibrium distribution functions also verify  the following properties: $\int f_{0}^{(0)} d\bm{v} = n_{0}$ and  $\int f_{z}^{(0)} d\bm{v} = m_{0}$, where $n_{0}$ and $m_{0}$ are, respectively, the electron density and magnetization at equilibrium.

In order to obtain a tractable system of equations, we need to make several assumptions. First, we neglect the effect of the magnetic field on the orbital motion: thus, there is no Lorentz force $\bm{v}\times \bm{B}$ in the Vlasov equations. Second, we consider purely longitudinal plasma waves, with wave vector $\bm{k} = k \bm{e}_x$.
Therefore, all physical quantities only depend on the spatial coordinate $x$, and the corresponding phase space $(x,v)$ is 2D, where $v \equiv v_{x}$ stands for the $x$-component of the velocity. As was mentioned before, the electron gas is polarized along the $z$ direction.
With these assumptions, the spin-Vlasov equations \eqref{vlasov equation f0 avec spin orbit}-\eqref{vlasov equation f avec spin orbit}, together with Maxwell's equations \eqref{maxwell equations}, become:
\begin{align}
&\frac{\partial f_{0}}{\partial t} + v ~\partial_{x}f_{0}
- \frac{1}{ m } \partial_{x}\left(- e \phi + V_{XC} \right) \partial_{v} f_{0}   - \frac{\mu_{B} }{m} \partial_{x}\left(B_{XC} + B_{z}\right)  \partial_{v} f_{z}  =  0,  \label{f0_evo_vlasov linear analysis}  \\
&\frac{\partial f_{z}}{\partial t} +  v ~\partial_{x}f_{z}
- \frac{1}{ m } \partial_{x} \left(- e \phi + V_{XC} \right) \partial_{v} f_{z}-
 \frac{\mu_{B}  }{ m}  \partial_{x}  \left(B_{XC} + B_{z}\right)  \partial_{v} f_{0}
  =0, \label{falpha_evo_vlasov linear analysis} \\
&\frac{\partial^{2} \phi}{\partial x^{2}} =  \frac{e}{\epsilon_{0}} \left( \int f_{0} dv -n_{0} \right),  ~~~~~~~
\frac{\partial B_{z}}{\partial x} = - \mu_B \mu_{0} \frac{\partial}{\partial x} \left(\int f_{z} dv\right), \label{poisson-Ampere linear analysis}
\end{align}
where $V_{XC}[n]$ and $B_{XC}[n]$ are the exchange and correlation functionals \cite{Maurat2009}, and $B_z$ is the self-consistent magnetic field.

In order to perform the linear wave analysis, we expand all quantities around the equilibrium configuration. For the distribution functions, we have
\be
f_{0}(x,v,t) = f^{(0)}_0(v) +  f_0^{(1)}(x,v,t)~~~\textrm{and}~~~f_z(x,v,t) = f^{(0)}_z(v) +  f_{z}^{(1)}(x,v,t),
\label{def delta f}
\ee
and for the exchange and correlation potentials
\be
V_{XC} = V_{XC}^{(0)} + \left[\partial_{n} V_{XC}\right]^{(0)} \int  f_{0}^{(1)} dv + \left[\partial_{m} V_{XC}\right]^{(0)} \int  f_{z}^{(1)}  dv
\equiv V_{XC}^{(0)} + V_{XC}^{(1)}, \label{V echange devloppement lineaire}
\ee
and similarly for $B_{XC}$.
Inserting Eqs. \eqref{def delta f}-\eqref{V echange devloppement lineaire} into Eqs. \eqref{f0_evo_vlasov linear analysis}-\eqref{poisson-Ampere linear analysis} and neglecting second-order terms, we obtain the linearized spin-Vlasov equations.
We then follow the standard Landau procedure of Fourier transforming in space and Laplace transforming in time. This leads to the dielectric function of the system
\begin{align}
\epsilon\left(\omega,k\right) &= 1+ \frac{\omega_{p}^{2}}{k n_{0}} \mathcal{I}_{0} - \frac{k \mu_{B}^{2} \mu_{0}}{m} \mathcal{I}_{0} \nonumber\\
&+ \frac{k}{m}  \left[ \left(\partial_{n}V_{XC} + \mu_{B} \partial_{m}B_{XC} \right)  \mathcal{I}_{0} +  \left(\partial_{m}V_{XC} + \mu_{B} \partial_{n}B_{XC} \right)  \mathcal{I}_{z}  \right]\nonumber \\
&+ \frac{\mu_{B}k^{2}}{m^{2}} \left[\left(\partial_{n}V_{XC}\right) \left(\partial_{m}B_{XC}\right)- \left(\partial_{n}B_{XC}\right) \left(\partial_{m}V_{XC}\right) \right]\left[  \mathcal{I}_{0}^{2} -  \mathcal{I}_{z}^{2} \right]\nonumber \\
&+ \left[ -\frac{\omega_{p}^{2} \mu_{B}^{2} \mu_{0}}{n_{0}m} + \frac{\mu_{B}\omega_{p}^{2}}{n_{0}m} \partial_{m}B_{XC} - \frac{k^{2} \mu_{B}^{2} \mu_{0}}{m^{2}} \partial_{n}V_{XC}\right] \left[\mathcal{I}_{0}^{2} -  \mathcal{I}_{z}^{2} \right],
\label{dielectric function}
\end{align}
where $\omega$ is the frequency of the perturbation and $k$ the wave number, and the integrals
\be
\mathcal{I}_{0,z}\left(\omega,k\right) = \int \frac{ \partial_{v} f_{0,z}^{(0)}}{\omega - kv} dv
\label{integrals i0 and iz def}
\ee
only depend on the ground-state properties.
The zeros of the dielectric function determine the dispersion relation of the system: $\omega=\omega(k)$.

Finally, one should specify the form of the exchange and correlation functionals. The minimal requirement is to neglect correlations and take an exchange functional that is spin dependent and local in space and time \cite{Gunnarsson1976}:
\begin{align}
V_{X}[n,m] &= -\frac{e^{2}}{4 \pi\epsilon_{0}}\left( \frac{3}{4 \pi} \right)^{1/3} \left[ \left(\frac{ n + m}{2} \right)^{1/3} +  \left( \frac{ n - m}{2}\right)^{1/3}\right], \label{echange local V} \\
\mu_{B}B_{X}[n,m] &= -\frac{e^{2}}{4 \pi\epsilon_0}\left( \frac{3}{4 \pi} \right)^{1/3} \left[ \left(\frac{ n + m}{2} \right)^{1/3} -  \left( \frac{ n - m}{2} \right)^{1/3}\right].  \label{echange local B}
\end{align}
The above functionals are the exact solutions of the Hartree-Fock equations in the case of homogeneous electronic densities. In our case, we are close to this situation, since we study perturbations around homogeneous ground states.

\subsection{Fermi-Dirac ground state}
\label{subsec:FD}
In the case of electrons in metals, the spin-up and spin-down ground states obey a Fermi-Dirac distribution in one dimension, obtained after integrating the 3D Fermi-Dirac function over the transversal velocities: $f^{1d} (v_x) = \int  \int f^{3d}(\bm{v}) dv_{y} dv_{z}$. Renaming $v \equiv v_x$, we obtain the ground state distributions:
\begin{align}
f^{\uparrow^{(0)}}(v) &= \frac{2 \pi k_{B} T}{m} \left(\frac{m}{2 \pi \hbar}\right)^{3} \ln\left[1 + \exp \left( -\frac{1}{k_{B}T}\left(\frac{m}{2}v^{2} + \mu_B B - \mu \right) \right) \right], \label{eq:fd1d_1}\\
f^{\downarrow^{(0)}}(v) &= \frac{2 \pi k_{B} T}{m} \left(\frac{m}{2 \pi \hbar}\right)^{3} \ln\left[1 + \exp \left( -\frac{1}{k_{B}T}\left(\frac{m}{2}v^{2} - \mu_B B - \mu \right) \right) \right],
\label{eq:fd1d_2}
\end{align}
where a magnetic field $\bm{B}=B \bm{e}_z$ is present in order to yield a spin-polarized system. The magnetic field can be either external or internal, i.e. originating from the local magnetic exchange interaction between the localized ions and the itinerant electrons. The chemical potential $\mu$ is set to fix the electron density to $n_0$, i.e. $\int \left( f^{\uparrow^{(0)}} + f^{\downarrow^{(0)}} \right) dv = n_{0}$. This integration can be performed exactly only in the zero-temperature limit, leading to the following identity:
\begin{align}
n_0 & = \frac{4\pi}{3}\left( \frac{m}{2\pi \hbar}\right)^3 \left(v_{F_{+}}^{3/2} + v_{F_{-}}^{3/2} \right),
\label{n0 FD T=0K}
\end{align}
where $v_{F_{\pm}} = \Re\left\{\sqrt{2\left[\mu_{T=0} \pm \mu_B B\right]/m}\right\}$, $\mu_{T=0}$ is the chemical potential at zero temperature, and $\Re$ denotes the real part. In the same way, we can express the magnetization at equilibrium as follows:
\begin{align}
m_0 = \int \left( f^{\uparrow^{(0)}} - f^{\downarrow^{(0)}} \right) dv = \frac{4\pi}{3}\left( \frac{m}{2\pi \hbar}\right)^3 \left(v_{F_{-}}^{3/2} - v_{F_{+}}^{3/2} \right).
\label{m0 FD T=0K}
\end{align}

\begin{figure}
	\centering \includegraphics[scale=0.25]{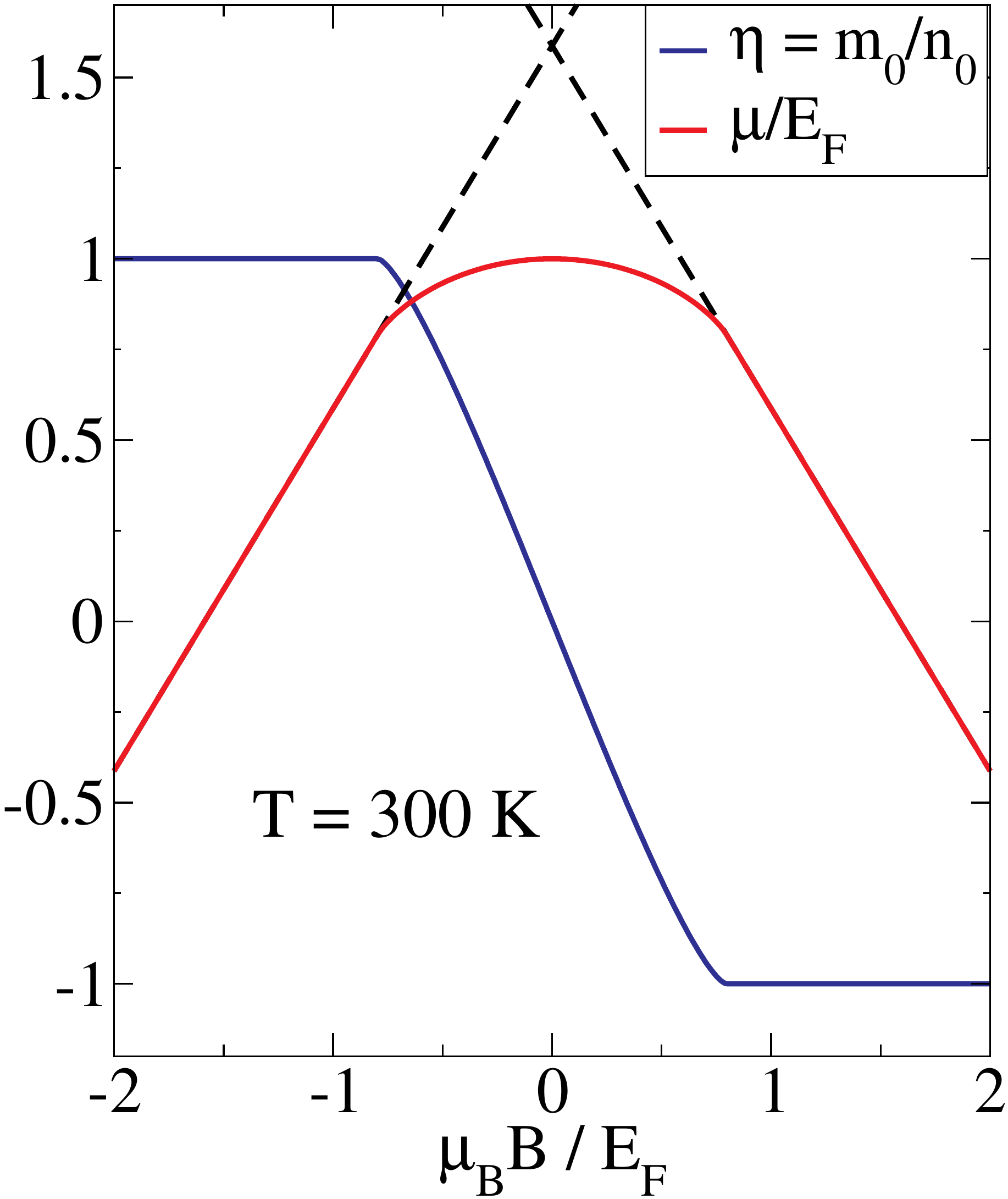}
    \caption{Chemical potential (red line) and electron magnetization (blue line) as a function of the external magnetic field in the case of a Fermi-Dirac ground state. The calculations were performed for an electron gas with density $n_{0} = 5.9 \times 10^{28}\,\rm m^{-3}$ (gold) and temperature $T=300\,\rm K$. The dashed lines represent analytical solutions valid for $|\mu_BB| \gg E_{\rm F}$. } \label{mu_vs_B}
\end{figure}

In Fig. \ref{mu_vs_B} we plot the polarization at equilibrium:
\begin{align}
\eta \equiv \frac{m_0}{n_0} = \frac{v_{F_{-}}^{3/2} - v_{F_{+}}^{3/2}}{v_{F_{-}}^{3/2} + v_{F_{+}}^{3/2}},
\end{align}
as a function of the magnetic field.
For a magnetic field larger than $E_{F}/\mu_B$, the electron gas is completely polarized.
Note that, for metals, one typically has $E_{F}/\mu_B \approx 10^5\,\rm T$, which is a huge value for a magnetic field generated by external coils. However, internal magnetic fields due to the interatomic exchange interaction can reach such large values, an effect that is at the basis of ferromagnetism.

In the case of a totally unpolarized ($\eta=0$) or fully polarized ($\eta=1$) electron gas, Eq. \eqref{n0 FD T=0K} can be inverted exactly, leading to the following expressions for the chemical potential:
\begin{align}
\mu_{T=0} &= E_{F} ~~~~\left(\textrm{for}~\eta=0 \right), \label{muT0a}\\
 \mu_{T=0}& = 2^{2/3}E_{F} \pm \mu_B B  ~~~~\left(\textrm{for}~|\eta|=1 \right).
 \label{muT0}
\end{align}
For intermediate cases ($0<|\eta|<1$), one has to find $\mu_{T=0}$ numerically.
In  Fig. \ref{mu_vs_B}, we plot the chemical potential as a function of the applied magnetic field in the case of a Fermi-Dirac distribution with a temperature equal to $300\,\rm K$ and a density relevant to metals.

In general, the full determination of the dispersion relation has to be done numerically. However in the limit of zero temperature and long wavelengths, i.e. $k \ll |\omega|/v_F$, one can develop the integrals $\mathcal{I}_0$ and $\mathcal{I}_z$, defined in Eq. \eqref{integrals i0 and iz def}, in a power series of $k v_{F_{\pm}}/\omega $, leading to:
\begin{align}
\mathcal{I}_{0} \left(\omega,k\right) &= -\frac{n_0k}{\omega^2} - \frac{3}{5}\frac{k^3}{\omega^4} n_0 v_{F}^2 \left(\frac{\mu_{T=0}}{E_F} - \frac{\mu_BB}{E_F}\eta\right) +\mathcal{O}\left(\left(\frac{k v_{F_{\mp}}}{\omega}\right)^{5}\right), \\
\mathcal{I}_{z} \left(\omega,k\right) &= -\frac{m_0k}{\omega^2} - \frac{3}{5}\frac{k^3}{\omega^4} m_0 v_{F}^2 \left( \frac{\mu_{T=0}}{E_F} - \frac{\mu_BB}{\eta E_F}\right) +\mathcal{O}\left(\left(\frac{k v_{F_{\mp}}}{\omega}\right)^{5}\right).
\end{align}
Using these expressions together with the dielectric function defined in Eq. \eqref{dielectric function}, we get the following dispersion relation:
\begin{align}
\omega^2 &=
\omega_p^2
+ \frac{3}{5}k^2 v_{F}^2 \left(\frac{\mu_{T=0}}{E_F} - \frac{\mu_B B}{E_F}\eta\right)
- \frac{e^2}{4\pi \epsilon_0}
\frac{k^2}{6m}\left(\frac{3n_0}{\pi}\right)^{1/3} \bigg{[}\left(1+\eta\right)^{4/3} \nonumber \\
&+
\left(1-\eta\right)^{4/3} \bigg{]}
- \frac{ n_0k^2}{m} \eta^2\mu_B^2 \mu_0.
\label{dispertion relation Fermi Dirac}
\end{align}
Equation \eqref{dispertion relation Fermi Dirac} constitutes an extension of the celebrated Bohm-Gross dispersion relation in the case of a fully degenerated ($T=0$) spin-polarized electron gas with exchange and magnetostatic interactions.
All the dispersive terms depend on the electron density $n_{0}$ and the spin polarization $\eta$. The second term on the right-hand side corresponds to the electron degeneracy pressure. The third and fourth terms represent, respectively, the exchange and the magnetostatic contributions.

In order to assess the relative importance of each term, it is useful to express
Eq. \eqref{dispertion relation Fermi Dirac} in non-dimensional form:
\be
{\omega^2 \over \omega_p^2} =
1+ k^2 \lambda_{TF}^2 \left\{\frac{3}{5}\frac{\mu_{T=0}}{E_F} - \frac{3}{5}\frac{\mu_B B}{E_F}\eta -C_X \, \frac{r_s}{a_0}\, f(\eta)
-{\eta^2 \over 4} \frac{\lambdabar_{C}^2}{\lambda_{TF}^2} \right\},
\label{dispertion relation Fermi Dirac_norm}
\ee
where $C_X=2/(12\pi^2)^{5/3}\approx 0.017$, $f(\eta)=(1+\eta)^{4/3}+(1-\eta)^{4/3}$, $\lambdabar_C=\hbar/(mc) =3.86 \times 10^{-4}\rm\, nm$ is the reduced Compton wavelength, and
$\lambda_{TF} = v_F/\omega_p$ is the Thomas-Fermi screening length.
The first term in the curly brackets is the dominant one. The next term is small except for very large magnetic fields, of the order of $E_F/\mu_B\approx 10^5\,\rm T$ for metallic densities. The third term is due to the spin-dependent electron-electron exchange interaction. It is a beyond-mean-field contribution, as is evidenced from the fact that it is proportional to the quantum coupling parameter $g_q \sim r_s/a_0$ (see Sec. \ref{sec:basic}). The fourth term originates from the self-consistent magnetostatic contribution, see Eq.  \eqref{poisson-Ampere linear analysis}, and it is almost always negligible, since for metals $\lambda_{TF}\approx 0.1\,\rm nm$ (Table \ref{tab:goldparam}).
All these corrections have an opposite sign with respect to the leading term, but they can never render the right-hand side of Eq. \eqref{dispertion relation Fermi Dirac_norm} negative. Indeed, when $r_s/a_0 \gg 1$, correlation effects (here neglected) start playing a role, so that the resulting frequency is always real.

Interestingly, even in the absence of magnetic interactions, the dispersion relation still depends on the spin polarization. Indeed, in the limit of an unpolarized $(\eta=0)$ or fully polarized $(|\eta|=1)$ electron gas and using Eqs. \eqref{muT0a}-\eqref{muT0}, the dispersion relation becomes:
\begin{align}
\omega^2 &= \omega_p^2 + \frac{3}{5}\,k^2 v_{F}^2 ~~~~\left(\eta = 0\right), \\
\omega^2 &= \omega_p^2 + 2^{2/3}\frac{3}{5}\,k^2 v_{F}^2 ~~~~ \left(|\eta|  = 1 \right).
\end{align}
This behavior is specific to the Fermi-Dirac distribution function and is absent in the case of a Maxwell-Boltzmann equilibrium, as we shall see in Sec. \ref{subsec:MB}. A spin-polarization dependence of the longitudinal Langmuir modes was predicted in Ref. \cite{Hurst2014fluid}, where it was shown, using a fluid model, that the pressure tensor is spin-dependent.

To obtain an explicit form of the dispersion relation, we have used the long wavelength approximation ($k \ll |\omega|/v_F$), which allowed us to avoid the singularity in the integrals of Eq. \eqref{integrals i0 and iz def}. As is well known, the correct treatment of the singularity leads to a complex $\omega$ and the appearance of Landau damping. However, this is a delicate mathematical issue when considering a Fermi-Dirac distribution \cite{Vladimirov2011}, due to the particular form of the Fermi-Dirac function, which has several singularities in the complex plane. We could not obtain a satisfactory solution to this problem and thus leave it open for the time being.
In the next section, we will study  a Maxwell-Boltzmann distribution, for which the calculations can be performed all along, at least numerically.

\subsection{Maxwell-Boltzmann ground state}
\label{subsec:MB}
The plasma dispersion relation for spinless systems, obtained from the Vlasov-Poisson equations with a Maxwell-Boltzmann equilibrium, was intensively  studied in the plasma physics literature since the seminal work of Landau. In that case, the dispersion relation can be fully evaluated even in the strong damping regime.

For a system of electrons endowed with spin, the Maxwell-Boltzmann distributions read as:
\begin{align}
f_{0}^{(0)}\left( v \right) &= \frac{n_{0}}{ v_{T} \sqrt{\pi}} \exp \left[-\frac{m}{2k_BT} v^{2}  \right], \\
f_{z}^{(0)}\left( v \right) &= \frac{n_{0}}{v_{T} \sqrt{\pi}} \tanh \left( \frac{\mu_B B}{k_BT} \right) \exp \left[-\frac{m}{2k_BT} v^{2} \right] \equiv \eta \,f_{0}^{(0)}\left( v \right) ,
\end{align}
where $v_{T} = \sqrt{2 k_{B} T / m}$ is the thermal speed and $\eta = \tanh \left[\mu_{B} B / \left(k_{B}T \right) \right]$ represents the spin polarization of the electrons gas.
In this case, the integrals $\mathcal{I}_{0}$ and $\mathcal{I}_{z}$ in Eq. \eqref{integrals i0 and iz def} become:
\begin{align}
\mathcal{I}_{0}
&=
-\frac{2n_{0}}{v_{T}^{3}\sqrt{\pi}} \int \frac{v \exp \left(-\frac{m}{2k_{B}T} v^{2}  \right)}{\omega -kv}\,dv , ~~~~~
\mathcal{I}_{z} = \eta\mathcal{I}_{0}.
\label{I0 and Iz maxwell boltzmann}
\end{align}
We use the Landau contour method \cite{Lyu2014}  to compute the integral in Eq.\ \eqref{I0 and Iz maxwell boltzmann}. For this purpose, we introduce the plasma dispersion function
\begin{align}
Z\left(\varrho\right) = \frac{1}{\sqrt{\pi}} \int_{\Gamma} \frac{\exp\left(-z ^{2} \right)}{z-\varrho} dz,
\label{plasma dispersion function}
\end{align}
where $\Gamma$ is a contour in the complex plane following the real axis at infinity and passing under the singularity $z=\varrho$. This function, originally introduced by Fried and Conte \cite{Fried1961}, is well defined for  $\Im \left( \omega \right)>0 $ and can be analytically continued in the lower part of the complex plane, i.e. for $\Im \left( \omega \right) <0 $. Then, the integrals $\mathcal{I}_{0}$ and $\mathcal{I}_{z}$ can be expressed in terms of the plasma dispersion function:
\begin{align}
\mathcal{I}_{0} &=  \frac{2n_{0}}{k v_{T}^{2}} \left[1 + \frac{\omega}{k v_{T}} Z\left(\frac{\omega}{k v_{T}}\right) \right] = - \frac{n_{0}}{k v_{T}^{2}}Z'\left(\frac{\omega}{k v_{T}}\right),
\label{I0 plasma function}
\\
\mathcal{I}_{z} &= - \eta  \frac{n_{0}}{k v_{T}^{2}}Z'\left(\frac{\omega}{k v_{T}}\right),
\label{Iz plasma function}
\end{align}
where the following property was used:
$
Z'\left(\varrho\right) =-2\varrho Z\left( \varrho \right) -2.
$
Using the above expressions for $\mathcal{I}_{0}$ and $\mathcal{I}_{z}$ and the local density approximation of $V_X$ and $B_X$ [see Eqs. \eqref{echange local V}-\eqref{echange local B}], one obtains the following dielectric function for the Maxwell-Boltzmann equilibrium:
\begin{align}
\epsilon\left(\omega,k\right)
&=
1- \frac{1}{v_{T}^{2}} \left[\frac{\omega_{p}^{2}}{k^{2}} - \frac{ \mu_B^{2} \mu_{0} n_{0}}{m } + 2 \frac{n_{0}}{m}  \left(\partial_{n}V_{X} + \eta \partial_{m}V_{X}  \right )\right] Z'\left(\frac{\omega}{k v_{T}}\right)  \nonumber \\
&~+
\left(\frac{n_{0}}{mv_{T}^{2}}\right)^{2} \left[\left(\partial_{n}V_{X}\right)^{2} - \left(\partial_{m}V_{X}\right)^{2} \right]
\left( 1 - \eta^{2} \right) Z'\left(\frac{\omega}{k v_{T}}\right)^{2}
\label{dielectric function mb I0 Iz} \\
&~+
\left(\frac{n_{0}}{kv_{T}^{2}}\right)^{2}
\left[  \frac{\omega_{p}^{2}}{n_{0}m} \partial_{n}V_{X} -\frac{\omega_{p}^{2} \mu_B^{2} \mu_{0}}{n_{0}m} - \frac{k^{2}  \mu_B^{2} \mu_{0}}{m^{2}} \partial_{n}V_{X}\right]
\left( 1 - \eta^{2} \right) Z'\left(\frac{\omega}{k v_{T}}\right)^{2}.
\nonumber
\end{align}

\begin{figure}
	\centering \includegraphics[scale=0.3]{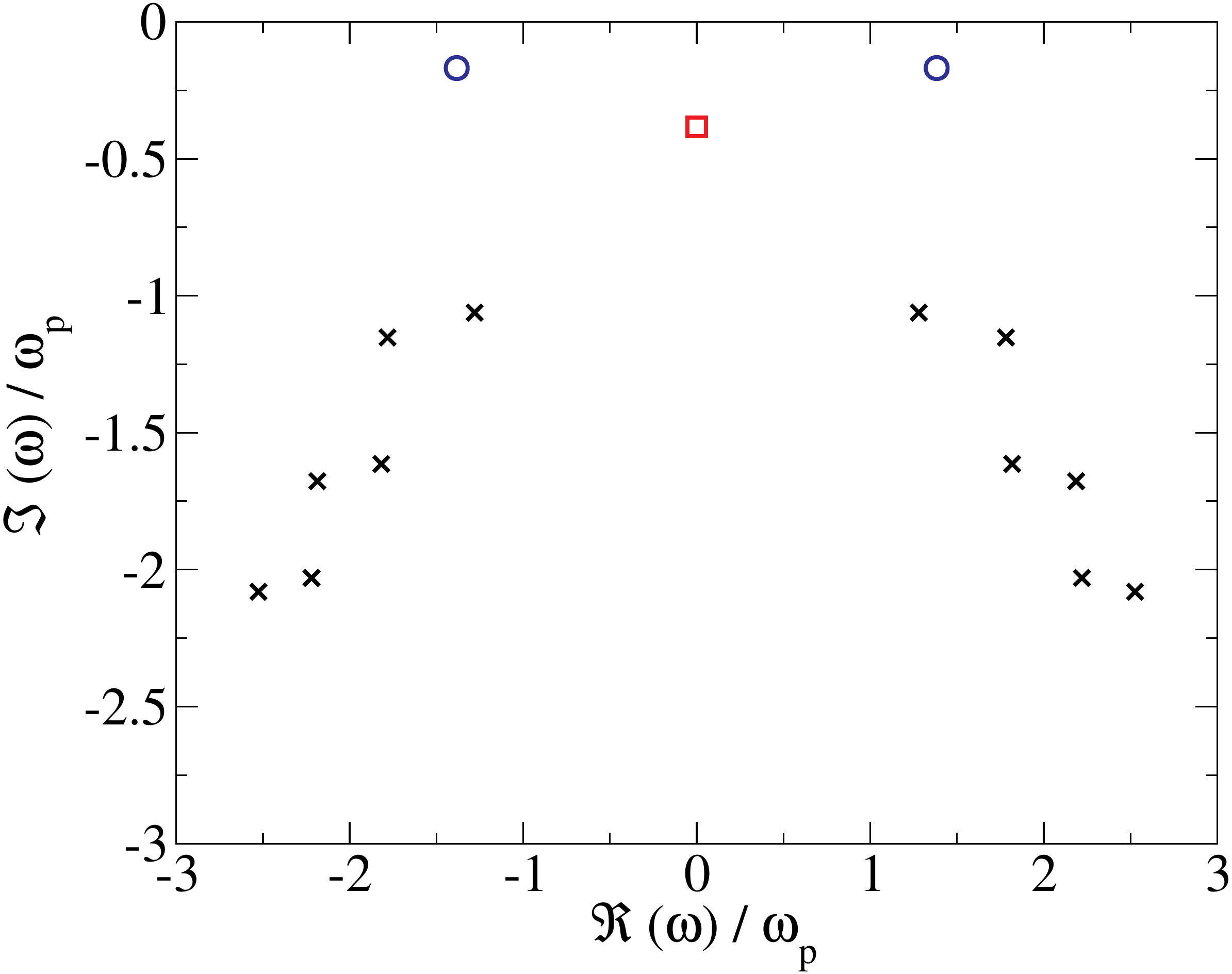}
    \caption{Zeros of the dielectric function, Eq. \eqref{dielectric function mb I0 Iz}, in the complex plane. The  symbols correspond to different types of modes: plasmon modes (blue circles), paramagnon mode (red square), and strongly damped modes (black crosses). The following parameters are used: $n_{0} = 5.9 \times 10^{28}\,\rm m^{-3}$, $T=64\,000\,\rm K$, and $k=0.4 k_{D}$, where $k_D= 2\pi/\lambda_D$. } \label{all zero mb}
\end{figure}

\begin{figure}
	\centering \includegraphics[scale=0.4]{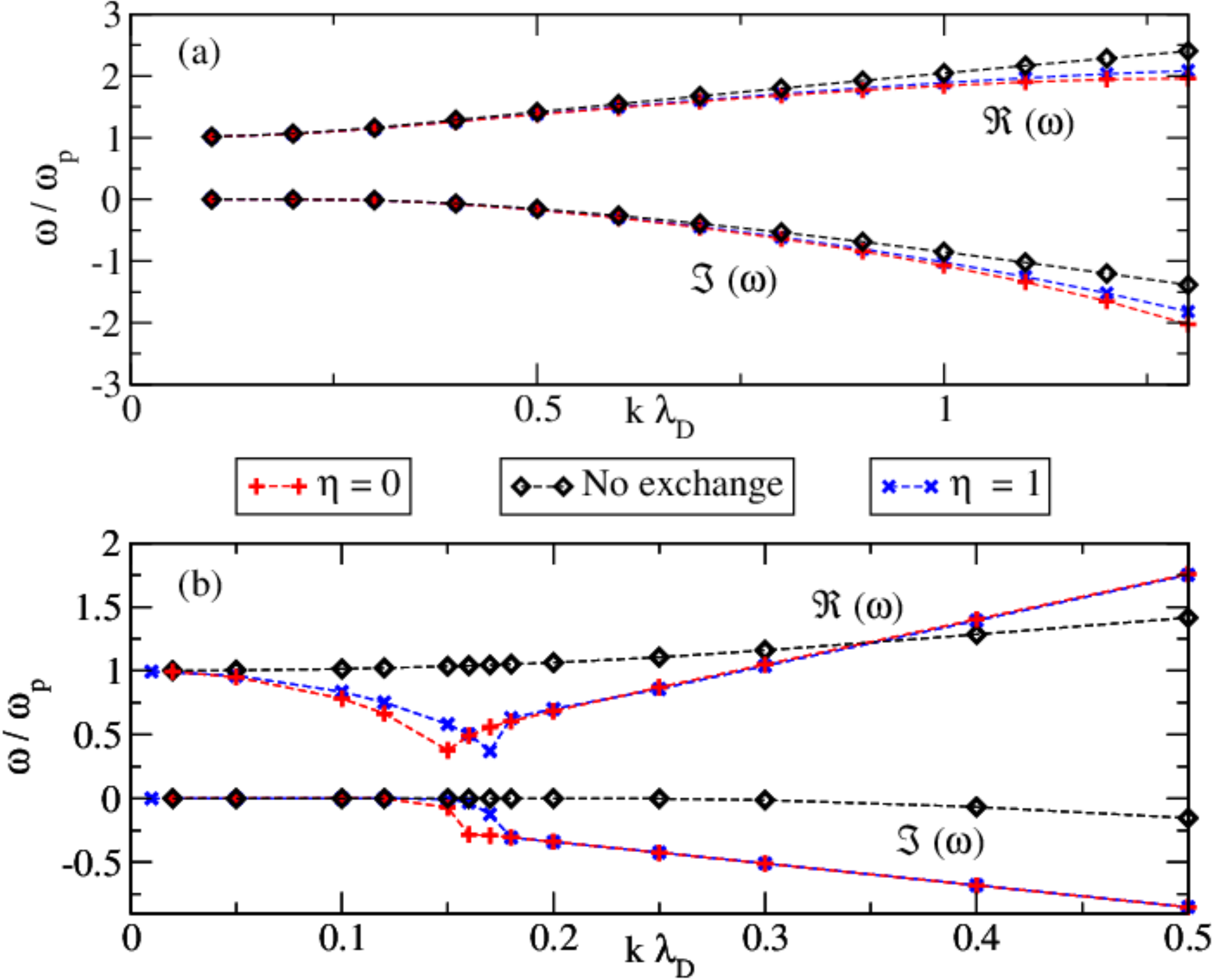}
    \caption{Complex-valued frequency $\omega(k)$ for two different physical systems. (a) Electron gas with metallic density $n_{0} = 5.9 \times 10^{28}\,\rm m^{-3}$ (gold) and high temperature $T= T_F=64\,000\,\rm K $; (b)  Low-density electron gas with $n_{0} = 10^{22}\,\rm m^{-3}$ and  $T=300\,\rm K$.     For both panels, the upper lines represent the real part of $\omega$, whereas the lower lines represent the imaginary part. The black lines refer to the case without exchange interactions, the blue lines to a fully polarized system ($\eta=1$), and the red lines to an unpolarized system ($\eta=0$).
 } \label{disp mb}
\end{figure}

The zeros of the dielectric function provide the dispersion relation $\omega \left( k\right)$ of the system. The latter can be obtained analytically in the long wavelength limit, i.e., for $k \ll |\omega|/v_T$ and $\omega \in \Re$. In that case, the derivative of the plasma dispersion function can be developed in a power series:
\begin{align}
Z'\left(\frac{\omega}{kv_T}\right) \simeq \frac{k^2 v_T^2}{\omega^2} + \frac{3}{2}\frac{k^4v_T^4}{\omega^4} + \mathcal{O}\left( \frac{k^6v_T^6}{\omega^6} \right),
\end{align}
leading to the following dispersion relation:
\begin{align}
\omega^2 &=
\omega_p^2
+ \frac{3k_B T}{m}k^2
- \frac{e^2}{4\pi \epsilon_0}
\frac{k^2}{6m}\left(\frac{3n_0}{\pi}\right)^{1 \over 3} \left[\left(1+\eta\right)^{4\over 3}
+ \left(1-\eta\right)^{4\over 3} \right]
- \frac{ n_0k^2}{m} \eta^2\mu_B^2 \mu_0.
\label{dispertion relation maxwell boltzmann}
\end{align}
We recover almost the same expression as with the Fermi-Dirac ground state, see Eq. \eqref{dispertion relation Fermi Dirac}.

To obtain the full dispersion relation, we  use ZEAL \cite{Kravanja2000}, a mathematical software package for computing the zeros of analytic functions. The dielectric function is parameterized by the electron density $n_{0}$ and the temperature $T$. We first consider an electron gas having a metallic density $n_{0} = 5.9 \times 10^{28}\,\rm m^{-3}$ (gold). Since the corresponding Fermi temperature is very high ($T_F=64\,000\,\rm K$), we need to use a temperature of the same order of magnitude for the Maxwell-Boltzmann approximation to be valid. Here, we take exactly $T=T_F$.
The zeros of the dielectric function, Eq. \eqref{dielectric function mb I0 Iz}, are shown in Fig. \ref{all zero mb}. They are symmetrically located with respect to the vertical axis, which is due to the symmetry properties of the plasma dispersion function. We also note that there exist an infinite number of zeros (black crosses) for which the imaginary part of the frequency is larger than the corresponding real part. These solutions are irrelevant for the dynamics, as they are rapidly damped away. Finally, there are three other zeros that have a different origin. Two of them, represented by blue circles, correspond to the plasmon response at the plasma frequency. For small value of $k$, they have a real part larger than the imaginary part, therefore they are weakly damped and govern the dynamics of the system according to the long-wavelength dispersion relation Eq. \eqref{dispertion relation maxwell boltzmann}.

The last zero (red square) has a vanishing real part and disappears when the magnetic exchange interactions are removed. This zero corresponds to a {\em paramagnon} mode, i.e., an oscillation in antiphase of the two spin populations (up and down), with no associated charge excitation.
For infinite systems, it was shown that paramagnons have a vanishing real frequency \cite{Jones1985}, which is in accordance with the result of Fig. \ref{all zero mb}. Their real frequency can become nonzero for systems with finite size \cite{Yin2009}.

Further, exchange interactions do not only lead to paramagnons, but also modify the plasmonic excitations.
In Fig. \ref{disp mb}, the plasma dispersion relation is depicted for two different physical systems. The first (Fig. \ref{disp mb}, top panel) corresponds to the case of electrons in a metal, with the density of gold $n_{0} = 5.9 \times 10^{28}\,\rm m^{-3}$ and temperature $T=64\,000\,\rm K$. As already mentioned, the temperature had to be set very high for the Maxwell-Boltzmann distribution to be a valid approximation. The second system (bottom panel) corresponds to an artificial case where the electron density is about $n_{0}  = 10^{22}\,\rm m^{-3}$. For such a system, the Fermi temperature is close to $2\,\rm K$ so that, at room temperature (here, we take $T=300\,\rm K$), it is justified to use a Maxwell-Boltzmann equilibrium.

In the metallic case (Fig. \ref{disp mb}, top panel), the resonant frequency is dominated by the electrostatic interactions, while the exchange interactions do not change significantly the dispersion relation. They may decrease slightly the value of the frequency for larger values of $k$, but in this case the plasma oscillations are strongly damped. Moreover, there are no appreciable changes in the dispersion relation depending on the spin polarization of the electron gas.
In contrast, in the idealized low-density system (Fig. \ref{disp mb}, bottom panel) exchange interactions play a significant role. For small wave numbers, the plasmon frequency decreases instead of increasing and we note a signature of the spin polarization, in particular a shift in $k$ for the minimum of the plasmon frequency.

\section{Application II: Nonlinear electron dynamics in thin metal films}
\label{sec:films}
Thin metal films have long been the object of intense investigations in many subfields of nanophysics.
They constitute key components in many modern technologies, ranging from integrated circuits to sensors. From a more fundamental point of view, due to the symmetry reduction and the associated electron confinement, thin metal films exhibit properties (both static and dynamic) remarkably different form their bulk counterparts. They are usually made of noble (Au, Ag, ...) or simple metals (Na, K, ...) and their sizes typically vary from a few nanometers to several hundred nanometers \cite{Maniyara2019}.
As a notable example, the first experimental signature of ultrafast demagnetization was observed in thin nickel films in 1996 \cite{Beaurepaire1996}.

In this section, we illustrate the usefulness of phase-space methods to study the ultrafast electron dynamics in thin metal films. The particular geometry of thin films, whereby their thickness is much smaller than the transversal size, means that they can be described by a one-dimensional (1D) model, at least as far as longitudinal modes are concerned. Here, we mainly report on earlier work of ours on this topic, spanning the years from the early 2000s until today. The simulations are of increasing complexity. In Sec. \ref{subsec:filmvlasov}, the electron motion is purely classical and follows the Vlasov equation, although the ground state is described by a Fermi-Dirac distribution. The main effect observed in these simulations is the emergence of a nonlinear ballistic mode due to bunches of electrons that travel across the film at the Fermi speed of the metal. Section \ref{subsec:filmwigner} revisits this effect using a fully quantum Wigner model and it is a good example of interplay between wave-function methods (used to compute the ground state) and phase-space methods (used to compute the dynamics). The Wigner simulations allow us to identify the parameter range where the ballistic oscillations may be observed. Finally, in Sec \ref{subsec:filmspin}, the description of the electron dynamics is augmented to include the spin degrees of freedom, according to the theoretical considerations presented in Sec. \ref{subsec:wigspin}. The inclusion of these effects allowed us to study ballistic modes that include a magnetic component, thus exploring
the possibility to generate a spin current in the film via an electromagnetic pulse.

\begin{figure}
  \centering \includegraphics[scale=0.2]{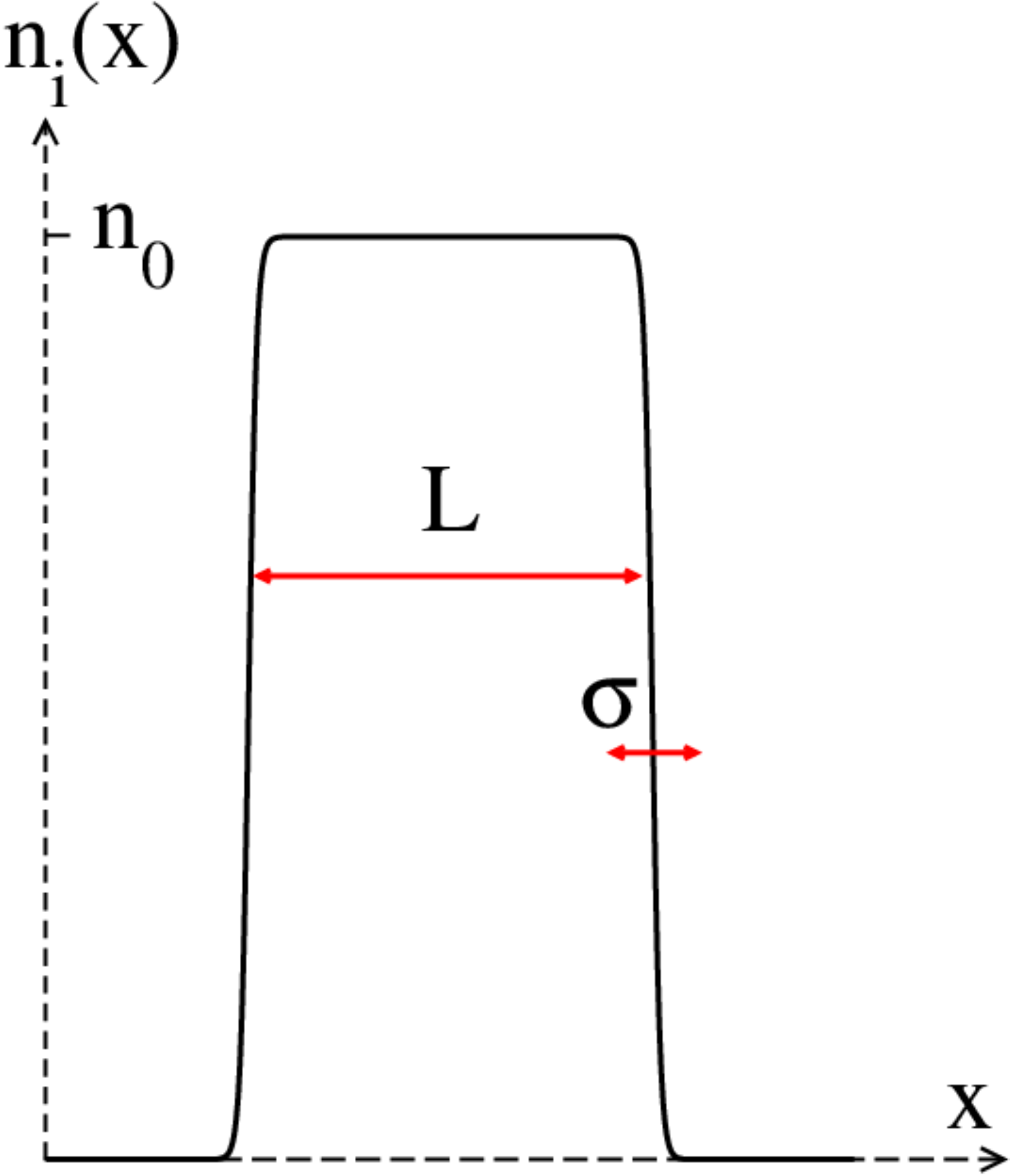}
    \caption{Schematic view of the ion density profile in a thin metal film.} \label{schema ni}
\end{figure}

\subsection{Thin metal films: Vlasov}
\label{subsec:filmvlasov}

Several experiments have shown \cite{Brorson1962,Suarez1995} that electron transport in thin metal films occurs on a femtosecond timescale and involves ballistic electrons traveling at the Fermi velocity of the metal $v_F$. Later, a regime of low-frequency nonlinear oscillations, corresponding to ballistic electrons bouncing back and forth on the film surfaces, was measured in transient reflection experiments on thin gold films \cite{Liu2005}. These experimental findings were corroborated by the Vlasov-Poisson simulations carried out in our research group \cite{Manfredi2004,Manfredi2005film,Manfredi2005optics}, which are summarized in this section.

In these simulations, the electrons are initially prepared in a 1D Fermi-Dirac equilibrium at finite (typically, room) temperature:
\begin{equation}
f_{e}(x,v_x,t=0) = \frac{3}{4}\frac{n_0}{ v_F}\frac{T_e}{T_F}\ln
\left(1+e^{ (\epsilon-\mu)/T_e} \right),
\label{fd1}
\end{equation}
where $\epsilon(x,v_x)=mv_x^2/2-e\phi(x)$ and $\phi$ is the electrostatic self-consistent potential, solution of the Poisson equation
\be
\label{poisson}
\frac{\partial^2 \phi}{\partial x^2} = -\frac{e}{\varepsilon_0}\left[n_i(x) - \int_{-\infty}^{+\infty} f_e dv_x \right] .
\ee
For the ion charge distribution, we use a jellium model with continuous density
$n_{i}(x) = n_{0}/ \left( 1+\exp \left[ \left(|x| -L/2 \right) \sigma\right]\right)$,
where  $L$ is the thickness of the film, $n_{0}$ is the ion density of the bulk metal and $\sigma \approx 0.1\, \rm nm$ represents the typical distance over which the density falls to zero on the border of the film.
The density profile is sketched in Fig. \ref{schema ni}.
The ions are supposed to be fixed during the electron motion because their typical timescale is significantly longer than that of the electrons. The effect of ionic vibrations (phonons) will be addressed in Sec. \ref{subsec:filmwigner}.

\begin{figure}
  \centering \includegraphics[scale=0.55]{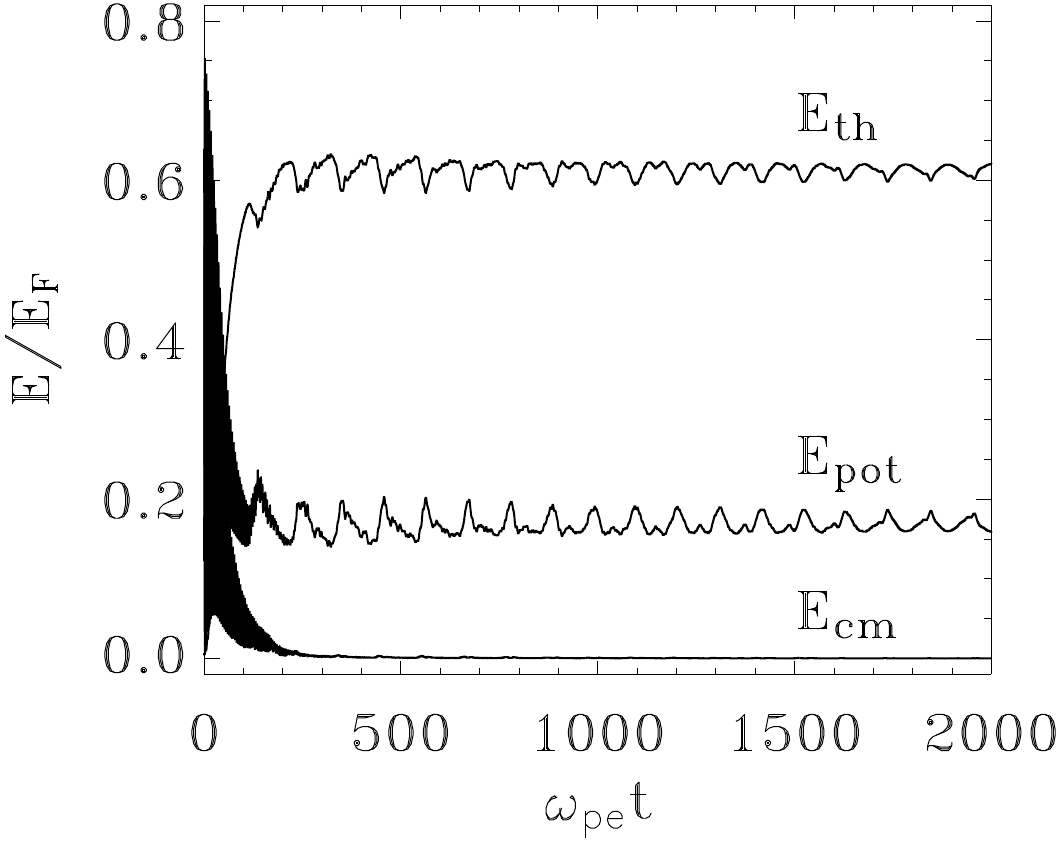}
    \caption{Time evolution of the thermal, potential (electrostatic) and center-of-mass energies. From \cite{Manfredi2010}.} \label{PAH:lowfreq}
\end{figure}

The electrons are subsequently excited by imposing a constant velocity shift $\delta v= 0.08 v_F$ to the initial equilibrium distribution, after which they evolve according to the 1D Vlasov-Poisson equations. This scenario is appropriate when no linear momentum is transferred parallel to the plane of the surface (i.e., $k_{\parallel} = 0$), so that only longitudinal modes can be excited. As a reference case, we studied a sodium film ($r_s=4a_0$, where $a_0$ is the Bohr radius) with initial temperature $T_e = 0.008\,T_F \approx 300$ K and thickness $L = 10 \lambda_{TF} \approx 12\,\rm nm $, where $\lambda_{TF}$ is the Thomas-Fermi screening length.

We consider the evolution of several energy quantities: (i) The self-consistent potential  energy $E_{pot} = \frac{\varepsilon_0}{2} \int (\partial_x \phi)^2 dx$, (ii) the center-of-mass energy $E_{cm}=\frac{1}{2}\int{\frac{j_e^2(x,t)}{n_e(x,t)}~dx}$
(where $j_e = \int v_x f_e dv_x$ is the electron current), and (iii) the thermal energy $E_{th}$.
The latter is defined as: $E_{th} = E_{kin} -E_{cm} -E_{TF}$, where the total kinetic energy is $ E_{kin}= {m\over 2}\int  v_x^2f_e dv_x$. The Thomas-Fermi energy, defined as $E_{TF}=\frac{E_F}{5}\int{n_e(x)^{5/3}dx}$, represents the kinetic energy of a state with same density but vanishing temperature and mean velocity, so that $E_{kin} (T_e=0) = E_{TF}$.

During an initial rapidly oscillating phase, the  center-of-mass energy is almost entirely converted into thermal energy, due to Landau damping, whereas the thermal and potential energies reach a plateau. This plateau is characterized by low-frequency oscillations with a period close to $50\omega_p^{-1} \approx 5.3$ fs. This period is close to the time of flight of electrons traveling ballistically through the film at the Fermi velocity of the metal.
The existence of such nonlinear ballistic mode is the main novel observation reported in Refs. \cite{Manfredi2004,Manfredi2005film}. Later, we proposed to exploit this ballistic mode to enhance the absorption of  infrared light in the film \cite{Manfredi2005optics}.

\begin{figure*}
  \centering \includegraphics[scale=0.5]{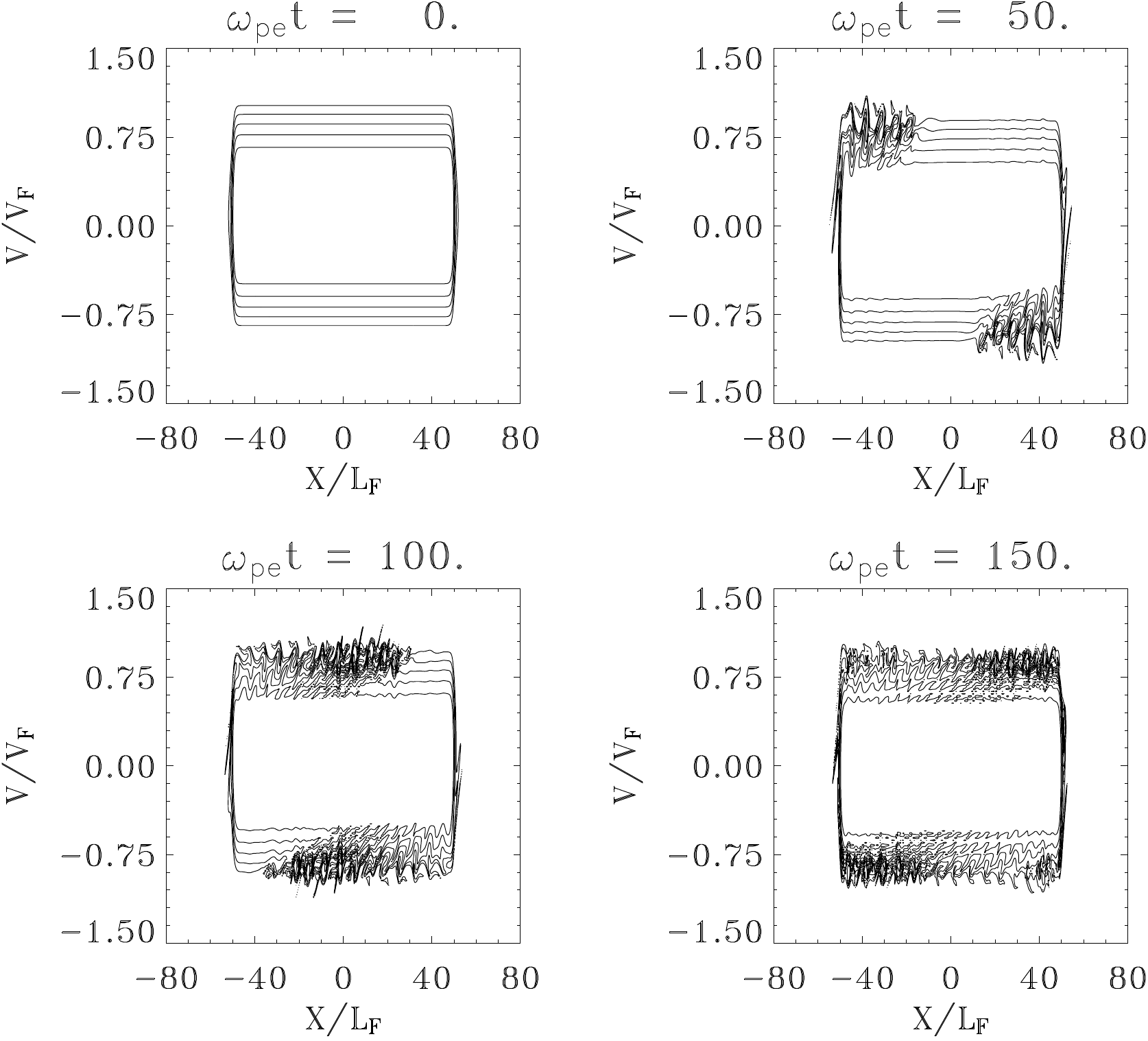}
    \caption{Phase-space portraits of the electron distribution function at different times. Here, and in the forthcoming figures, $L_F$ is the same as the Thomas-Fermi screening length $\lambda_{TF}$. From \cite{Manfredi2010}.} \label{PAH:phasespace}
\end{figure*}

The phase-space portraits of the electron distribution, shown in Fig. \ref{PAH:phasespace}, clearly reveal that the perturbation starts at the film surfaces and then proceeds inward at the Fermi velocity of the metal. The structure formation at the Fermi surface, which has spread over the entire film for $\omega_p t > 150$, is responsible for the increase of the thermal energy (and thus the electron temperature) observed in Fig. \ref{PAH:lowfreq}. As no collisional terms are present, the low-frequency oscillations of Fig. \ref{PAH:lowfreq} are never completely damped away.

\begin{figure}
\centering
\includegraphics[scale=0.65]{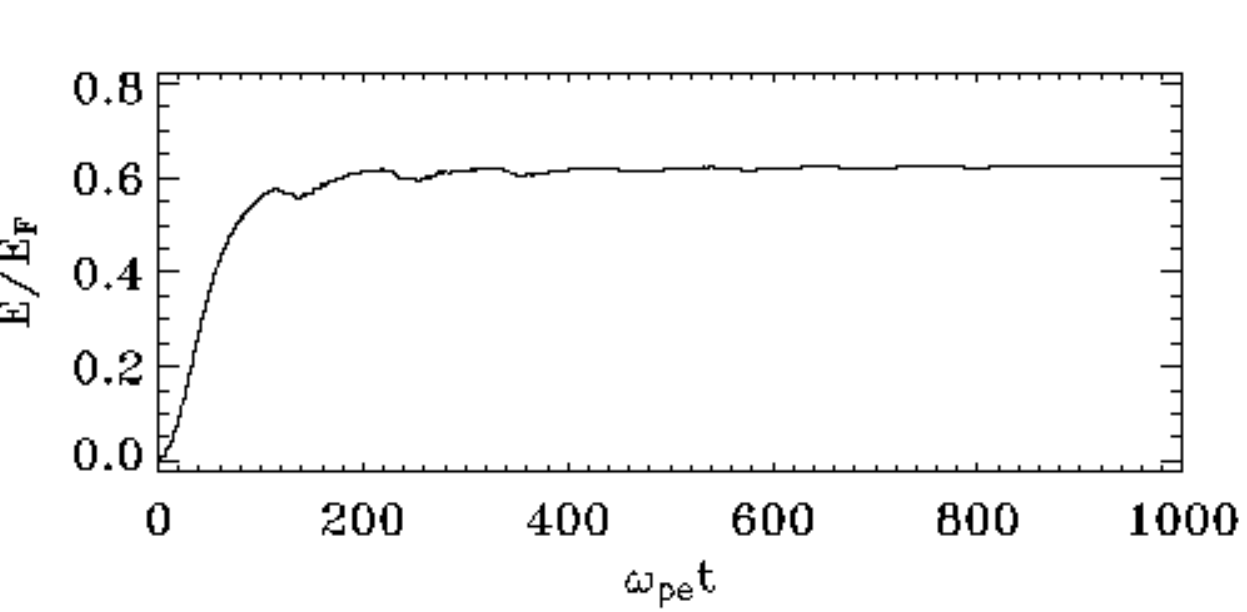}
\caption{Evolution of the thermal energy for a case with e-e
collisions. From \cite{Manfredi2010}.} \label{fig:eth-coll}
\end{figure}

In order to include the effect of e-e collisions in a simple manner \cite{Manfredi2010}, we have augmented the Vlasov equation by a relaxation term, as described in Sec. \ref{sec:collisions}. The constant in Eq. \eqref{nuee1} is taken to be $\approx 0.4~{\rm fs^{-1}eV^{-2}}$, which was obtained from first-principles numerical simulations of the electron dynamics in sodium clusters \cite{Domps1998}. The electron temperature is computed instantaneously
during the simulation, and plugged into the expression for the
collision rate (\ref{nuee}).
When this collision model is applied to our case, the slow ballistic oscillations of Fig. \ref{PAH:lowfreq} are still observed, but they are damped away on a timescale of the
order of $500\omega_{pe}^{-1} \simeq 50\rm fs$ (see Fig. \ref{fig:eth-coll}).

\subsection{Thin metal films: Wigner}
\label{subsec:filmwigner}

One of the most important properties of metal films, when their thickness is reduced to only a few nanometers, is the quantization of the energy levels along the direction perpendicular to the film surface (quantum size effects). The continuum that characterizes the bulk materials is replaced by a discrete spectrum of quantum states with their energies and wave functions depending on the boundary conditions \cite{Eguiluz1984}. The consequent modifications of the electronic structure with respect to a classical description lead to significant changes in their physical properties and, as shown in the following, their dynamical behaviour.

Our numerical study of the quantum electron dynamics in thin metal films is performed in two steps: first, the ground state of the electron population (possibly at finite temperature) is determined self-consistently; subsequently, the equilibrium distribution is perturbed by injecting some energy into the system. The quantum dynamics is then described by solving the time-dependent Wigner equation (\ref{wignerequation}) coupled to the Poisson equation. Although the Wigner approach is a fully quantum-mechanical description, it is intrinsically ill-suited to deal with stationary states (indeed, quantization rules are overlooked in the Wigner formalism and must be imposed as additional constraints). Therefore it is more convenient to determine the ground-state from a standard density functional approach (Kohn-Sham equations) and then to construct the corresponding Wigner function from the computed Kohn-Sham wave functions.

\begin{figure}
\centering
\includegraphics[scale=0.3]{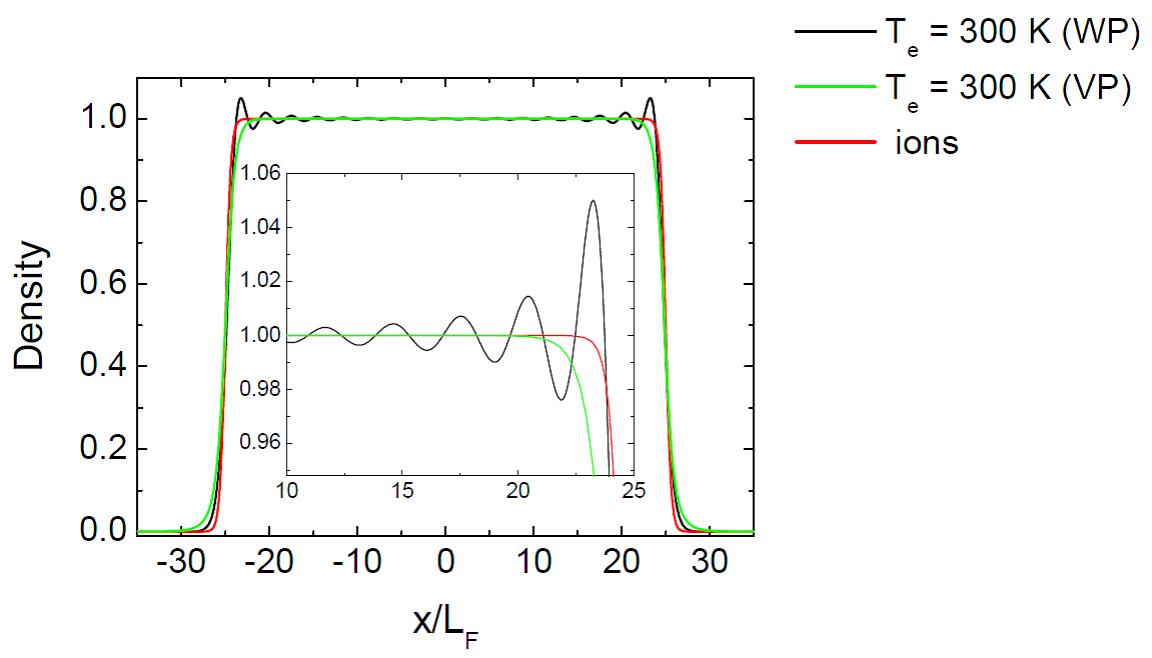}
\caption{Ground-state electron and ion densities for a sodium film with $L = 50\lambda_{TF}$ and $T_e=300\,\rm K$ in the classical (VP) and quantum (WP) regimes. The inset shows a zoom near the film surface.} \label{fig:grounddensity}
\end{figure}

The electron gas is assumed to be at thermal equilibrium with temperature $T_e$ and, as in the Vlasov approach described above in Sec. \ref{subsec:filmvlasov}, while the ions are described by a continuous, immobile density with soft edges, as in Fig. \ref{schema ni}. The electrons obey a Fermi-Dirac distribution at the corresponding temperature (for more details about the model, see \cite{Jasiak2009}). The Kohn-Sham equations (\ref{Kohn-Sham}) are solved numerically using a finite-difference iterative method \cite{Eguiluz1984}. Some typical density profiles are shown in Fig. \ref{fig:grounddensity}. In this example $N=18$ Kohn-Sham orbitals are occupied. The classical Vlasov-Poisson result (VP) yields a smooth density profile, without spatial oscillations. In the quantum regime (labeled WP), the standard Friedel oscillations are observed, particularly near the film surface (see the inset). Note that the profiles are weakly dependent on the electron temperature, because all considered cases are in the strongly degenerate regime $T _e \ll T_F$.

\begin{figure}
\centering
\includegraphics[scale=0.22]{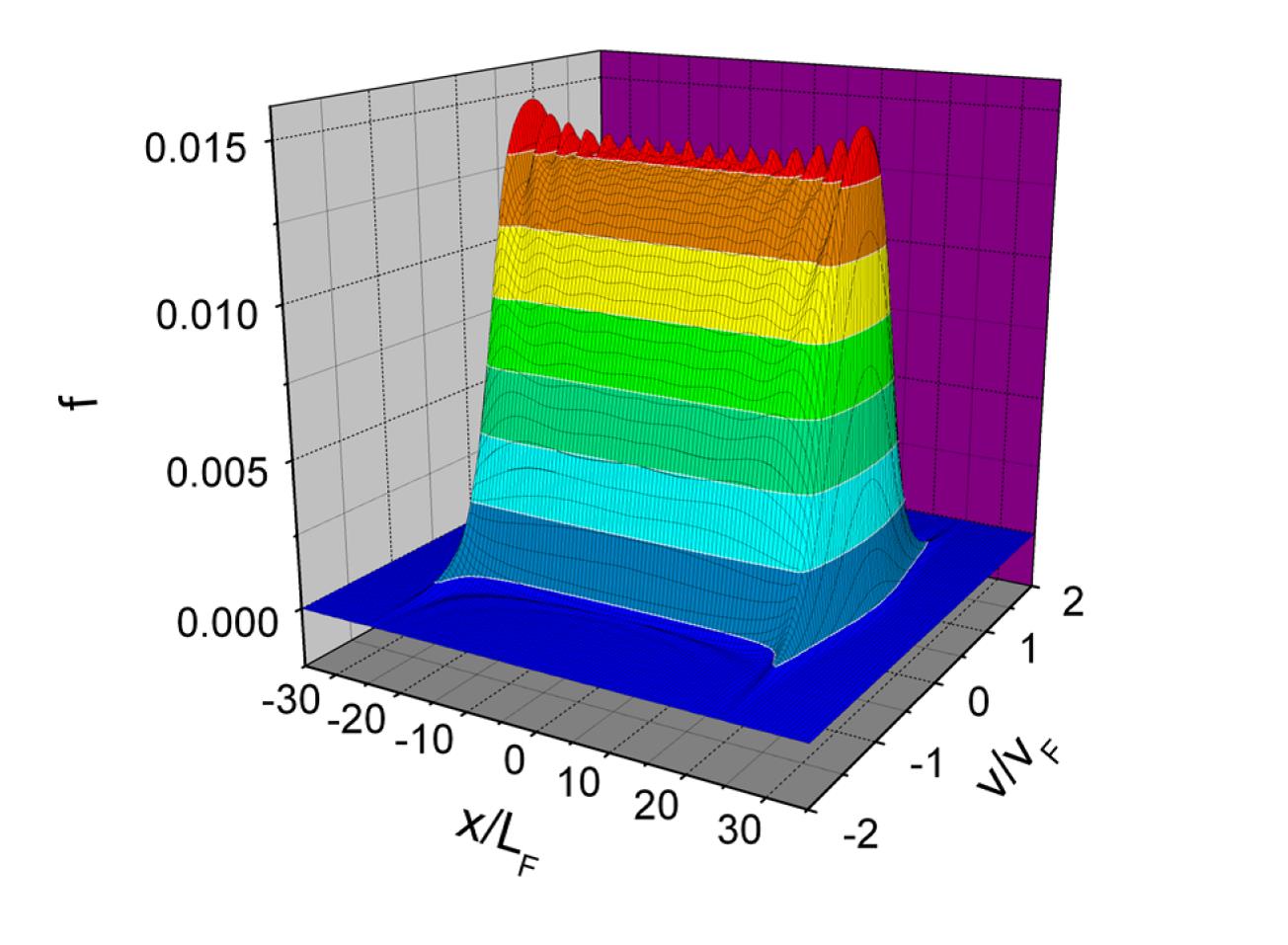}
\caption{Ground-state Wigner function $f(x,v)$ for a sodium film with $L = 50\lambda_{TF}$ at $T_e = 300\,\rm K$.} \label{fig:groundstate}
\end{figure}

After computing the $N$ Kohn-Sham wave functions $\phi_k(x)$ and their respective occupation numbers $p_k$ \cite{Jasiak2009}, we can proceed to construct the ground-state Wigner function $f(x,v)$. We first compute the partial Wigner functions corresponding to each Kohn-Sham wave function $\phi_k$ of the ground-state
\begin{equation}
f_k(x,v)= \frac{m}{2 \pi \hbar} \int_{-\infty}^{+\infty} d\lambda\, \phi^*_k \left(x+\frac{\lambda}{2}\right) \phi_k\left(x-\frac{\lambda}{2}\right) e^{imv}  \;.
\end{equation}
The total Wigner function (Fig. \ref{fig:groundstate}) is given by the sum of the partial Wigner functions, weighed by the occupation numbers $p_k$, $f(x,v) = \sum_{k=1}^N p_k f_k(x,v)$.

To study the electron dynamics, we make use of the evolution equation for the Wigner function \eqref{wignerequation}. In the classical limit (i.e., taking $\hbar \rightarrow 0$) this equation reduces to the Vlasov equation that was used in our previous works, summarized in Sec. \ref{subsec:filmvlasov}. The Wigner results were obtained with a numerical code based on a uniform grid covering the relevant phase space \cite{Suh1991}. The method combines a splitting technique with fast Fourier transforms in the velocity coordinate.

\begin{figure}
\centering
\includegraphics[scale=0.15]{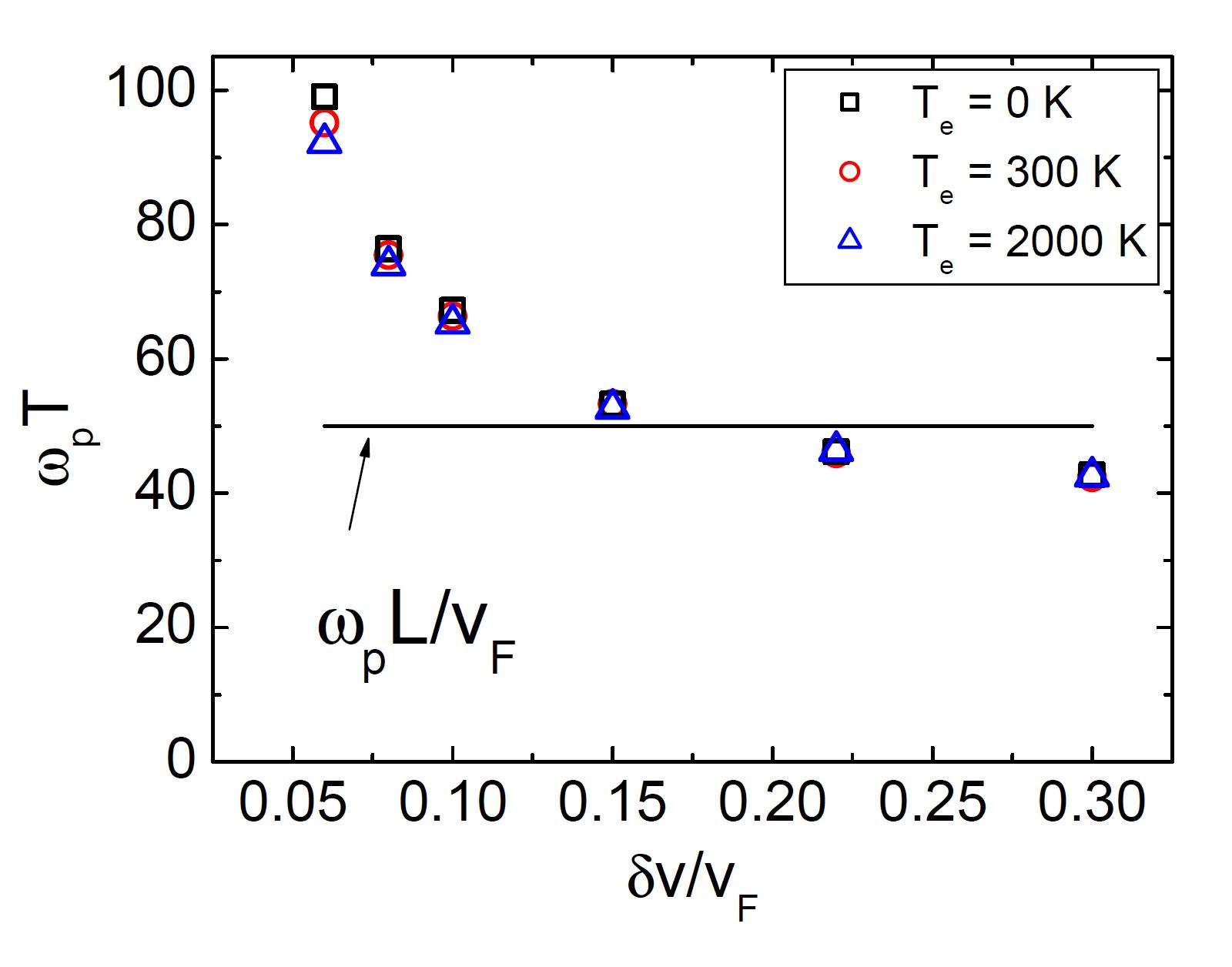}
\caption{Period of the low-frequency ballistic oscillations  as a function of the perturbation $\delta v$, for Wigner-Poisson simulations of a thin film of thickness $L = 50\lambda_{TF}$ and different temperatures in the range $T_e = 0-2000$ K.} \label{fig:period}
\end{figure}

Similarly to the Vlasov simulations described in Sec. \ref{subsec:filmvlasov}, the electron dynamics is excited by perturbing the ground-state equilibrium. Again, this is done by imparting a uniform velocity shift $\delta v$ to the initial distribution.
The ensuing electron dynamics also reveals some low-frequency oscillations for the thermal and potential energies, as in the Vlasov case.
By plotting the observed period $T$ of the low-frequency oscillations against the value of the perturbation (see Fig. \ref{fig:period}), we observe an interesting feature. While, for a large excitation $\delta v$, the period is close to the ballistic value $T = L/v_F$ (as in the Vlasov simulations, see Sec. \ref{subsec:filmvlasov}), it diverges from this classical value below a certain excitation threshold $\delta v_{th}$

\begin{figure}
\centering
\includegraphics[scale=0.35]{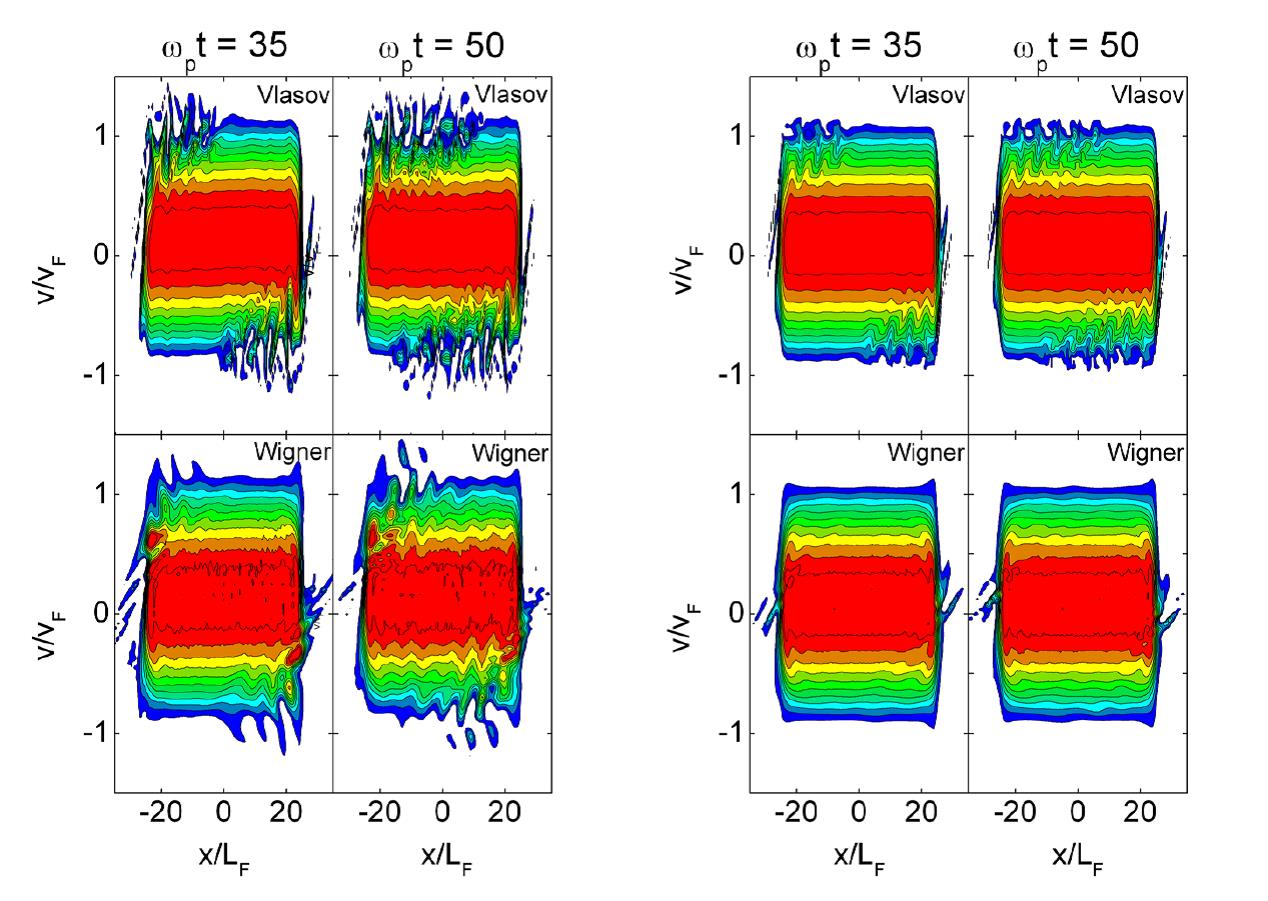}
\caption{Plots of the electron distribution function in the phase space at different times, $\omega_p t = 35$ and $\omega_p t = 50$, for a large excitation $\delta v = 0.15v_F$ (left panel) and a small excitation $\delta v = 0.06 v_F$ (right panel). Top panels: classical results; bottom panels: quantum results.} \label{fig:phasespace}
\end{figure}

This departure from the ballistic oscillation period constitutes a clear transition between the classical and the quantum regime. The estimation of the threshold value requires the investigation of the microscopic electron dynamics in the phase space.
In the classical case, we saw that the ballistic oscillations are due to bunches of nonequlibrium electrons traveling at the Fermi speed between the two surfaces of the film (Fig. \ref{PAH:phasespace}).
These bunches (i.e., vortices in the phase space) have a spatial extension roughly equal to $2\pi \lambda_{TF}$ and a width of the order of $\delta v$ in velocity space. The surface of the these vortices in the phase space (which has the dimension of an action) is thus approximately $A \approx 2\pi \lambda_{TF} m \delta v$. Quantum effects are expected to become significant when this action is of the same order as Planck's constant, i.e. $A \approx \hbar$. For sodium, this leads to the following estimate for the threshold: $\delta v_{th} \approx 0.15 v_F$, which is fairly close to the observed value (see Fig. \ref{fig:period}).
This is confirmed by the inspection of the classical and quantum phase-space portraits, as seen in Fig. \ref{fig:phasespace}. For the large excitation $\delta v = 0.15v_F$ the phase-space vortices are clearly visible in the quantum Wigner simulations and they resemble the analog vortices of the Vlasov case. In contrast, the vortices have completely disappeared in the low-excitation case ($\delta v = 0.06 v_F$), whereas they are still present in the corresponding Vlasov portraits.

So far, we focussed exclusively on the ultrafast (up to $\approx 100 \,\rm fs$) electron dynamics, while the ionic background was assumed to be frozen. In order to describe the long-time dynamics ($ >1\,\rm ps$) it is mandatory to include  electron-phonon interactions in the model (see Fig. \ref{fig:timescales} for a visual representation of the various timescales). In Ref. \cite{Jasiak2010}, we developed a microscopic phase-space model, based on the quantum Wigner distribution, which encompasses all relevant timescales from the femtosecond plasmon oscillations up to the phonon-mediated coupling to the ionic lattice, which occurs on a picosecond time scale. The main ingredients of the model are summarized in Sec. \ref{sec:collisions} in the paragraph devoted to Fokker-Planck methods.

For a system of quantum particles interacting with an environment, two distinct timescales are particularly important.
The {\em relaxation time} measures the speed at which the energy is exchanged between the electrons and the lattice, while the {\em decoherence time} represents the typical time over which quantum correlations are lost to the external environment (here, represented by the ion lattice).
In a density matrix language, the relaxation time corresponds to the decay of the diagonal terms, whereas the decoherence time is related to the nondiagonal terms of the density matrix.

Our simulations (see Ref. \cite{Jasiak2010} for details) showed that the relaxation time is of the order of $\tau_R \approx 2-3$ ps,  for the sodium films considered here. It also increases with the amplitude of the excitation, in agreement with measurements  performed on sodium clusters \cite{Maier2006}.

\begin{figure}
\centering
\includegraphics[scale=0.45]{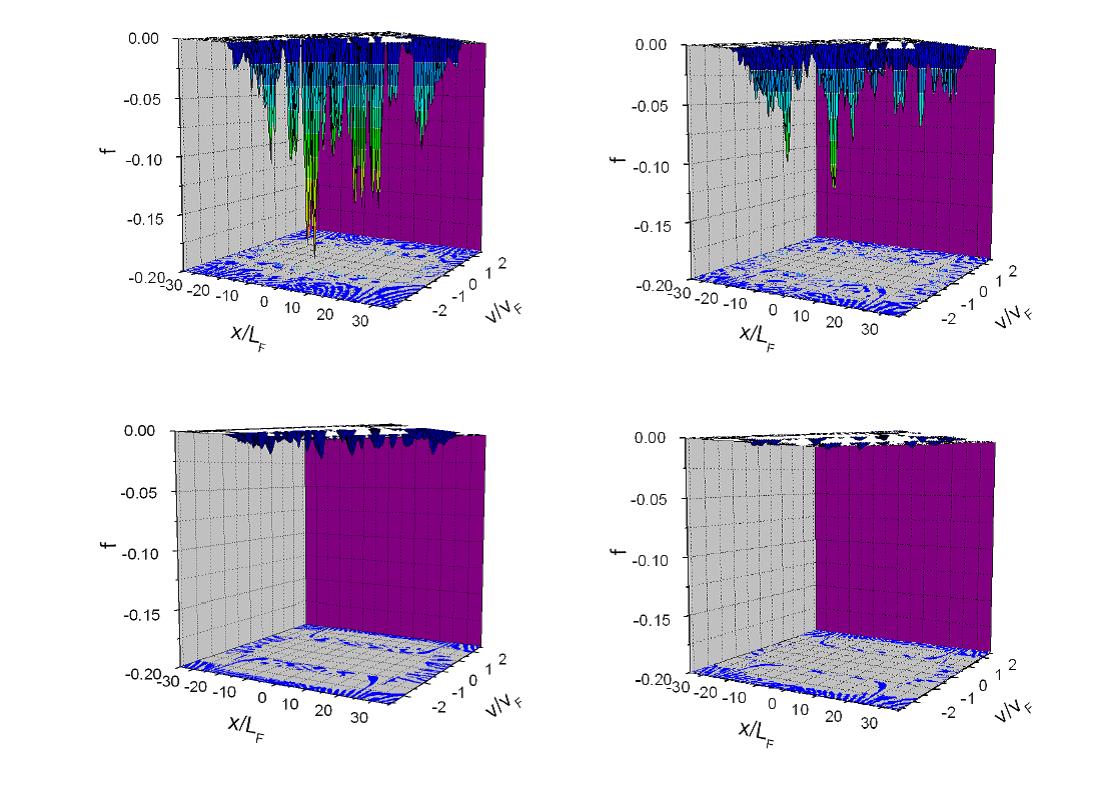}
\caption{Negative values of the Wigner distribution function in the phase space, at four different times $\omega_p t=1\,000, \,5\,000, \,10\,000$, and $20\,000$.} \label{fig:negative}
\end{figure}

In our phase-space representation, the decoherence time $\tau_D$ is defined in a different way compared to the density matrix approach.
We recall that the Wigner distribution function $f(x, v, t)$ can take negative values. The degree of ``classicality" of a Wigner distribution can thus be estimated from the weight of its negative parts \cite{Kenfack2004, Deleglise2008}.
The negative part of the Wigner function in the phase space is shown in Fig. \ref{fig:negative} at different times, for a typical case. As time goes on, the quantum distribution progressively loses its negative values and becomes more and more classical.

\begin{figure}
\centering
\includegraphics[scale=0.3]{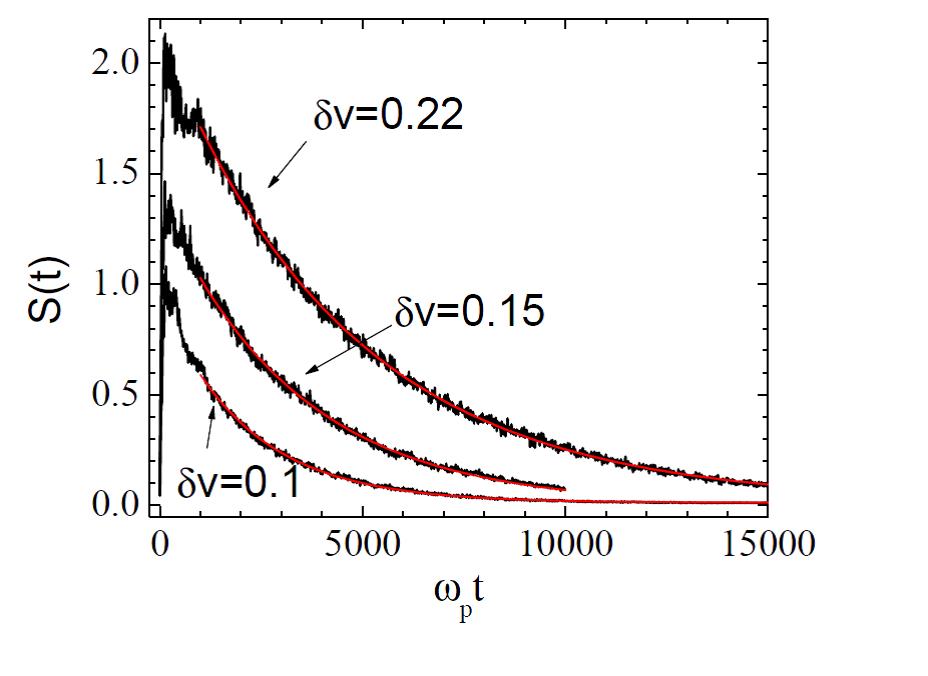}
\caption{Evolution of the total negative part of the Wigner function $S(t)$, for different excitations $\delta v$ (in units of $v_F$), with the corresponding exponential fits (red curves).} \label{fig:negtotal}
\end{figure}

In order to quantify more precisely this loss of classicality, we define the ``negativity" of the distribution function as $S(t) = \int\int f_<(x,v,t) dx dv$, where $f_<=-f$ if $f<0$ and $f_<=0$ elsewhere ($S$ is then a positive quantity).
The typical behavior of $S(t)$ is displayed in Fig. \ref{fig:negtotal}. The initial equilibrium distribution possesses only small negative parts, so the initial value of $S(0)$ is rather small. Soon after the excitation, when the system is driven out of equilibrium, $S$ grows very rapidly, then it decays more slowly to zero for longer times.

\begin{figure*}
\centering
\includegraphics[scale=0.6]{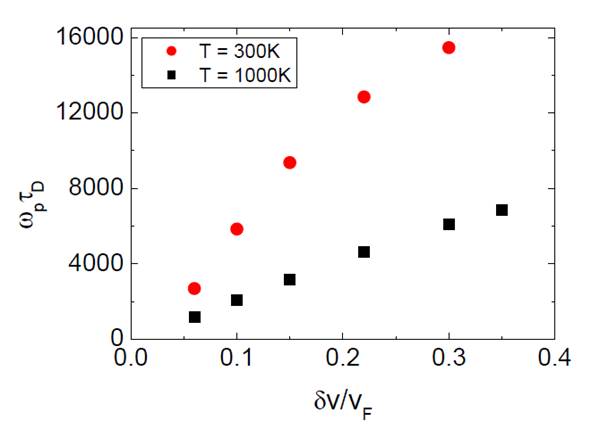}
\includegraphics[scale=0.5]{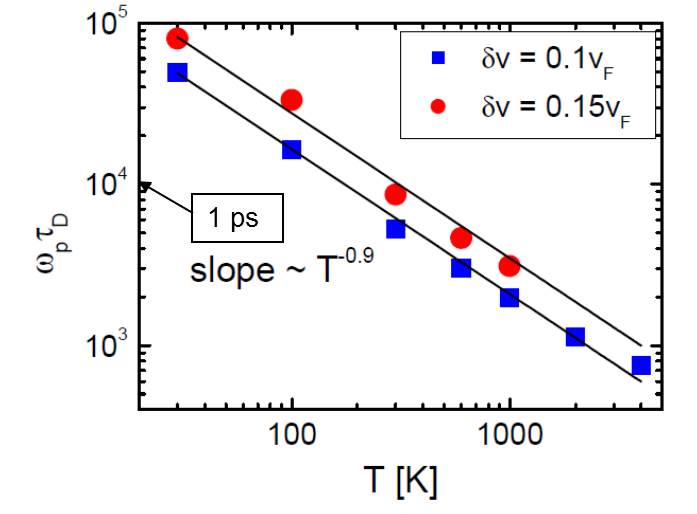}
\caption{Left panel: Decoherence time as a function of the excitation amplitude for two values of the lattice temperature. Right panel: Decoherence time as a function of the lattice temperature for two values of the excitation amplitude.} \label{fig:deco}
\end{figure*}

The decoherence time $\tau_D$ can be estimated by fitting $S(t)$ (after the initial sudden growth) with a simple decaying exponential function $S_{fit}(t) = S_0 \exp(-t/\tau_D)$.
We observed that the the decoherence time $\tau_D$ increases with the excitation energy and decreases with the lattice temperature (Fig. \ref{fig:deco}), also in agreement with measurements on thin metal films \cite{Komori1987}.
At room temperature ($T=300\,\rm K$) the decoherence time is roughly $\tau_D \approx 0.4\,{\rm ps}$, shorter than the relaxation time $\tau_R \approx 3\,\rm ps$.

\subsection{Thin metal films: Spin}
\label{subsec:filmspin}

In this section, we extend the above investigation of the electron dynamics in thin metal films by including the spin degrees of freedom. To this end, we turn to ferromagnetic materials and study the coupled charge and spin dynamics in nickel films.  We will use the  self-consistent spin-Vlasov model defined in Eqs. \eqref{vlasov equation f0 avec spin orbit}-\eqref{vlasov equation f avec spin orbit}. To keep the model as simple as possible, we restrict our problem to a reduced phase space ($x, v$), where the electrons are only allowed to move in the direction normal to the surfaces of the film ($x$). We only take into account electrostatic interactions through the Poisson equation and disregard effects due to exchange and correlations. The spin-orbit interaction, a higher order effect in $c^{-2}$, is also neglected.
With these assumptions, the set of spin-Vlasov-Poisson equations that we solve is the following (see Sec. \ref{subsec:wigspin}):
\begin{align}
\begin{array}{lll}
\displaystyle
\frac{\partial f_{0}}{\partial t}
+ v   \partial_{x} f_{0}
- \frac{e}{m} E_{x} \partial_{v} f_{0} - \frac{\mu_{B}}{m} \left(\partial_{x} B_{i}\right) \left( \partial_{v} f_{i} \right)  = 0,  \\  \\
\displaystyle
\frac{\partial f_{k}}{\partial t}
+ v  \partial_{x}f_{k}
-\frac{e}{m} E_{x} \partial_{v} f_{k} - \frac{\mu_{B}}{m} \left(\partial_{x} B_{k}\right) \left( \partial_{v} f_{0} \right)~-\frac{e}{m}  \left[\bm{B}  \times \bm{f}\right]_{k}
= 0,  \\  \\
\displaystyle
{\partial E_x \over {\partial x}}  = - \frac{e}{\epsilon_{0}} \left( \int f_{0}(x,v) dv -n_{i}(x) \right),
\end{array}
\label{spin-vlasov for nickel}
\end{align}
where $k=(x,y,z)$ and $E_x = -\partial_x \phi$.

The typical parameters of the nickel films under consideration, expressed in both SI and atomic units (a.u.), are as follows: $r_{s}=2.6~\rm a.u.= 0.14\,\rm nm$, $n_{0} = 0.0136 \rm~a.u. = 91.8 \rm~nm^{-3}$,  $T_{p} = 2 \pi \omega_p^{-1} = 15.32\rm~a.u. = 0.37\rm~fs$, $v_{F} = 0.74\rm ~a.u. = 1.62 \rm~nm. fs^{-1}$, $ \lambda_{TF} = 1.80\rm~a.u. =  0.095\rm~ nm$ and $E_{\textrm{F}} =  0.27\rm~a.u. = 7.34\rm~eV$.
We have studied films of thickness $L = 5\rm~ nm$.

In a realistic model of ferromagnetism, one should take into account the electronic structure of nickel, which comprises ten valence electrons of which eight are in the $3d$ shell and the remaining two are in the $4s$ shell.
The $3d$ electrons are more localized around the ions than the $4s$ electrons. Therefore, we model the $4s$ (``itinerant") electrons with the spin-Vlasov equations \eqref{spin-vlasov for nickel}, whereas the localized $3d$ electrons and the corresponding ions form a motionless positively-charged background.
The localized ions interact among themselves and with the itinerant electrons through  magnetic exchange. The internal magnetic fields involved in these interactions can be very strong, of the order of $10^3-10^4\, \rm T$, and are responsible for the ferromagnetic properties of nickel. A study of laser-matter interaction using such realistic model of itinerant and localized magnetism was used recently to study spin-current generation in a nickel film \cite{Hurst2018}.

Here, to illustrate the main effects of the spin dynamics, we shall use a simplified ``toy" model, where the internal magnetic field (generated by the magnetic ions) is replaced by a uniform external field of similar intensity directed along the $z$ axis: $\bm{B}=B \bm{e}_z$. This field will be responsible for the spin polarization of the electron gas. Hence, in this toy model the ions have no magnetic properties and are simply taken as a positively-charged density $n_i(x)$.
\begin{figure}
	\centering \includegraphics[scale=0.35]{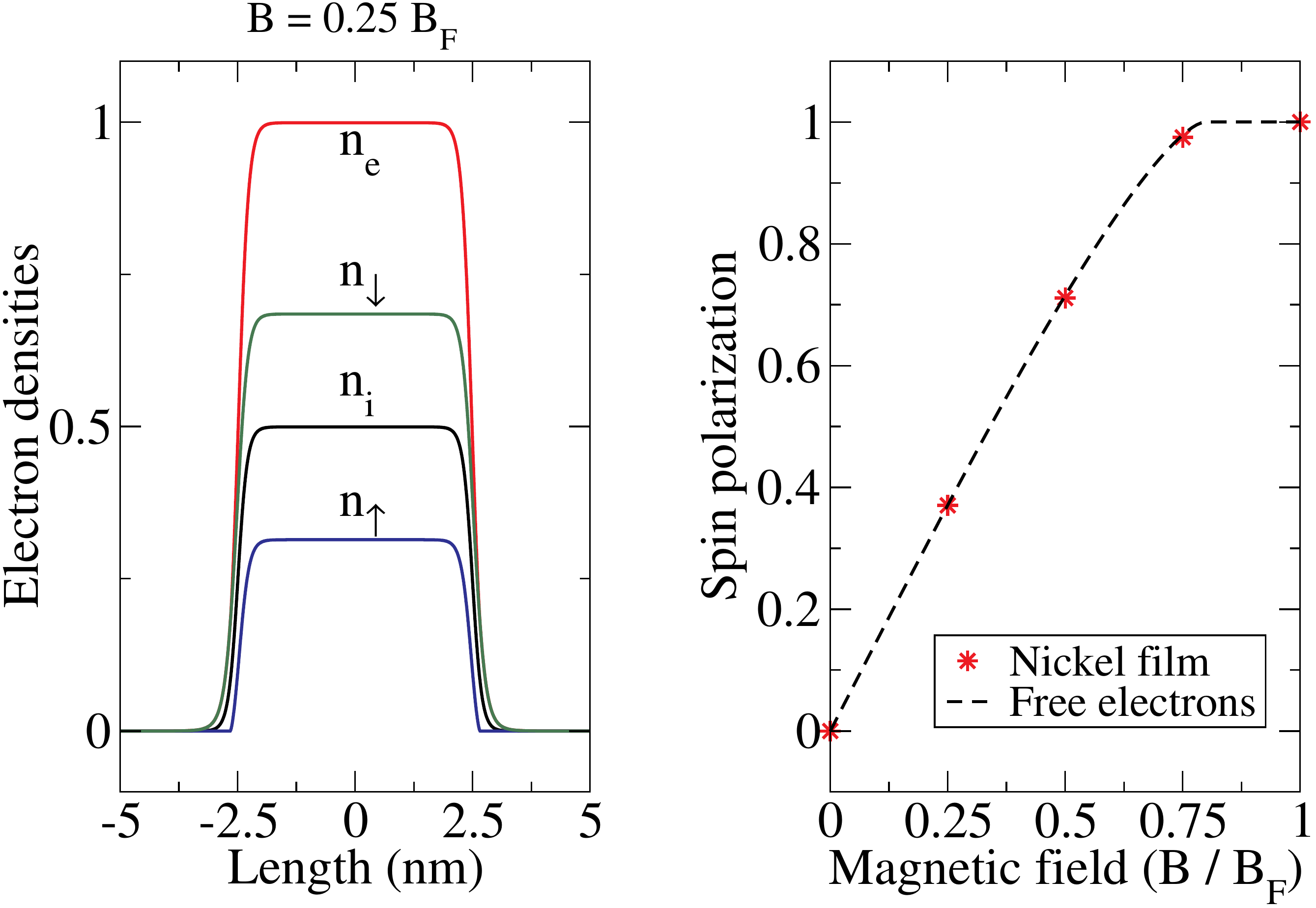}
    \caption{Self-consistent ground state for a 5\,nm nickel film. Left panel: Spin-resolved electron and ion densities for an external magnetic field $B = 0.25 B_F$. Right panel: Spin polarization of the electrons in a nickel film (red stars) as a function of the magnetic field; the black dashed line represents the theoretical prediction for a free electron gas at zero temperature.} \label{figure spin polarized ground state}
\end{figure}

The ground state of the system is given by a 1D Fermi-Dirac equilibrium, see Eqs. \eqref{eq:fd1d_1}-\eqref{eq:fd1d_2}. In practice, the ground state is obtained self-consistently by plugging Eqs. \eqref{eq:fd1d_1}-\eqref{eq:fd1d_2} into Poisson's equation and solving for $\phi$ (for instance, iteratively) and then injecting the obtained potential back into the Fermi-Dirac distributions. This procedure can be performed numerically and the resulting ground state is shown in Fig. \ref{figure spin polarized ground state} for different magnetic fields. As expected, when $B=0$ the electron gas is totally unpolarized, meaning that the spin-up and spin-down densities are identical. Increasing the magnetic field induces a partial polarization of the electron gas, which becomes fully polarized when $B$ approaches $B_F=E_F/\mu_B$. The relation between the magnetic field and the degree of polarization is in agreement with the free electron gas model at zero temperature.

To simulate a laser pulse excitation, we shift the ground state distributions $f_\uparrow$ and $f_\downarrow$ in the velocity space by an amount $\delta v = 0.05 v_F$. This is equivalent to applying an instantaneous electric field in the direction normal to the film surfaces. We performed several simulations starting from different ground states with increasing spin polarization. To characterize the spin dynamics, we study the time evolution of the magnetic dipole
\begin{align}
\langle X \rangle_m = \int\int x f_z(x,v,t)\, dx dv,
\end{align}
as well as the electric dipole
\begin{align}
\langle X \rangle_e = \int\int x f_0(x,v,t)\, dx dv .
\end{align}
Both dipoles vanish in the ground state.

\begin{figure}
	\centering \includegraphics[scale=0.35]{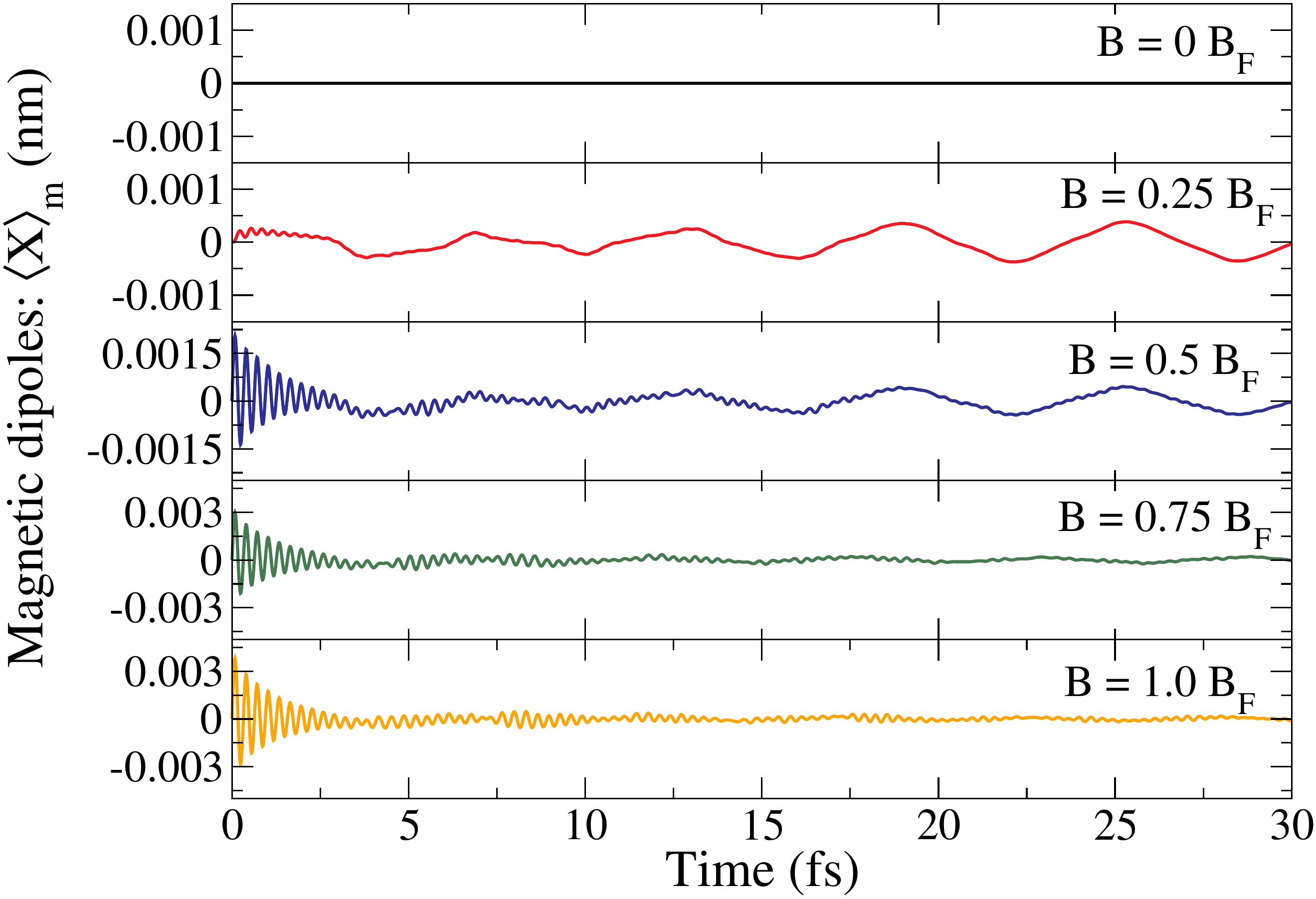}
    \caption{Time evolution of the magnetic dipole for several values of the external magnetic field $B$.} \label{magnetic dipole}
\end{figure}

The dynamics of the magnetic dipole is shown in Fig. \ref{magnetic dipole} for the different spin-polarized ground states. For an unpolarized electron gas ($B=0$), the magnetic dipole remains equal to zero, as the spin-up and spin-down distribution functions are identical for all times. For a partially polarized ground state, the magnetic dipole mainly oscillates at the ballistic frequency $\omega_b = 2\pi v_{F} /2L  = 2\pi/6.17$ fs$^{-1}$. These oscillations are similar to the ballistic oscillations observed in non-magnetic materials (see Secs. \ref{subsec:filmvlasov} and \ref{subsec:filmwigner}) and correspond to electrons traveling back and forth in the film at the Fermi speed.

The presence of a magnetic dipole implies that the spin-up and spin-down electron populations oscillate out of phase, because:
\[
\langle X \rangle_m =  \int\int x \left[f_\uparrow(x,v,t)-f_\downarrow(x,v,t)\right] \,dx dv.
\]
Since the spin-up and spin-down components were excited in phase in the initial excitation, we deduce that their dephasing occurs during the first instants of the dynamics (roughly the first 5\,fs in Fig. \ref{magnetic dipole}, before the magnetic dipole is clearly formed). One can show that this dephasing is a subtle effect due to the interplay of the self-consistent dynamics with the presence of strong electric fields at the boundaries of the film \cite{Hurst2018}.
When the electron gas is fully polarized ($B=B_F$), no magnetic dipole is observed, simply because only one component (spin-up or spin-down) is present.
Some faster oscillations at the plasma frequency $\omega_p = 2\pi/0.37$ fs$^{-1}$ are also observed, but are rapidly removed by Landau damping.
It is interesting to stress that, while the plasmon mode is quickly damped away, the magnetic mode at intermediate values of $B$ persists for much longer times.

Importantly, these ballistic oscillations of the magnetic dipole are still observed if one uses a more realistic model for both the magnetic material and the laser-film interaction \cite{Hurst2018}.
We also stress that an oscillating magnetic dipole is equivalent to an AC {\em spin current}, since: $d\langle  X \rangle _{m}/dt = \int v_{x} f_{z} dx dv_x  = \int v_x (f_{\uparrow}-f_{\downarrow})dx dv_x = J_{\uparrow}-J_{\downarrow} \equiv J^{S}_{xz}$,
where $J^{S}_{xz}$ is a spin current that propagates in the direction $x$ and is polarized along $z$ (in the general case ${\bm J}^S$ is a tensor, see \cite{Hurst2014}).
Thus the generation of a time-dependent magnetic dipole amounts to the generation of a spin current, which is an important issue for modern spintronic devices \cite{Alekhin2017}.

\begin{figure}
	\centering \includegraphics[scale=0.3]{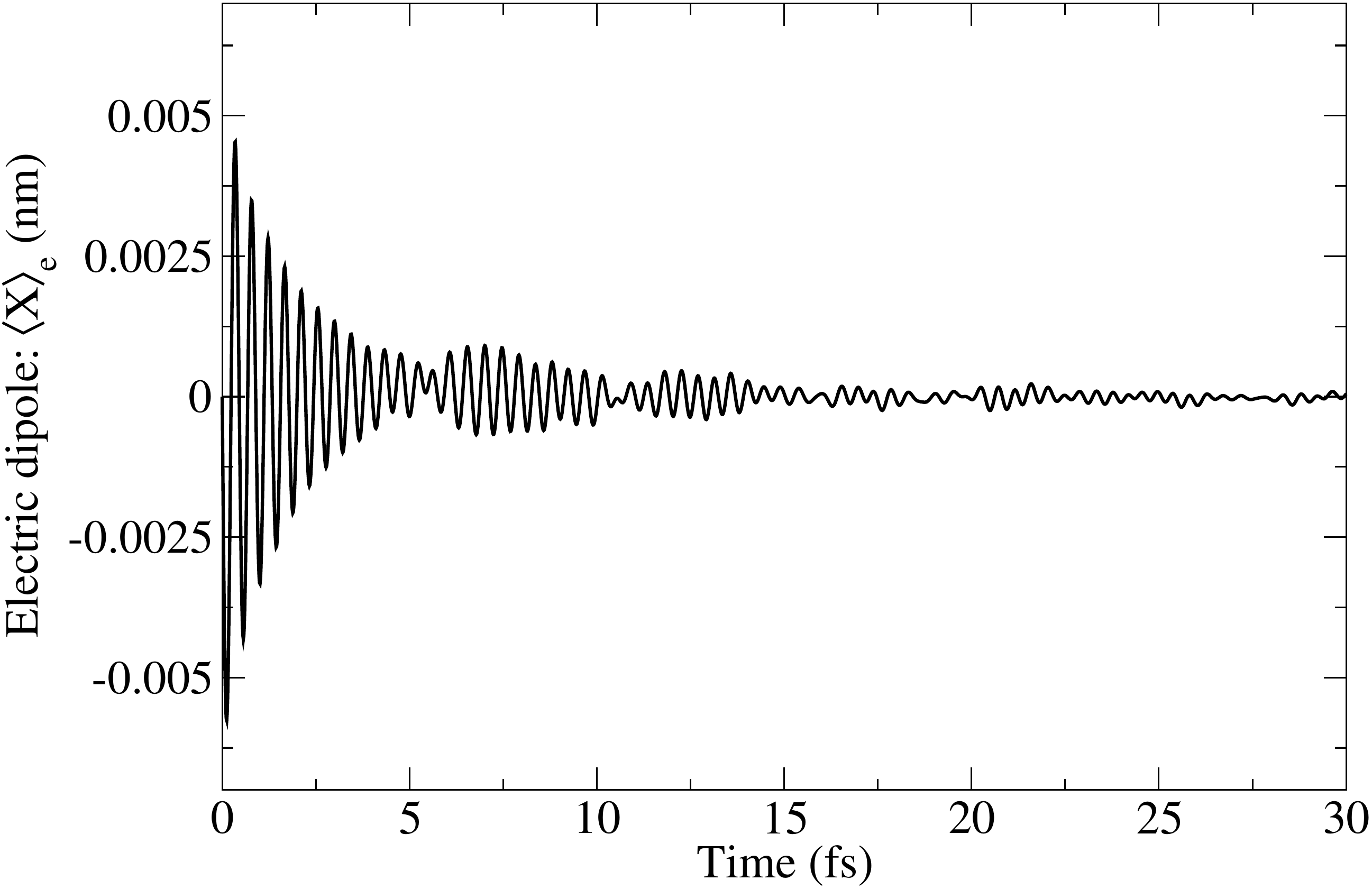}
    \caption{Time evolution of the electric dipole.} \label{edipole}
\end{figure}

Finally, in Fig. \ref{edipole} we show the time evolution of the electric dipole. Note that this is the same for all values of the magnetic field, because there is no backreaction of the spin dynamics on the charge dynamics in this toy model -- see Eqs. \eqref{spin-vlasov for nickel} with a constant magnetic field. The electric dipole shows prominent plasmon oscillations that decay away through Landau damping. A weak signature of the ballistic mode can be seen between 5--15\,fs.

\section{Conclusions}
\label{sec:conclusion}

A number of advanced methods to model the many-electron dynamics in nanometric condensed-matter systems have been developed over the years. In particular, TD-DFT has been developed to such a high level of accuracy to be capable of reproducing quantitatively many experimental results on large molecules and small nano-objects. Recent developments have extended its range of application to the spin dynamics \cite{Krieger2015}, shedding new light on old, but never properly understood, problems such as the ultrafast demagnetization occurring in ferromagnetic nano-objects illuminated with a laser pulse \cite{Beaurepaire1996}. All these methods are based on the propagation of a set of wave functions according to some Schr\"odinger-like equations.
Notwithstanding these successes, TD-DFT and Hartree-Fock methods remain computationally very costly for large nano-objects containing more than a few hundred electrons. They are also not easily compatible with dissipative dynamics, because of their intrinsically Hamiltonian (hence, unitary) character, in spite of recent attempts to include dissipation in a mean-field formalism \cite{Dinh2018}.

Here, we have made a case for an alternative approach based on Wigner's representation of quantum mechanics in the classical phase space.
Phase-space models are certainly not a panacea and remain computationally costly for large systems. But they do have some advantages, in particular: (i) they are better adapted for the semiclassical limit and to compare to classical results, (ii) they can profit from existing numerical methods, either grid-based or particle-based, which have been developed over the years most notably in the plasma physics community, (iii) they can be adapted to include non-unitary evolutions in order to model dissipative effects, and, last but not least, (iv) the phase-space representation can be a powerful visual aid to help intuition and to display the numerical results.

The main scope of this review was to summarize the theoretical basis of phase-space methods in condensed matter physics, particularly at the nanoscale (Secs. \ref{sec:intro}-\ref{sec:collisions}) and to illustrate these concepts through two relevant examples, namely the propagation of linear waves in an infinite medium (Sec. \ref{sec:linear}) and the nonlinear electron dynamics in thin metal films (Se. \ref{sec:films}).
Beyond this, we would like to increase the awareness of plasma physicists for the emerging field of nanoplasmonics, i.e.,
the study of collective effect arising from the interaction between electromagnetic radiation and free electrons in metallic nano-objects  \cite{Stockman2011,manfredi2018}, which is evolving  very rapidly since its beginnings in the 1980s.
The topics treated here show that the methods and tools of plasmas physics can be fruitfully extended and adapted to model typical plasma phenomena occurring in solid-state nanometric devices.
Given these common goals and approaches, the present review hopes to encourage future exchanges between the plasmonics and plasma physics communities.



\bibliographystyle{spmpsci}
\bibliography{biblio1}

\begin{thebibliography}{100}
\providecommand{\url}[1]{{#1}}
\providecommand{\urlprefix}{URL }
\expandafter\ifx\csname urlstyle\endcsname\relax
  \providecommand{\doi}[1]{DOI~\discretionary{}{}{}#1}\else
  \providecommand{\doi}{DOI~\discretionary{}{}{}\begingroup
  \urlstyle{rm}\Url}\fi

\bibitem{Aeschlimann2000}
Aeschlimann, M., Bauer, M., Pawlik, S., Knorren, R., Bouzerar, G., Bennemann,
  K.: {Transport and dynamics of optically excited electrons in metals}.
\newblock Applied Physics A: Materials Science {\&} Processing \textbf{71}(5),
  485--491 (2000).
\newblock \doi{10.1007/s003390000704}.
\newblock \urlprefix\url{http://link.springer.com/10.1007/s003390000704}

\bibitem{Alekhin2017}
Alekhin, A., Razdolski, I., Ilin, N., Meyburg, J.P., Diesing, D., Roddatis, V.,
  Rungger, I., Stamenova, M., Sanvito, S., Bovensiepen, U., Melnikov, A.:
  Femtosecond spin current pulses generated by the nonthermal spin-dependent
  seebeck effect and interacting with ferromagnets in spin valves.
\newblock Phys. Rev. Lett. \textbf{119}, 017202 (2017).
\newblock \doi{10.1103/PhysRevLett.119.017202}

\bibitem{Arnold1989}
Arnold, A., Steinr\"{u}ck, H.: {The 'electromagnetic' Wigner equation for an
  electron with spin}.
\newblock ZAMP Zeitschrift f\"{u}r angewandte Mathematik und Physik
  \textbf{40}(6), 793--815 (1989).
\newblock \doi{10.1007/BF00945803}.
\newblock \urlprefix\url{http://link.springer.com/10.1007/BF00945803}

\bibitem{Asenjo2012}
Asenjo, F.A., Zamanian, J., Marklund, M., Brodin, G., Johansson, P.:
  {Semi-relativistic effects in spin-1/2 quantum plasmas}.
\newblock New Journal of Physics \textbf{14}(7), 073042 (2012).
\newblock \doi{10.1088/1367-2630/14/7/073042}.
\newblock
  \urlprefix\url{http://stacks.iop.org/1367-2630/14/i=7/a=073042?key=crossref.dbb3da4cb34edf4c4dee78fc680140fb}

\bibitem{Banerjee2000}
Banerjee, A., Harbola, M.K.: {Hydrodynamic approach to time-dependent density
  functional theory; Response properties of metal clusters}.
\newblock http://dx.doi.org/10.1063/1.1290610  (2000).
\newblock \doi{10.1063/1.1290610}

\bibitem{Barletti2003}
Barletti, L.: {Wigner Envelope Functions for Electron Transport in
  Semiconductor Devices}.
\newblock Transport Theory and Statistical Physics \textbf{32}(3-4), 253--277
  (2003).
\newblock \doi{10.1081/TT-120024764}.
\newblock
  \urlprefix\url{http://www.tandfonline.com/doi/abs/10.1081/TT-120024764}

\bibitem{Battiato2010}
Battiato, M., Carva, K., Oppeneer, P.M.: {Superdiffusive Spin Transport as a
  Mechanism of Ultrafast Demagnetization}.
\newblock Physical Review Letters \textbf{105}(2), 027203 (2010).
\newblock \doi{10.1103/PhysRevLett.105.027203}.
\newblock
  \urlprefix\url{http://link.aps.org/doi/10.1103/PhysRevLett.105.027203}

\bibitem{Beaurepaire1996}
Beaurepaire, E., Merle, J.C., Daunois, A., Bigot, J.Y.: {Ultrafast Spin
  Dynamics in Ferromagnetic Nickel}.
\newblock Physical Review Letters \textbf{76}(22), 4250--4253 (1996).
\newblock \doi{10.1103/PhysRevLett.76.4250}.
\newblock \urlprefix\url{http://link.aps.org/doi/10.1103/PhysRevLett.76.4250}

\bibitem{Bertoni1999}
Bertoni, A., Bordone, P., Brunetti, R., Jacoboni, C.: {The Wigner function for
  electron transport in mesoscopic systems}.
\newblock Journal of Physics: Condensed Matter \textbf{11}(31), 5999--6012
  (1999).
\newblock \doi{10.1088/0953-8984/11/31/308}.
\newblock
  \urlprefix\url{http://stacks.iop.org/0953-8984/11/i=31/a=308?key=crossref.c57ce4e002585e0e87dab31014e4f108}

\bibitem{Bialynicki-Birula2014}
Bialynicki-Birula, I.: {Relativistic Wigner functions}.
\newblock EPJ Web of Conferences \textbf{78}, 01001 (2014).
\newblock \doi{10.1051/epjconf/20147801001}.
\newblock
  \urlprefix\url{http://www.epj-conferences.org/10.1051/epjconf/20147801001}

\bibitem{Bigot2000}
Bigot, J.Y., Halt{\'{e}}, V., Merle, J.C., Daunois, A.: {Electron dynamics in
  metallic nanoparticles}.
\newblock Chemical Physics \textbf{251}(1), 181--203 (2000).
\newblock \doi{10.1016/S0301-0104(99)00298-0}

\bibitem{Bigot2009}
Bigot, J.Y., Vomir, M., Beaurepaire, E.: {Coherent ultrafast magnetism induced
  by femtosecond laser pulses}.
\newblock Nature Physics \textbf{5}(7), 515--520 (2009).
\newblock \doi{10.1038/nphys1285}.
\newblock \urlprefix\url{http://www.nature.com/doifinder/10.1038/nphys1285}

\bibitem{Brewczyk1997}
Brewczyk, M., Rzazewski, K., Clark, C.W.: {Multielectron Dissociative
  Ionization of Molecules by Intense Laser Radiation}.
\newblock Physical Review Letters \textbf{78}(2), 191--194 (1997).
\newblock \doi{10.1103/PhysRevLett.78.191}.
\newblock \urlprefix\url{http://link.aps.org/doi/10.1103/PhysRevLett.78.191}

\bibitem{Brorson1962}
Brorson, S.D., Fujimoto, J.G., Ippen, E.P.: Femtosecond electronic
  heat-transport dynamics in thin gold films.
\newblock Phys. Rev. Lett. \textbf{59}, 1962--1965 (1987).
\newblock \doi{10.1103/PhysRevLett.59.1962}

\bibitem{Butet2010}
Butet, J., Duboisset, J., Bachelier, G., Russier-Antoine, I., Benichou, E.,
  Jonin, C., Brevet, P.F.: {Optical Second Harmonic Generation of Single
  Metallic Nanoparticles Embedded in a Homogeneous Medium}.
\newblock Nano Letters \textbf{10}(5), 1717--1721 (2010).
\newblock \doi{10.1021/nl1000949}.
\newblock \urlprefix\url{http://pubs.acs.org/doi/abs/10.1021/nl1000949}

\bibitem{Calvayrac2000}
Calvayrac, F., Reinhard, P.G., Suraud, E., Ullrich, C.: {Nonlinear electron
  dynamics in metal clusters}.
\newblock Physics Reports \textbf{337}(6), 493--578 (2000).
\newblock \doi{10.1016/S0370-1573(00)00043-0}

\bibitem{Cobley2009}
Cobley, C.M., Skrabalak, S.E., Campbell, D.J., Xia, Y.: Shape-controlled
  synthesis of silver nanoparticles for plasmonic and sensing applications.
\newblock Plasmonics \textbf{4}(2), 171--179 (2009).
\newblock \doi{10.1007/s11468-009-9088-0}.
\newblock \urlprefix\url{https://doi.org/10.1007/s11468-009-9088-0}

\bibitem{Cohen2013}
Cohen, L.: {The Weyl Operator and its Generalization}.
\newblock Springer Basel, Basel (2013).
\newblock \doi{10.1007/978-3-0348-0294-9}.
\newblock \urlprefix\url{http://link.springer.com/10.1007/978-3-0348-0294-9}

\bibitem{Crouseilles2008}
Crouseilles, N., Hervieux, P.A., Manfredi, G.: {Quantum hydrodynamic model for
  the nonlinear electron dynamics in thin metal films}.
\newblock Physical Review B \textbf{78}(15), 155412 (2008).
\newblock \doi{10.1103/PhysRevB.78.155412}.
\newblock \urlprefix\url{http://link.aps.org/doi/10.1103/PhysRevB.78.155412}

\bibitem{Daligault2016}
Daligault, J.: On the quantum landau collision operator and electron collisions
  in dense plasmas.
\newblock Physics of Plasmas \textbf{23}(3), 032706 (2016).
\newblock \doi{10.1063/1.4944392}

\bibitem{Daligault2003}
Daligault, J., Guet, C.: {Large amplitude femtosecond electron dynamics in
  metal clusters}.
\newblock Journal of Physics A: Mathematical and General \textbf{36}(22),
  5847--5855 (2003).
\newblock \doi{10.1088/0305-4470/36/22/304}.
\newblock
  \urlprefix\url{http://stacks.iop.org/0305-4470/36/i=22/a=304?key=crossref.b4ae0e27352e953b77967a36b50b0c6e}

\bibitem{Daniel2004}
Daniel, M.C., Astruc, D.: {Gold Nanoparticles: Assembly, Supramolecular
  Chemistry, Quantum-Size-Related Properties, and Applications toward Biology,
  Catalysis, and Nanotechnology}.
\newblock Chem. Rev. \textbf{104}, 239--346 (2004).
\newblock \doi{10.1021/cr030698+}

\bibitem{Deleglise2008}
Del{\'{e}}glise, S., Dotsenko, I., Sayrin, C., Bernu, J., Brune, M., Raimond,
  J.M., Haroche, S.: Reconstruction of non-classical cavity field states with
  snapshots of their decoherence.
\newblock Nature \textbf{455}(7212), 510--514 (2008).
\newblock \doi{10.1038/nature07288}

\bibitem{Dinh2018}
Dinh, P.M., Lacombe, L., Reinhard, P.G., Suraud, {\'E}., Vincendon, M.: On the
  inclusion of dissipation on top of mean-field approaches.
\newblock The European Physical Journal B \textbf{91}(10), 246 (2018).
\newblock \doi{10.1140/epjb/e2018-90147-0}

\bibitem{Dittrich2010}
Dittrich, T., G{\'{o}}mez, E.A., Pach{\'{o}}n, L.A.: {Semiclassical propagation
  of Wigner functions}.
\newblock The Journal of Chemical Physics \textbf{132}(21), 214102 (2010).
\newblock \doi{10.1063/1.3425881}.
\newblock \urlprefix\url{http://aip.scitation.org/doi/10.1063/1.3425881}

\bibitem{Dixit2013}
Dixit, A., Hinschberger, Y., Zamanian, J., Manfredi, G., Hervieux, P.A.:
  {Lagrangian approach to the semirelativistic electron dynamics in the
  mean-field approximation}.
\newblock Physical Review A \textbf{88}(3), 032117 (2013).
\newblock \doi{10.1103/PhysRevA.88.032117}.
\newblock
  \urlprefix\url{http://journals.aps.org/pra/abstract/10.1103/PhysRevA.88.032117}

\bibitem{Domps1998}
Domps, A., Reinhard, P.G., Suraud, E.: {Theoretical Estimation of the
  Importance of Two-Electron Collisions for Relaxation in Metal Clusters}.
\newblock Physical Review Letters \textbf{81}(25), 5524--5527 (1998).
\newblock \doi{10.1103/PhysRevLett.81.5524}.
\newblock \urlprefix\url{http://link.aps.org/doi/10.1103/PhysRevLett.81.5524}

\bibitem{Dragan2013}
Dragan, A., Odrzyg{\'{o}}{\'{z}}d{\'{z}}, T.: {A half-page derivation of the
  Thomas precession}.
\newblock American Journal of Physics \textbf{81}(8), 631 (2013).
\newblock \doi{10.1119/1.4807564}.
\newblock
  \urlprefix\url{http://link.aip.org/link/AJPIAS/v81/i8/p631/s1{\&}Agg=doi}

\bibitem{Drude1900}
Drude, P.: {Zur Elektronentheorie der Metalle}.
\newblock Annalen der Physik \textbf{306}(3), 566--613 (1900).
\newblock \doi{10.1002/andp.19003060312}.
\newblock \urlprefix\url{http://doi.wiley.com/10.1002/andp.19003060312}

\bibitem{Eguiluz1984}
Eguiluz, A.G., Campbell, D.A., Maradudin, A.A., Wallis, R.F.: Static response
  of a jellium surface: The image potential and indirect interaction between
  two charges.
\newblock Phys. Rev. B \textbf{30}, 5449--5459 (1984).
\newblock \doi{10.1103/PhysRevB.30.5449}

\bibitem{Ekici2008}
Ekici, O., Harrison, R.K., Durr, N.J., Eversole, D.S., Lee, M., Ben-Yakar, A.:
  Thermal analysis of gold nanorods heated with femtosecond laser pulses.
\newblock Journal of Physics D: Applied Physics \textbf{41}(18), 185501 (2008).
\newblock \doi{10.1088/0022-3727/41/18/185501}.
\newblock
  \urlprefix\url{https://doi.org/10.1088%2F0022-3727%2F41%2F18%2F185501}

\bibitem{Ekman2017}
Ekman, R., Asenjo, F.A., Zamanian, J.: Relativistic kinetic equation for
  spin-1/2 particles in the long-scale-length approximation.
\newblock Phys. Rev. E \textbf{96}, 023207 (2017).
\newblock \doi{10.1103/PhysRevE.96.023207}.
\newblock \urlprefix\url{https://link.aps.org/doi/10.1103/PhysRevE.96.023207}

\bibitem{Fock1930}
Fock, V.: {N{\"{a}}herungsmethode zur L{\"{o}}sung des quantenmechanischen
  Mehrk{\"{o}}rperproblems}.
\newblock Zeitschrift f{\"{u}}r Physik \textbf{61}(1-2), 126--148 (1930).
\newblock \doi{10.1007/BF01340294}.
\newblock \urlprefix\url{http://link.springer.com/10.1007/BF01340294}

\bibitem{Foldy1950}
Foldy, L.L., Wouthuysen, S.A.: {On the Dirac Theory of Spin 1/2 Particles and
  Its Non-Relativistic Limit}.
\newblock Physical Review \textbf{78}(1), 29--36 (1950).
\newblock \doi{10.1103/PhysRev.78.29}.
\newblock \urlprefix\url{http://link.aps.org/doi/10.1103/PhysRev.78.29}

\bibitem{Fomichev1999}
Fomichev, S.V., Zaretsky, D.F.: {Vlasov theory of Mie resonance broadening in
  metal clusters}.
\newblock Journal of Physics B: Atomic, Molecular and Optical Physics
  \textbf{32}(21), 5083--5102 (1999).
\newblock \doi{10.1088/0953-4075/32/21/303}.
\newblock
  \urlprefix\url{http://stacks.iop.org/0953-4075/32/i=21/a=303?key=crossref.4ee03a258a162ffb65776a3b9304d174}

\bibitem{Fourment2014}
Fourment, C., Deneuville, F., Descamps, D., Dorchies, F., Petit, S., Peyrusse,
  O., Holst, B., Recoules, V.: {Experimental determination of
  temperature-dependent electron-electron collision frequency in isochorically
  heated warm dense gold}.
\newblock PHYSICAL REVIEW B \textbf{89} (2014).
\newblock \doi{10.1103/PhysRevB.89.161110}

\bibitem{Fried1961}
Fried, B.D., Conte, S.D.: {The plasma dispersion function : the Hilbert
  transform of the Gaussian}.
\newblock Academic Press (1961)

\bibitem{Gerlach1922}
Gerlach, W., Stern, O.: {Der experimentelle Nachweis der Richtungsquantelung im
  Magnetfeld}.
\newblock Zeitschrift f{\"{u}}r Physik \textbf{9}(1), 349--352 (1922).
\newblock \doi{10.1007/BF01326983}.
\newblock \urlprefix\url{http://link.springer.com/10.1007/BF01326983}

\bibitem{Guillon2003}
Guillon, C., Langot, P., Fatti, N.D., Vall{\'{e}}e, F.: {Nonequilibrium
  electron energy-loss kinetics in metal clusters}.
\newblock New Journal of Physics \textbf{5}(1), 13--13 (2003).
\newblock \doi{10.1088/1367-2630/5/1/313}.
\newblock
  \urlprefix\url{http://stacks.iop.org/1367-2630/5/i=1/a=313?key=crossref.3da95a1c614ebea110e3b81d49cf137b}

\bibitem{Gunnarsson1976}
Gunnarsson, O., Lundqvist, B.I.: {Exchange and correlation in atoms, molecules,
  and solids by the spin-density-functional formalism}.
\newblock Physical Review B \textbf{13}(10), 4274--4298 (1976).
\newblock \doi{10.1103/PhysRevB.13.4274}.
\newblock \urlprefix\url{http://link.aps.org/doi/10.1103/PhysRevB.13.4274}

\bibitem{Haas2011}
Haas, F.: {Quantum plasmas : an hydrodynamic approach}.
\newblock Springer (2011)

\bibitem{Haas2009}
Haas, F., Manfredi, G., Shukla, P.K., Hervieux, P.A.: {Breather mode in the
  many-electron dynamics of semiconductor quantum wells}.
\newblock Physical Review B \textbf{80}(7), 073301 (2009).
\newblock \doi{10.1103/PhysRevB.80.073301}.
\newblock \urlprefix\url{http://link.aps.org/doi/10.1103/PhysRevB.80.073301}

\bibitem{Hainfeld2004}
Hainfeld, J.F., Slatkin, D.N., Smilowitz, H.M.: {The use of gold nanoparticles
  to enhance radiotherapy in mice}.
\newblock Phys. Med. Biol. Phys. Med. Biol \textbf{49}(4904), 309--315 (2004).
\newblock \doi{10.1088/0031-9155/49/18/N03}.
\newblock \urlprefix\url{http://iopscience.iop.org/0031-9155/49/18/N03}

\bibitem{Hartree1928}
Hartree, D.R.: {The Wave Mechanics of an Atom with a Non-Coulomb Central Field.
  Part I. Theory and Methods}.
\newblock Mathematical Proceedings of the Cambridge Philosophical Society
  \textbf{24}(01), 89 (1928).
\newblock \doi{10.1017/S0305004100011919}.
\newblock
  \urlprefix\url{http://www.journals.cambridge.org/abstract{\_}S0305004100011919}

\bibitem{Heller1976}
Heller, E.J.: {Wigner phase space method: Analysis for semiclassical
  applications}.
\newblock The Journal of Chemical Physics \textbf{65}(4), 1289--1298 (1976).
\newblock \doi{10.1063/1.433238}.
\newblock \urlprefix\url{http://aip.scitation.org/doi/10.1063/1.433238}

\bibitem{Hinschberger2012}
Hinschberger, Y., Hervieux, P.A.: {Foldy–Wouthuysen transformation applied to
  the interaction of an electron with ultrafast electromagnetic fields}.
\newblock Physics Letters A \textbf{376}(6), 813--819 (2012).
\newblock \doi{10.1016/j.physleta.2012.01.023}

\bibitem{Hohenberg1964}
Hohenberg, P., Kohn, W.: {Inhomogeneous Electron Gas}.
\newblock Physical Review \textbf{136}(3B), B864--B871 (1964).
\newblock \doi{10.1103/PhysRev.136.B864}.
\newblock \urlprefix\url{http://link.aps.org/doi/10.1103/PhysRev.136.B864}

\bibitem{Hurst2017}
Hurst, J., Hervieux, P.A., Manfredi, G.: Phase-space methods for the spin
  dynamics in condensed matter systems.
\newblock Philosophical Transactions of the Royal Society A: Mathematical,
  Physical and Engineering Sciences \textbf{375}, 20160199 (2017).
\newblock \doi{10.1098/rsta.2016.0199}.
\newblock \urlprefix\url{https://doi.org/10.1098/rsta.2016.0199}

\bibitem{Hurst2018}
Hurst, J., Hervieux, P.A., Manfredi, G.: Spin current generation by ultrafast
  laser pulses in ferromagnetic nickel films.
\newblock Phys. Rev. B \textbf{97}, 014424 (2018).
\newblock \doi{10.1103/PhysRevB.97.014424}.
\newblock \urlprefix\url{https://link.aps.org/doi/10.1103/PhysRevB.97.014424}

\bibitem{Hurst2014}
Hurst, J., Morandi, O., Manfredi, G., Hervieux, P.A.: {Semiclassical Vlasov and
  fluid models for an electron gas with spin effects}.
\newblock The European Physical Journal D \textbf{68}(6), 176 (2014).
\newblock \doi{10.1140/epjd/e2014-50205-5}.
\newblock \urlprefix\url{http://arxiv.org/abs/1405.1184}

\bibitem{Hurst2014fluid}
Hurst, J., Morandi, O., Manfredi, G., Hervieux, P.A.: {Semiclassical Vlasov and
  fluid models for an electron gas with spin effects}.
\newblock The European Physical Journal D \textbf{68}(6), 176 (2014).
\newblock \doi{10.1140/epjd/e2014-50205-5}.
\newblock \urlprefix\url{http://arxiv.org/abs/1405.1184}

\bibitem{Jasiak2009}
Jasiak, R., Manfredi, G., Hervieux, P.A.: {Quantum-classical transition in the
  electron dynamics of thin metal films}.
\newblock New Journal of Physics \textbf{11}(6), 063042 (2009).
\newblock \doi{10.1088/1367-2630/11/6/063042}

\bibitem{Jasiak2010}
Jasiak, R., Manfredi, G., Hervieux, P.A.: {Electron thermalization and quantum
  decoherence in metal nanostructures}.
\newblock Physical Review B \textbf{81}(24), 241401 (2010).
\newblock \doi{10.1103/PhysRevB.81.241401}.
\newblock \urlprefix\url{http://link.aps.org/doi/10.1103/PhysRevB.81.241401}

\bibitem{Jones2015}
Jones, R.O.: {Density functional theory: Its origins, rise to prominence, and
  future}.
\newblock Reviews of Modern Physics \textbf{87}(3), 897--923 (2015).
\newblock \doi{10.1103/RevModPhys.87.897}.
\newblock \urlprefix\url{http://link.aps.org/doi/10.1103/RevModPhys.87.897}

\bibitem{Jones1985}
Jones, W., March, N.H.N.H.: {Theoretical solid state physics}.
\newblock Dover Publications (1985)

\bibitem{Kania}
Kaniadakis, G., Quarati, P.: Kinetic equation for classical particles obeying
  an exclusion principle.
\newblock Phys. Rev. E \textbf{48}, 4263--4270 (1993).
\newblock \doi{10.1103/PhysRevE.48.4263}.
\newblock \urlprefix\url{https://link.aps.org/doi/10.1103/PhysRevE.48.4263}

\bibitem{Kenfack2004}
Kenfack, A., yczkowski, K.: Negativity of the wigner function as an indicator
  of non-classicality.
\newblock Journal of Optics B: Quantum and Semiclassical Optics \textbf{6}(10),
  396--404 (2004).
\newblock \doi{10.1088/1464-4266/6/10/003}

\bibitem{Kohn1965}
Kohn, W., Sham, L.J.: {Self-Consistent Equations Including Exchange and
  Correlation Effects}.
\newblock Physical Review \textbf{140}(4A), A1133--A1138 (1965).
\newblock \doi{10.1103/PhysRev.140.A1133}.
\newblock \urlprefix\url{http://link.aps.org/doi/10.1103/PhysRev.140.A1133}

\bibitem{Komori1987}
Komori, F., Okuma, S., ichi Kobayashi, S.: Inelastic scattering time and
  metal-insulator transition in thick disordered bismuth films.
\newblock Journal of the Physical Society of Japan \textbf{56}(2), 691--696
  (1987).
\newblock \doi{10.1143/jpsj.56.691}

\bibitem{Kravanja2000}
Kravanja, P., {Van Barel}, M., Ragos, O., Vrahatis, M.N., Zafiropoulos, F.A.:
  {ZEAL: A mathematical software package for computing zeros of analytic
  functions}.
\newblock Computer Physics Communications \textbf{124}(124), 212--232 (2000).
\newblock \urlprefix\url{www.elsevier.nl/locate/cpc
  http://cpc.cs.qub.ac.uk/summaries/ADKW}

\bibitem{Kreibig1995}
Kreibig, U., Vollmer, M.: {Optical properties of metal clusters}.
\newblock Springer (1995)

\bibitem{Krieger2015}
Krieger, K., Dewhurst, J.K., Elliott, P., Sharma, S., Gross, E.K.U.:
  {Laser-Induced Demagnetization at Ultrashort Time Scales: Predictions of
  TDDFT}.
\newblock Journal of Chemical Theory and Computation \textbf{11}(10),
  4870--4874 (2015).
\newblock \doi{10.1021/acs.jctc.5b00621}.
\newblock \urlprefix\url{http://pubs.acs.org/doi/10.1021/acs.jctc.5b00621}

\bibitem{Lamprecht1999}
Lamprecht, B., Krenn, J.R., Leitner, A., Aussenegg, F.R.: {Resonant and
  Off-Resonant Light-Driven Plasmons in Metal Nanoparticles Studied by
  Femtosecond-Resolution Third-Harmonic Generation}.
\newblock Physical Review Letters \textbf{83}(21), 4421--4424 (1999).
\newblock \doi{10.1103/PhysRevLett.83.4421}.
\newblock \urlprefix\url{http://link.aps.org/doi/10.1103/PhysRevLett.83.4421}

\bibitem{Landau1946}
Landau, L.D.: {On the vibrations of the electronic plasma}.
\newblock Zh. Eksp. Teor. Fiz. \textbf{10}, 25 (1946)

\bibitem{Lindblad1976}
Lindblad, G.: On the generators of quantum dynamical semigroups.
\newblock Communications in Mathematical Physics \textbf{48}(2), 119--130
  (1976).
\newblock \doi{10.1007/BF01608499}.
\newblock \urlprefix\url{https://doi.org/10.1007/BF01608499}

\bibitem{Liu2005}
Liu, X., Stock, R., Rudolph, W.: Ballistic electron transport in au films.
\newblock Phys. Rev. B \textbf{72}, 195431 (2005).
\newblock \doi{10.1103/PhysRevB.72.195431}

\bibitem{Loomba2013}
Loomba, L., Scarabelli, T.: {Metallic nanoparticles and their medicinal
  potential. Part II: aluminosilicates, nanobiomagnets, quantum dots and
  cochleates}.
\newblock Therapeutic Delivery \textbf{4}(9), 1179--1196 (2013).
\newblock \doi{10.4155/tde.13.74}.
\newblock \urlprefix\url{http://www.ncbi.nlm.nih.gov/pubmed/24024515
  http://www.future-science.com/doi/10.4155/tde.13.74}

\bibitem{Luo2013}
Luo, Y., Fernandez-Dominguez, A.I., Wiener, A., Maier, S.A., Pendry, J.B.:
  {Surface Plasmons and Nonlocality: A Simple Model}.
\newblock Physical Review Letters \textbf{111}(9), 093901 (2013).
\newblock \doi{10.1103/PhysRevLett.111.093901}.
\newblock
  \urlprefix\url{http://link.aps.org/doi/10.1103/PhysRevLett.111.093901}

\bibitem{Lyu2014}
Lyu, L.H.: {Elementary Space Plasma Physics}, second edi edn.
\newblock Taiwan, R.O.C. (2014)

\bibitem{Maier2006}
Maier, M., Wrigge, G., Hoffmann, M.A., Didier, P., Issendorff, B.v.:
  {Observation of Electron Gas Cooling in Free Sodium Clusters}.
\newblock Physical Review Letters \textbf{96}(11), 117405 (2006).
\newblock \doi{10.1103/PhysRevLett.96.117405}.
\newblock \urlprefix\url{http://link.aps.org/doi/10.1103/PhysRevLett.96.117405}

\bibitem{Manfredi2005}
Manfredi, G.: {How to model quantum plasmas}.
\newblock Fields Institute Communications Series \textbf{46}, 263--287 (2005).
\newblock \urlprefix\url{http://arxiv.org/abs/quant-ph/0505004}

\bibitem{Manfredi2013}
Manfredi, G.: {Non-relativistic limits of Maxwell's equations}.
\newblock European Journal of Physics \textbf{34}(4), 859--871 (2013).
\newblock \doi{10.1088/0143-0807/34/4/859}.
\newblock
  \urlprefix\url{http://stacks.iop.org/0143-0807/34/i=4/a=859?key=crossref.9f18d8a7a26a9511afdbb216fb09b214}

\bibitem{manfredi2018}
Manfredi, G.: Preface to special topic: Plasmonics and solid state plasmas.
\newblock Physics of Plasmas \textbf{25}(3), 031701 (2018).
\newblock \doi{10.1063/1.5026653}.
\newblock \urlprefix\url{https://doi.org/10.1063/1.5026653}

\bibitem{Manfredi2004}
Manfredi, G., Hervieux, P.A.: Vlasov simulations of ultrafast electron dynamics
  and transport in thin metal films.
\newblock Phys. Rev. B \textbf{70}, 201402 (2004).
\newblock \doi{10.1103/PhysRevB.70.201402}

\bibitem{Manfredi2005film}
Manfredi, G., Hervieux, P.A.: {Finite-size and nonlinear effects on the
  ultrafast electron transport in thin metal films}.
\newblock Physical Review B \textbf{72}(15), 155421 (2005).
\newblock \doi{10.1103/PhysRevB.72.155421}.
\newblock \urlprefix\url{http://link.aps.org/doi/10.1103/PhysRevB.72.155421}

\bibitem{Manfredi2005optics}
Manfredi, G., Hervieux, P.A.: Nonlinear absorption of ultrashort laser pulses
  in thin metal films.
\newblock Opt. Lett. \textbf{30}(22), 3090--3092 (2005).
\newblock \doi{10.1364/OL.30.003090}

\bibitem{Manfredi2012}
Manfredi, G., Hervieux, P.A., Haas, F.: {Nonlinear dynamics of
  electron-positron clusters}.
\newblock New Journal of Physics \textbf{14}(7), 075012 (2012).
\newblock \doi{10.1088/1367-2630/14/7/075012}.
\newblock
  \urlprefix\url{http://stacks.iop.org/1367-2630/14/i=7/a=075012?key=crossref.08f4b57085c2e5ac81fdb64d2cc0b36e}

\bibitem{Manfredi2010}
Manfredi, G., Hervieux, P.A., Yin, Y., Crouseilles, N.: Collective Electron
  Dynamics in Metallic and Semiconductor Nanostructures, pp. 1--44.
\newblock Springer Berlin Heidelberg, Berlin, Heidelberg (2010).
\newblock \doi{10.1007/978-3-642-04650-6_1}.
\newblock \urlprefix\url{https://doi.org/10.1007/978-3-642-04650-6_1}

\bibitem{Maniyara2019}
Maniyara, R.A., Rodrigo, D., Yu, R., Canet-Ferrer, J., Ghosh, D.S.,
  Yongsunthon, R., Baker, D.E., Rezikyan, A., de~Abajo, F.J.G., Pruneri, V.:
  Tunable plasmons in ultrathin metal films.
\newblock Nature Photonics \textbf{13}(5), 328--333 (2019).
\newblock \doi{10.1038/s41566-019-0366-x}.
\newblock \urlprefix\url{https://doi.org/10.1038/s41566-019-0366-x}

\bibitem{Marklund2010}
Marklund, M., Zamanian, J., Brodin, G.: {Spin Kinetic Theory—Quantum Kinetic
  Theory in Extended Phase Space}.
\newblock Transport Theory and Statistical Physics \textbf{39}(5-7), 502--523
  (2010).
\newblock \doi{10.1080/00411450.2011.566502}.
\newblock
  \urlprefix\url{http://www.tandfonline.com/doi/abs/10.1080/00411450.2011.566502}

\bibitem{Maurat2009}
Maurat, E., Hervieux, P.A.: Thermal properties of open-shell metal clusters.
\newblock New Journal of Physics \textbf{11}(10), 103031 (2009).
\newblock \doi{10.1088/1367-2630/11/10/103031}

\bibitem{Molina2002}
Molina, R.A., Weinmann, D., Jalabert, R.A.: {Oscillatory size dependence of the
  surface plasmon linewidth in metallic nanoparticles}.
\newblock Physical Review B \textbf{65}(15), 155427 (2002).
\newblock \doi{10.1103/PhysRevB.65.155427}.
\newblock \urlprefix\url{http://link.aps.org/doi/10.1103/PhysRevB.65.155427}

\bibitem{Morandi2010}
Morandi, O.: {Effective classical Liouville-like evolution equation for the
  quantum phase-space dynamics}.
\newblock Journal of Physics A: Mathematical and Theoretical \textbf{43}(36),
  365302 (2010).
\newblock \doi{10.1088/1751-8113/43/36/365302}.
\newblock
  \urlprefix\url{http://stacks.iop.org/1751-8121/43/i=36/a=365302?key=crossref.5a018cfc502129aa820183a577121009}

\bibitem{Morandi2009}
Morandi, O., Hervieux, P.A., Manfredi, G.: {Ultrafast magnetization dynamics in
  diluted magnetic semiconductors}.
\newblock New Journal of Physics \textbf{11}(7), 073010 (2009).
\newblock \doi{10.1088/1367-2630/11/7/073010}.
\newblock
  \urlprefix\url{http://stacks.iop.org/1367-2630/11/i=7/a=073010?key=crossref.c01ba3ff1fbda7b5b9b049075a1d9abf}

\bibitem{Morandi2011}
Morandi, O., Sch{\"{u}}rrer, F.: {Wigner model for quantum transport in
  graphene}.
\newblock Journal of Physics A: Mathematical and Theoretical \textbf{44}(26),
  265301 (2011).
\newblock \doi{10.1088/1751-8113/44/26/265301}.
\newblock
  \urlprefix\url{http://stacks.iop.org/1751-8121/44/i=26/a=265301?key=crossref.9c38a6baa4753b3171f66c02867efa02}

\bibitem{Moreau2012}
Moreau, A., Cirac{\`{i}}, C., Mock, J.J., Hill, R.T., Wang, Q., Wiley, B.J.,
  Chilkoti, A., Smith, D.R.: {Controlled-reflectance surfaces with film-coupled
  colloidal nanoantennas}.
\newblock Nature \textbf{492}(7427), 86--89 (2012).
\newblock \doi{10.1038/nature11615}.
\newblock \urlprefix\url{http://www.nature.com/doifinder/10.1038/nature11615}

\bibitem{Muller2004}
M{\"{u}}ller, T., Parz, W., Strasser, G., Unterrainer, K.: {Influence of
  carrier-carrier interaction on time-dependent intersubband absorption in a
  semiconductor quantum well}.
\newblock Physical Review B \textbf{70}(15), 155324 (2004).
\newblock \doi{10.1103/PhysRevB.70.155324}.
\newblock \urlprefix\url{http://link.aps.org/doi/10.1103/PhysRevB.70.155324}

\bibitem{Pereira2004}
Pereira, M., Wenzel, H.: {Interplay of Coulomb and nonparabolicity effects in
  the intersubband absorption of electrons and holes in quantum wells}.
\newblock Physical Review B \textbf{70}(20), 205331 (2004).
\newblock \doi{10.1103/PhysRevB.70.205331}.
\newblock \urlprefix\url{http://link.aps.org/doi/10.1103/PhysRevB.70.205331}

\bibitem{Pines1995}
Pines, D., Nozi\`{e}res, P.P.: {Theory of quantum liquids}.
\newblock Addison-Wesley Pub. Co., Advanced Book Program (1995)

\bibitem{Puente2000}
Puente, A., Casas, M., Serra, L.: {A semiclassical approach to the ground state
  and density oscillations of quantum dots}.
\newblock Physica E: Low-dimensional Systems and Nanostructures \textbf{8}(4),
  387--397 (2000).
\newblock \doi{10.1016/S1386-9477(99)00042-9}

\bibitem{Raza2013}
Raza, S., Stenger, N., Kadkhodazadeh, S., Fischer, S.V., Kostesha, N., Jauho,
  A.P., Burrows, A., Wubs, M., Mortensen, N.A.: {Blueshift of the surface
  plasmon resonance in silver nanoparticles studied with EELS}.
\newblock Nanophotonics \textbf{2}(2), 131--138 (2013).
\newblock \doi{10.1515/nanoph-2012-0032}.
\newblock
  \urlprefix\url{http://www.degruyter.com/view/j/nanoph.2013.2.issue-2/nanoph-2012-0032/nanoph-2012-0032.xml}

\bibitem{Rethfeld2002}
Rethfeld, B., Kaiser, A., Vicanek, M., Simon, G.: {Ultrafast dynamics of
  nonequilibrium electrons in metals under femtosecond laser irradiation}.
\newblock Physical Review B \textbf{65}(21), 214303 (2002).
\newblock \doi{10.1103/PhysRevB.65.214303}.
\newblock \urlprefix\url{http://link.aps.org/doi/10.1103/PhysRevB.65.214303}

\bibitem{Runge1984}
Runge, E., Gross, E.K.U.: {Density Functional Theory for Time Dependent
  Systems}.
\newblock Physical Review Letters \textbf{52}(12), 997--1000 (1984).
\newblock \doi{10.1103/PhysRevLett.52.997}.
\newblock \urlprefix\url{http://link.aps.org/doi/10.1103/PhysRevLett.52.997}

\bibitem{Salata2004}
Salata, O.: {Applications of nanoparticles in biology and medicine.}
\newblock Journal of nanobiotechnology \textbf{2}(1), 3 (2004).
\newblock \doi{10.1186/1477-3155-2-3}.
\newblock \urlprefix\url{http://www.ncbi.nlm.nih.gov/pubmed/15119954
  http://www.pubmedcentral.nih.gov/articlerender.fcgi?artid=PMC419715}

\bibitem{Scholl2012}
Scholl, J.A., Koh, A.L., Dionne, J.A.: {Quantum plasmon resonances of
  individual metallic nanoparticles}.
\newblock Nature \textbf{483}(7390), 421--427 (2012).
\newblock \doi{10.1038/nature10904}.
\newblock \urlprefix\url{http://www.nature.com/doifinder/10.1038/nature10904}

\bibitem{Schwengelbeck2000}
Schwengelbeck, U., Plaja, L., Roso, L., Jarque, E.C.: {Plasmon-induced photon
  emission from thin metal films}.
\newblock Journal of Physics B: Atomic, Molecular and Optical Physics
  \textbf{33}(8), 1653--1661 (2000).
\newblock \doi{10.1088/0953-4075/33/8/314}.
\newblock
  \urlprefix\url{http://stacks.iop.org/0953-4075/33/i=8/a=314?key=crossref.2a4aafb012676b13d9dd8431abeb8be2}

\bibitem{Serimaa1986}
Serimaa, O.T., Javanainen, J., Varr{\'{o}}, S.: {Gauge-independent Wigner
  functions: General formulation}.
\newblock Physical Review A \textbf{33}(5), 2913--2927 (1986).
\newblock \doi{10.1103/PhysRevA.33.2913}.
\newblock
  \urlprefix\url{http://journals.aps.org/pra/abstract/10.1103/PhysRevA.33.2913}

\bibitem{Serra2001}
Serra, L., Puente, A.: {Magnetic Thomas-Fermi-Weizs{\"{a}}cker model for
  quantum dots: A comparison with Kohn-Sham ground states}.
\newblock The European Physical Journal D \textbf{14}(1), 77--81 (2001).
\newblock \doi{10.1007/s100530170237}.
\newblock
  \urlprefix\url{http://www.springerlink.com/index/10.1007/s100530170237}

\bibitem{Shahbazyan2016}
Shahbazyan, T.V.: {Landau damping of surface plasmons in metal nanostructures}.
\newblock Physical Review B \textbf{94}(23), 235431 (2016).
\newblock \doi{10.1103/PhysRevB.94.235431}.
\newblock \urlprefix\url{http://link.aps.org/doi/10.1103/PhysRevB.94.235431}

\bibitem{Shukla2006}
Shukla, P.K., Eliasson, B.: {Formation and Dynamics of Dark Solitons and
  Vortices in Quantum Electron Plasmas}.
\newblock Physical Review Letters \textbf{96}(24), 245001 (2006).
\newblock \doi{10.1103/PhysRevLett.96.245001}.
\newblock \urlprefix\url{http://link.aps.org/doi/10.1103/PhysRevLett.96.245001}

\bibitem{Shukla2010}
Shukla, P.K., Eliasson, B.: {Nonlinear aspects of quantum plasma physics}.
\newblock Physics-Uspekhi \textbf{53}(1), 51--76 (2010).
\newblock \doi{10.3367/UFNe.0180.201001b.0055}.
\newblock
  \urlprefix\url{http://stacks.iop.org/1063-7869/53/i=1/a=R02?key=crossref.ddb065ad70b08d77f0fd46e65dfd923c}

\bibitem{Singhal2010}
Singhal, H., Ganeev, R.A., Naik, P.A., Srivastava, A.K., Singh, A., Chari, R.,
  Khan, R.A., Chakera, J.A., Gupta, P.D.: {Study of high-order harmonic
  generation from nanoparticles}.
\newblock Journal of Physics B: Atomic, Molecular and Optical Physics
  \textbf{43}(2), 025603 (2010).
\newblock \doi{10.1088/0953-4075/43/2/025603}.
\newblock
  \urlprefix\url{http://stacks.iop.org/0953-4075/43/i=2/a=025603?key=crossref.49a4d77b18731634172632dc777fb0db}

\bibitem{Slater1929}
Slater, J.C.: {The Theory of Complex Spectra}.
\newblock Physical Review \textbf{34}(10), 1293--1322 (1929).
\newblock \doi{10.1103/PhysRev.34.1293}.
\newblock \urlprefix\url{http://link.aps.org/doi/10.1103/PhysRev.34.1293}

\bibitem{Smithey1993}
Smithey, D.T., Beck, M., Raymer, M.G., Faridani, A.: {Measurement of the Wigner
  distribution and the density matrix of a light mode using optical homodyne
  tomography: Application to squeezed states and the vacuum}.
\newblock Physical Review Letters \textbf{70}(9), 1244--1247 (1993).
\newblock \doi{10.1103/PhysRevLett.70.1244}.
\newblock \urlprefix\url{http://link.aps.org/doi/10.1103/PhysRevLett.70.1244}

\bibitem{Snider1967}
Snider, R.F., Lewchuk, K.S.: {Irreversible Thermodynamics of a Fluid System
  with Spin}.
\newblock The Journal of Chemical Physics \textbf{46}(8), 3163--3172 (1967).
\newblock \doi{10.1063/1.1841187}.
\newblock \urlprefix\url{http://aip.scitation.org/doi/10.1063/1.1841187}

\bibitem{Stamenova2016}
Stamenova, M., Simoni, J., Sanvito, S.: {Role of spin-orbit interaction in the
  ultrafast demagnetization of small iron clusters}.
\newblock Physical Review B \textbf{94}(1), 014423 (2016).
\newblock \doi{10.1103/PhysRevB.94.014423}.
\newblock \urlprefix\url{http://link.aps.org/doi/10.1103/PhysRevB.94.014423}

\bibitem{Stockman2011}
Stockman, M.I.: {Nanoplasmonics: The physics behind the applications}.
\newblock Physics Today \textbf{64}(2), 39--44 (2011).
\newblock \doi{10.1063/1.3554315}.
\newblock
  \urlprefix\url{http://physicstoday.scitation.org/doi/10.1063/1.3554315}

\bibitem{Strange1998}
Strange, P.: {Relativistic quantum mechanics : with applications in condensed
  matter and atomic physics}.
\newblock Cambridge University Press (1998)

\bibitem{Stratonovich1957}
Stratonovich, R.L.: {On The Statistical Interpretation of Quantum Theory}
  \textbf{5}(32), 1483--1495 (1957)

\bibitem{Suarez1995}
Su{\'{a}}rez, C., Bron, W.E., Juhasz, T.: {Dynamics and Transport of Electronic
  Carriers in Thin Gold Films}.
\newblock Physical Review Letters \textbf{75}(24), 4536--4539 (1995).
\newblock \doi{10.1103/PhysRevLett.75.4536}.
\newblock \urlprefix\url{http://link.aps.org/doi/10.1103/PhysRevLett.75.4536}

\bibitem{Suh1991}
Suh, N.D., Feix, M.R., Bertrand, P.: Numerical simulation of the quantum
  liouville-poisson system.
\newblock Journal of Computational Physics \textbf{94}(2), 403 -- 418 (1991).
\newblock \doi{https://doi.org/10.1016/0021-9991(91)90227-C}

\bibitem{Sun16}
Sun, L., Chen, P., Lin, L.: Enhanced molecular spectroscopy via localized
  surface plasmon resonance.
\newblock In: M.T. Stauffer (ed.) Applications of Molecular Spectroscopy to
  Current Research in the Chemical and Biological Sciences, chap.~18.
  IntechOpen, Rijeka (2016).
\newblock \doi{10.5772/64380}.
\newblock \urlprefix\url{https://doi.org/10.5772/64380}

\bibitem{Tame2013}
Tame, M.S., McEnery, K.R., {\"{O}}zdemir, K., Lee, J., Maier, S.A., Kim, M.S.:
  {Quantum plasmonics}.
\newblock Nature Physics \textbf{9}(6), 329--340 (2013).
\newblock \doi{10.1038/nphys2615}.
\newblock \urlprefix\url{http://www.nature.com/doifinder/10.1038/nphys2615}

\bibitem{Tanjia2018}
Tanjia, F., Hurst, J., Hervieux, P.A., Manfredi, G.: Plasmonic breathing modes
  in ${\mathrm{c}}_{60}$ molecules: A quantum hydrodynamic approach.
\newblock Phys. Rev. A \textbf{98}, 043430 (2018).
\newblock \doi{10.1103/PhysRevA.98.043430}

\bibitem{TatsuroEndo2006}
Tatsuro, E., Kagan, K., Naoki, N., Ha~Minh, H., Do-Kyun, K., Yuji, Y., Koichi,
  N., Eiichi, T.: {Multiple LabelFree Detection of Antigen Antibody Reaction
  Using Localized Surface Plasmon Resonance Based Core Shell Structured
  Nanoparticle Layer Nanochip}  (2006).
\newblock \doi{10.1021/AC0608321}

\bibitem{Teperik2013}
Teperik, T.V., Nordlander, P., Aizpurua, J., Borisov, A.G.: {Robust
  Subnanometric Plasmon Ruler by Rescaling of the Nonlocal Optical Response}.
\newblock Physical Review Letters \textbf{110}(26), 263901 (2013).
\newblock \doi{10.1103/PhysRevLett.110.263901}.
\newblock
  \urlprefix\url{http://link.aps.org/doi/10.1103/PhysRevLett.110.263901}

\bibitem{Thomas1926}
Thomas, L.H.: {The Motion of the Spinning Electron}.
\newblock Nature \textbf{117}(2945), 514--514 (1926).
\newblock \doi{10.1038/117514a0}

\bibitem{Thomas1970}
Thomas, M.W., Snider, R.F.: {Boltzmann equation and angular momentum
  conservation}.
\newblock Journal of Statistical Physics \textbf{2}(1), 61--81 (1970).
\newblock \doi{10.1007/BF01009711}.
\newblock \urlprefix\url{http://link.springer.com/10.1007/BF01009711}

\bibitem{UU}
Uehling, E.A., Uhlenbeck, G.E.: Transport phenomena in einstein-bose and
  fermi-dirac gases. i.
\newblock Phys. Rev. \textbf{43}, 552--561 (1933).
\newblock \doi{10.1103/PhysRev.43.552}.
\newblock \urlprefix\url{https://link.aps.org/doi/10.1103/PhysRev.43.552}

\bibitem{Ullrich2000}
Ullrich, C.A., Reinhard, P.G., Suraud, E.: {Simplified implementation of
  self-interaction correction in sodium clusters}.
\newblock Physical Review A \textbf{62}(5), 053202 (2000).
\newblock \doi{10.1103/PhysRevA.62.053202}.
\newblock \urlprefix\url{http://link.aps.org/doi/10.1103/PhysRevA.62.053202}

\bibitem{Vignale97}
Vignale, G., Kohn, W.: Current-dependent exchange-correlation potential for
  dynamical linear response theory.
\newblock Phys. Rev. Lett. \textbf{77}, 2037--2040 (1996).
\newblock \doi{10.1103/PhysRevLett.77.2037}.
\newblock \urlprefix\url{https://link.aps.org/doi/10.1103/PhysRevLett.77.2037}

\bibitem{Vladimirov2011}
Vladimirov, S.V., Tyshetskiy, Y.O.: {On description of a collisionless quantum
  plasma}.
\newblock Physics-Uspekhi \textbf{54}(12), 1243--1256 (2011).
\newblock \doi{10.3367/UFNe.0181.201112g.1313}.
\newblock
  \urlprefix\url{http://stacks.iop.org/1063-7869/54/i=12/a=A03?key=crossref.acff246829e4ae4dbdf3f44959125b77}

\bibitem{vlasov1938}
Vlasov, A.A.: {On the oscillation properties of an electron gas}.
\newblock Zh. Eksp. Teor. Fiz. \textbf{8}, 291--318 (1938)

\bibitem{Voisin2000}
Voisin, C., Christofilos, D., {Del Fatti}, N., Vall{\'{e}}e, F., Pr{\'{e}}vel,
  B., Cottancin, E., Lerm{\'{e}}, J., Pellarin, M., Broyer, M.: {Size-Dependent
  Electron-Electron Interactions in Metal Nanoparticles}.
\newblock Physical Review Letters \textbf{85}(10), 2200--2203 (2000).
\newblock \doi{10.1103/PhysRevLett.85.2200}.
\newblock \urlprefix\url{http://link.aps.org/doi/10.1103/PhysRevLett.85.2200}

\bibitem{WeylHermann1928}
Weyl, H.: {Quantenmechanik und Gruppentheorie}.
\newblock Zeitschrift f{\"{u}}r Physik \textbf{46}(1-2), 1--46 (1927).
\newblock \doi{10.1007/BF02055756}

\bibitem{Wigner1932}
Wigner, E.: {On the Quantum Correction For Thermodynamic Equilibrium}.
\newblock Physical Review \textbf{40}(5), 749--759 (1932).
\newblock \doi{10.1103/PhysRev.40.749}.
\newblock
  \urlprefix\url{http://journals.aps.org/pr/abstract/10.1103/PhysRev.40.749}

\bibitem{Yin2009}
Yin, Y., Hervieux, P.A., Jalabert, R.A., Manfredi, G., Maurat, E., Weinmann,
  D.: {Spin-dependent dipole excitation in alkali-metal nanoparticles}.
\newblock Physical Review B \textbf{80}(11), 115416 (2009).
\newblock \doi{10.1103/PhysRevB.80.115416}.
\newblock \urlprefix\url{http://link.aps.org/doi/10.1103/PhysRevB.80.115416}

\bibitem{Zamanian2010nj}
Zamanian, J., Marklund, M., Brodin, G.: {Scalar quantum kinetic theory for
  spin-1/2 particles: mean field theory}.
\newblock New Journal of Physics \textbf{12}(4), 043019 (2010).
\newblock \doi{10.1088/1367-2630/12/4/043019}.
\newblock
  \urlprefix\url{http://stacks.iop.org/1367-2630/12/i=4/a=043019?key=crossref.153368f55cfe1c0f5f8e618f46552dfd}

\bibitem{Zamanian2010}
Zamanian, J., Stefan, M., Marklund, M., Brodin, G.: {From extended phase space
  dynamics to fluid theory}.
\newblock Physics of Plasmas \textbf{17}(10), 102109 (2010).
\newblock \doi{10.1063/1.3496053}.
\newblock
  \urlprefix\url{http://scitation.aip.org/content/aip/journal/pop/17/10/10.1063/1.3496053}

\bibitem{Zaretsky2004}
Zaretsky, D.F., Korneev, P.A., Popruzhenko, S.V., Becker, W.: {Landau damping
  in thin films irradiated by a strong laser field}.
\newblock Journal of Physics B: Atomic, Molecular and Optical Physics
  \textbf{37}(24), 4817--4830 (2004).
\newblock \doi{10.1088/0953-4075/37/24/008}.
\newblock
  \urlprefix\url{http://stacks.iop.org/0953-4075/37/i=24/a=008?key=crossref.01fff65f63af7fb5e0c44ca491a20489}

\bibitem{Zurek2003}
Zurek, W.H.: Decoherence, einselection, and the quantum origins of the
  classical.
\newblock Rev. Mod. Phys. \textbf{75}, 715--775 (2003).
\newblock \doi{10.1103/RevModPhys.75.715}.
\newblock \urlprefix\url{https://link.aps.org/doi/10.1103/RevModPhys.75.715}

\end{thebibliography}

%
%

\end{document}